# Universität Bonn
## Physikalisches Institut

## Supersymmetric Integrable Hierarchies
## and
## String Theory

by

## Sonia Stanciu

### Abstract

This thesis is roughly organized into two parts. The first one (the first three chapters), expository in nature, attempts to place the current work in context: at first historically, but then focusing on the Lax formalism and the Adler–Gel'fand–Dickey scheme for hierarchies of the KdV type. The second part (the last four chapters) comprises the main body of this work. It begins by developing the supersymmetric Lax formalism, introducing the ring of formal superpseudodifferential operators and the associated Poisson structures. We discuss three supersymmetric extensions of the KP hierarchy (MRSKP, SKP$_2$, and JSKP). We define and compute their additional symmetries and we find that the algebra of additional symmetries are in all three cases isomorphic to the Lie algebra of superdifferential operators. We discuss a new reduction of SKP$_2$ and the relation between MRSKP and SKP$_2$ is clarified. Finally we consider the (so far) only integrable hierarchy to have played a role in noncritical superstring theory (sKdV-B). We identify it, prove its bihamiltonian integrability, and extend it by odd flows. We close with a discussion of new integrable supersymmetrizations of the KdV-like hierarchies suggested by the study of sKdV-B.

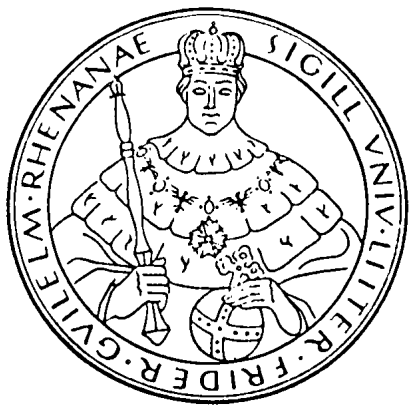

Post address:
Nußallee 12
D-53115 Bonn
Germany
e-mail:
sonia@avzw02.physik.uni-bonn.de

# SUMMARY OF CONTENTS

This thesis centers around the topics of integrable hierarchies and string theory. It is based on my papers [**69**], [**70**], [**52**], and [**65**] written in 1993 and 1994 at the University of Bonn, where I am graduate student, and at Queen Mary and Westfield of College (University of London) where I am visiting during the present academic year. It also contains some new material not yet published.

The thesis is organized as follows. The first three chapters are expository in nature. They attempt to place the current work in context: at first historically, but then focusing on more technical aspects. Thus, Chapter One briefly recounts the history of KdV-like systems from the time of its inception at the end of the last century, until its most recent avatar in two-dimensional quantum gravity and string theory. Chapter Two illustrates how the formalism used in the main body of the thesis fits within the conceptual framework of hamiltonian dynamical systems on (formal) Poisson manifolds. Then in Chapter Three we describe in detail the Lax formalism and the Adler–Gel'fand–Dickey scheme for hierarchies of KdV-type. Its purpose is mostly motivational but also serves to illustrate the difference between the supersymmetric and nonsupersymmetric theories.

The last four chapters comprise the main body of this work. Chapter Four develops the supersymmetric Lax formalism. It introduces the ring of formal superpseudodifferential operators and the associated Poisson structures. It also introduces three supersymmetric extensions of the KP hierarchy (MRSKP, SKP$_2$, and JSKP) to whose study Chapters Five and Six are devoted. In Chapter Five we find the additional symmetries of these supersymmetric KP hierarchies. We find that the algebra of additional symmetries are in all three cases isomorphic to the Lie algebra of superdifferential operators (also known as $\mathsf{SW}_{1+\infty}$). In Chapter Six we discuss a new reduction of SKP$_2$ and the relation between MRSKP and SKP$_2$ is clarified. Finally Chapter Seven is devoted to the study of sKdV-B—the (so far) only integrable hierarchy to have played a role in non-critical superstring theory. We identify the hierarchy, prove its bihamiltonian integrability, and extend it by odd flows. We close with a discussion of new integrable supersymmetrizations of the KdV-like hierarchies suggested by the study of sKdV-B.

# ACKNOWLEDGEMENTS

I would like to thank all those who have made it possible for me to write this thesis: my parents and my brother, Alexandru Petrescu who converted me to Physics, Petre Diţă and Vladimir Rittenberg for the chance given to study in Germany, Werner Nahm and Don Zagier for my first paper, Eduardo Ramos for my latest one, Chris Hull for offering me the possibility to visit Queen Mary and Westfield College. I am grateful to the theory group in Bucharest for encouragement and support, to the theory groups in Bonn and London for their hospitality, and to all my colleagues and friends who have been so helpful all these years. In particular, I would like to thank José M. Figueroa-O'Farrill who taught me most of what I know about integrable systems.

# TABLE OF CONTENTS

# Chapter One

# INTRODUCTION AND MOTIVATION

## 1.1. A BRIEF HISTORY OF THE KDV EQUATION

The KdV equation is the quintessential integrable system. In the twenty-seven years since computer experimentation began to reveal traces of its integrability, the KdV equation has served as an abundant source of results and inspiration to physicists and mathematicians in fields once as far apart as high energy physics and algebraic geometry. In its richness of structure it is comparable only to string theory, to which it is happily related. It is this very relation that motivates the present work; but the story of the KdV equation itself takes us back a while earlier...

### THE FIRST SOLITON

1834 was a remarkable year, for it represents the inception of solitons into recorded science. And the event could not have been more fortuitous:

> ' I was observing the motion of a boat which was rapidly drawn along a narrow channel by a pair of horses, when the boat suddenly stopped—not so the mass of water in the channel which it had put in motion; it accumulated round the prow of the vessel in a state of violent agitation, then suddenly leaving it behind, rolled forward with great velocity, assuming the form of a large solitary elevation, a rounded smooth and well-defined heap of water, which continued its course along the channel apparently without change of form or diminution of speed. I followed it on horseback, and overtook it still rolling on at a rate of some eight or nine miles an hour, preserving its original figure some thirty feet long and a foot to a foot and a half in height. Its height gradually diminished, and a after a chase of one or two miles I lost in in the windings of the channel. Such in the month of August 1834, was my first chance interview with that singular and beautiful phenomenon...'

Thus wrote John Scott-Russell ten years later in his report to the British Association for the Advancement of Science [1]. It would not be his last 'interview' with what are now termed solitons, for legend has it that managed managed to consistently reproduce this phenomenon in the Union canal, by having two horses drag a large wooden barge and then suddenly stopping.

### THE KDV EQUATION

Scott-Russell's excitement about his 'Wave of Translation' was not shared at first by his contemporaries. In fact his discovery was treated with scepticism—if not outright hostility—by Airy and by Stokes, who in 1849 published a 'proof' that such a wave could not exist (he later retracted). It was not until the 1870s that Scott-Russell's work became to be accepted by prominent scientists like Boussinesq and Rayleigh, both of whom knew—at least qualitatively—what we now know as the one-soliton solution to the KdV equation. The KdV equation itself appeared in 1895 in a paper by Korteweg and de Vries [2] who, apparently unaware of the work of Boussinesq and Rayleigh, offered their equation as a rebuttal to the early criticisms of Airy and Stokes. Korteweg and de Vries introduced the equation that now bears their name in order to model solitonic behavior mathematically. Their equation reads

$$\dot{u} = 6uu' + u''' \, , \qquad (1.1.1)$$

where $u = u(x,t)$ is a real valued function with faster than polynomial decay at spatial infinity $x \to \pm\infty$ and where $'$ and $\dot{}$ denote derivatives with respect to $x$ and $t$ respectively. Physically $u$ is the height of a water wave in a long and shallow canal. As it was intended from the start, the KdV equation does indeed possess solitonic solutions. Indeed, if we make the Ansatz $u(x,t) = w(x + c^2 t)$, then we find that $w(x) = \frac{1}{2}c^2 \operatorname{sech}^2(\frac{1}{2}cx)$ which gives us a pictorial idea of what Scott-Russell saw in the channel (see Fig. 1.1). The above solution is called the one-soliton solution. Notice that as the wave evolves in time its form does not change. Moreover the time evolution of the 'peak' of the soliton is linear with a speed proportional to the height. The effective dynamics of one KdV soliton are therefore (trivially) completely integrable.

### EARLY EVIDENCE FOR INTEGRABILITY

An analogous solution exists for two solitons. This is not a trivial fact, because the nonlinearity of the KdV equation destroys the superposition principle; but one can argue as follows. Since solitons decay fast at infinity and the bigger the soliton the faster it moves, it makes sense to consider an initial configuration (say, at large negative time) of two solitons of different sizes—the larger one to the right of the smaller one—and sufficiently spatially separated, that we may consider them as non-interacting. As we start the clock, the two solitons start to move independently: the fast one striving to overtake the slow one. As the solitons get closer, the behavior of the solution becomes complicated due to the nonlinearity; but the astonishing fact which emerged from computer simulations in the 1960s [**3**], is that if one waits long enough, the original solitons reappear with their original shapes and speeds (see Fig. 1.2). Apart from the complicated interacting behavior when the solitons meet, the only other remnant of the nonlinearity is the following. If the evolution had been linear, then for large positive $t$, the positions of the solitons would be the same as if there had been no interaction; but in the nonlinear case, the positions of the solitons are actually shifted: the larger soliton having gained some ground and the smaller soliton having lost some. In fact a closer look at Fig. 1.2 reveals that as the solitons merge, the larger one decreases in size and the smaller one increases in size in such a way that the interaction looks like a classical scattering process in which the solitons have exchanged their momentum.

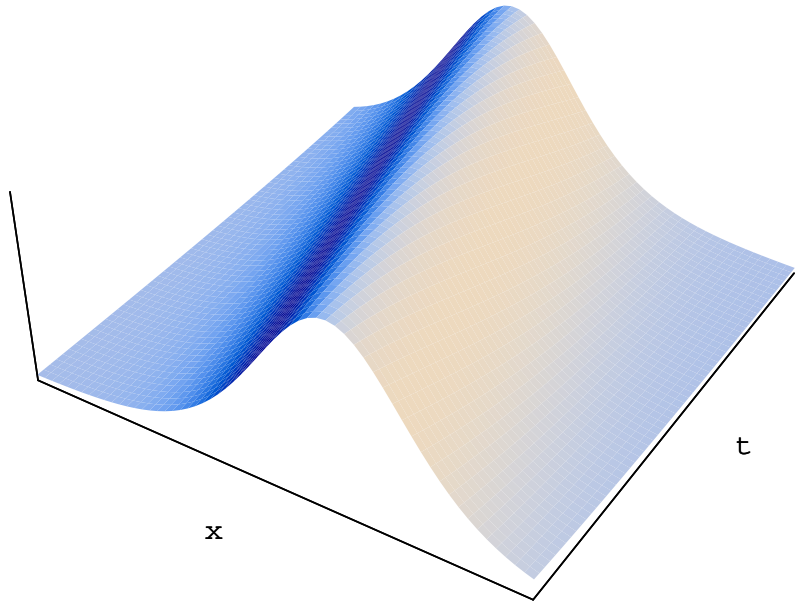

**Figure 1.1**    *The time evolution of a KdV soliton*

It is clear that the initial configurations of the two-soliton solution are parametrized by four numbers: the positions and the heights of the two solitons at a fixed large negative time. In fact, the effective dynamics of the two-soliton solution are governed by a completely integrable system in a four-dimensional phase space. The same is true for arbitrary $N$: there exist $N$-soliton solutions to the KdV equation which are effectively described by a completely integrable

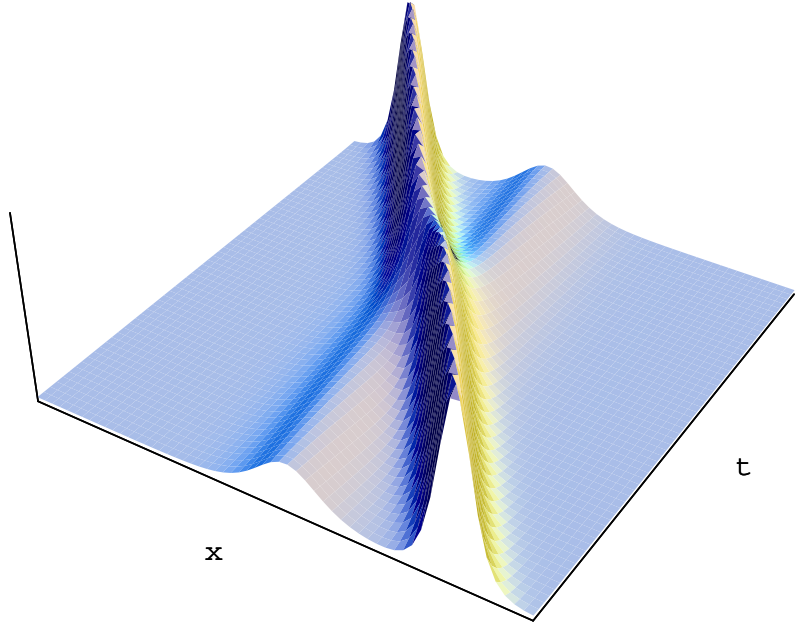

**Figure 1.2**   *A 2-soliton solution of the KdV equation*

system in a $2N$-dimensional phase space.

### THE MIURA TRANSFORMATION

Urged on by the existing numerical results, a number of people (Gardner, Kruskal and Miura among others) started the systematic investigation of the KdV equation as a potentially integrable system. It is clear that $H_1 = \int u$ is conserved, since the right-hand side of the KdV equation (1.1.1) is a total derivative. Similarly, multiplying (1.1.1) by $u$, one can conclude that $H_2 = \int u^2$ is also conserved. By the summer of 1967 there were three more charges known, all of which sharing the property that $H_n = \int p_n(u)$ where $p_n(u) = u^n + \cdots$ is a polynomial in $u$ and its spatial derivatives. That same summer, Robert Miura (then a graduate student) was sent out to find more conserved quantities for the KdV equation. Miura found a few more charges by hand before he discovered a remarkable transformation [**4**] relating solutions of the KdV equation to solutions of another nonlinear differential equation

$$\dot{v} = -6v^2 v' + v''' \, , \tag{1.1.2}$$

nowadays known as the modified KdV equation (mKdV). If $v(x,t)$ is a solution of the mKdV equation, then

$$u = -v^2 - v' \tag{1.1.3}$$

is a solution of the KdV equation. This transformation was subsequently general-ized by Gardner, exploiting the formal Galilean invariance of the KdV equation[1], to derive an infinite number of polynomial conserved charges [**5**] [**6**].

In that same series of papers, a remarkable observation was made that was to have profound implications in the field. If we understand the Miura transformation (1.1.3) as a Riccati equation for $v$ and we linearize it by defining $v = \psi'/\psi$, then $u = -\psi''/\psi$. Performing a Galilean transformation with $c = 6\lambda$ (see previous footnote) on $u$, we find that $\psi$ obeys the one-dimensional time-independent Schrödinger equation with potential $-u$:

$$\psi'' + u\psi = \lambda\psi \ . \tag{1.1.4}$$

If we now let $u$ evolve in time according to the KdV equation, then it makes sense to ask how $\lambda$ and $\psi$ evolve. Remarkably, it turns out that $\lambda$ remains constant! In other words, the KdV flow (1.1.1) is an **isospectral deformation** of the Sturm-Liouville operator $\partial^2 + u$, where $\partial = \partial/\partial x$.

This observation gave rise to the inverse scattering method, by which $N$-soliton solutions of the KdV equation were found from scattering data [**7**]. It turns out that the scattering data associated to the potential has a very simple time evolution. To solve the KdV equation with boundary conditions $u(x)$ one simply solves the scattering problem for the potential $-u(x)$, one then evolves the scattering data, and finally one applies the inverse method to determine the new potential. Using the inverse scattering method, Zakharov and Faddeev [**8**] were the first to demonstrate that the KdV equation was completely integrable in the sense that it possesses action-angle variable, namely the scattering data. The inverse scattering method and its quantum counterpart has been the primary source of the modern theory of quantum groups (see, for example, [**9**]).

### THE KDV EQUATION AS A HAMILTONIAN SYSTEM

Parallel to these developments the hamiltonian side of the story was starting to unfold. In [**10**], Gardner proved that the KdV equation could be written in hamiltonian form relative to one of the polynomial conserved charges that had been known for some time. In fact, relative to the following Poisson bracket in the space of initial configurations $u(x)$:

$$\{u(x)\,,\,u(y)\}_1 = \delta'(x - y) \ , \tag{1.1.5}$$

and taking the conserved charge $H_3 = \int u^3 - \frac{1}{2}(u')^2$ as hamiltonian, the KdV

---

[1] One can see that the transformations $u \mapsto u - \frac{1}{6}c$, $x \mapsto x + ct$, $t \mapsto t$ leave the KdV equation (1.1.1) invariant. We call this a formal invariance, however, since it does not preserve the boundary conditions at $x \to \pm\infty$.

equation can be written in hamiltonian form:

$$\dot{u} = \{u\,,\,H_3\} = \left(\frac{\delta H_3}{\delta u}\right)' \ . \tag{1.1.6}$$

Furthermore, relative to the Gardner bracket (1.1.5) all the polynomial conserved charges $H_n$ are in involution $\{H_i\,,\,H_j\} = 0$. This fact established the (formal) integrability of the KdV equation.

There is more, however. In [**11**] Magri discovered that the KdV equation could be written in hamiltonian form relative to a second bracket and relative to a second hamiltonian. In this case the bracket is given by

$$\{u(x)\,,\,u(y)\}_2 = (\tfrac{1}{2}\partial^3 + 2u\partial + u') \cdot \delta(x - y) \ , \tag{1.1.7}$$

and the hamiltonian is simply $H_2 = \int u^2$. Magri noticed that in addition to having a flow in common, both Poisson brackets are coordinated: that is, that any linear combination $\alpha\,\{-\,,\,-\}_1 + \beta\,\{-\,,\,-\}_2$ is again a Poisson bracket. This condition is of course nontrivial, since the Jacobi identities are quadratic. This bihamiltonian structure for the KdV equation implies a series of relations between the conserved charges. In fact, for all $n \geq 1$, one has the following relation:

$$\partial \cdot \frac{\delta H_{n+1}}{\delta u} = (\tfrac{1}{2}\partial^3 + 2u\partial + u') \cdot \frac{\delta H_n}{\delta u} \ , \tag{1.1.8}$$

which for $n = 2$ is precisely the fact that both sides of the equation equal the KdV equation. These relations, originally due to Lenard, can be used to recursively compute the conserved quantities starting from the trivial one $H_1 = \int u$.

It is interesting to notice that the Magri bracket is nothing but a representation of the (symmetric algebra of the) Virasoro algebra, as can be trivially seen by assuming that the field $u(x)$ lives on the circle and writing the induced Poisson brackets on the modes. In fact, this simple realization lies at the heart of much of the present research in this topic; especially in its relation with conformal field theory and string theory.

### THE HIROTA EQUATION

In 1972, Hirota introduced a remarkable trick to obtain soliton solutions to the KdV equation. This method has since shown itself very deep, but the idea is simple enough [**12**]. One introduces a potential $\tau$ related to $u$ as follows

$$u(x) = 2\frac{\partial^2}{\partial x^2} \log \tau(x) \ . \tag{1.1.9}$$

In terms of $\tau$ the KdV equation (1.1.1) becomes quadratic:

$$\tau'\dot{\tau} - \tau\dot{\tau}' + \tau\tau^{(4)} - 4\tau'\tau''' + (\tau'')^2 = 0 \ .\qquad(1.1.10)$$

This is known as the KdV equation in Hirota bilinear form and $\tau$ is known as the 'tau'-function. This equation may seem more complicated, but it has the nice property that it almost linearizes the KdV equation. Notice that $\tau(x) = 1$ is trivially a solution: it corresponds to $u(x) = 0$. Suppose that we now alter this solution by adding the exponential of an affine linear term: $\tau = 1 + \exp\phi(x,t)$ with $\phi(x,t) = kx + \omega t + \theta$. Then we find that $\tau$ obeys Hirota's equation (1.1.10) provided that $\omega = k^3$. Plugging this solution back into (1.1.9) we find that it is precisely the one-soliton solution. We can try to obtain a 2-soliton solution by adding another term $\tau = 1 + \exp\phi_1 + \exp\phi_2$, with $\phi_i(x,t) = k_i x + k_i^3 t + \theta_i$. We find that this is not a solution: the terms in $\exp 2\phi_i$ cancel but not the mixed terms $\exp(\phi_1 + \phi_2)$. We can try to cancel this term by adding yet an extra term to $\tau$ of the form $\exp(\phi_1 + \phi_2 + \Theta_{12})$, for some constant $\Theta_{12}$. Remarkably the new tau function

$$\tau(x,t) = 1 + e^{\phi_1} + e^{\phi_2} + e^{\phi_1 + \phi_2 + \Theta_{12}}\qquad(1.1.11)$$

is an exact solution of (1.1.10) provided that

$$\exp\Theta_{12} = \left(\frac{k_1 - k_2}{k_1 + k_2}\right)^2 \ .\qquad(1.1.12)$$

This goes on and one can obtain all $N$-soliton solutions in this fashion. The approach of Hirota not only facilitates the description of the soliton solutions, but it also makes contact—after the work of the Kyoto school—with the theory of infinite-dimensional Lie algebras. We shall comment briefly about this later on.

## 1.2. THE LAX FORMALISM

### LAX'S OBSERVATION

The fact that the eigenvalues of the Sturm-Liouville operator $L = \partial^2 + u$ remain constant provided $u$ evolves according to the KdV equation, was explained conceptually by Peter Lax [**13**] by showing that the KdV equation itself could be written in a manifestly isospectral form. If we take $P = 4\partial^3 + 6u\partial + 3u'$, then we can write the KdV equation in the form:

$$\dot{L} = \dot{u} = [P, L] \ . \tag{1.2.1}$$

This equation is remarkable in many ways. First of all, it displays the isospectral nature manifestly: $P$ is a skew-hermitian operator and equation (1.2.1) can be understood as the infinitesimal version of a unitary evolution

$$L(t) = U(t)^{-1}L(0)U(t) \tag{1.2.2}$$

with $P = -\dot{U}(0)$. But more importantly, it constitutes the ideal point from which to generalize. First of all, one can find other operators $P$ for which (1.2.1) makes sense. It is clear that not every operator $P$ will be consistent with (1.2.1), because $P$ must be such that its commutator with $L$ is a zeroth order differential operator. This is a highly restrictive fact; but nevertheless, one can find an infinite number of such operators—a fact that is intimately linked with the complete integrability of the KdV equation. Similarly, one can generalize this problem by considering other Lax operators $L$. In fact, one can define in this way a vast number of integrable hierarchies as isospectral deformations of a given Lax operator. For $L$ a differential operator of the form $L = \partial^n + \cdots$, the resulting hierarchy is known as the generalized $n$th order KdV hierarchy or simply $n$-KdV. We will have ample opportunity to discuss these hierarchies in Chapter Three; but let us just mention now the following beautiful and deep fact. The Miura transformation (1.1.3) can be understood in terms of the Lax operator $L$ as simply a formal factorization: $L = \partial^2 + u = (\partial + v)(\partial - v)$. This result, which appeared for the first time in [**14**], has now been generalized in a variety of ways—see, for example, [**15**], [**16**], [**17**], [**18**].

### the adler–gel'fand–dickey scheme

All the information concerning the spectrum of an operator is contained in its resolvent. For $\lambda$ not in the spectrum of the Lax operator $L = \partial^2 + u$, we define the resolvent by $R(\lambda) = (L-\lambda)^{-1}$. Gel'fand and Dickey, in a remarkable series of papers [**19**] [**20**] [**21**] demonstrated that the polynomial conserved charges of the KdV equation could be recovered from the asymptotic expansion (as $\lambda \to \infty$) of the resolvent $R(\lambda)$; and, in doing so, introduced the extremely useful concept of the fractional powers of $L$. They furthermore generalized the Gardner bracket (1.1.5) to other KdV-type equations. This generalization now bears their name: the first Gel'fand–Dickey bracket. Perhaps the only shortcoming of the approach of Gel'fand and Dickey is that they treated $L$ as an honest operator and as a result their work was full of the unavoidable analytic subtleties. It was Adler [**22**] who first noticed that their results could also be obtained in a completely algebraic fashion if one considered $L$ to be a formal differential operator instead.

Adler introduced the formal inverse $\partial^{-1}$ of the derivative operator $\partial$, which obeys

$$\partial^{-1} f = f \partial^{-1} - f' \partial^{-2} + f'' \partial^{-3} - \cdots \qquad (1.2.3)$$

for consistence with the Leibniz rule $\partial f = f \partial + f'$. One is then forced to extend the ring of differential operators to objects containing negative powers of $\partial$. In this ring of **(formal) pseudodifferential operators**, one can take the square root of $L$; that is, there exists a unique operator

$$L^{1/2} = \partial + v_1 + v_2 \partial^{-1} + v_3 \partial^{-2} + \cdots \qquad (1.2.4)$$

satisfying $L^{1/2} L^{1/2} = L$ and such that the $v_i$ are polynomials in $u$ and its derivatives. Furthermore Adler introduced a trace on the ring of formal pseudodifferential operators as follows. If $P = \sum p_i \partial^i$, then $\mathrm{Tr}\, P = \int p_{-1}$, which as the notation suggests, annihilates commutators. In terms of the Adler trace, one can write down all the conserved charges of the KdV equation simply as traces of fractional powers of the Lax operator

$$H_i = \mathrm{Tr}\, L^{i-1/2} \;, \qquad (1.2.5)$$

which are manifestly conserved since the evolution of $L$ and of any fractional power is given by a commutator (1.2.1). Moreover the possible operators $P$ in (1.2.1) defining isospectral deformations of $L$, can be written in terms of the fractional powers; in particular the KdV equation is given (up to trivial rescalings of $u$, $x$ and $t$) by

$$\dot{L} = [L_+^{3/2}, L] \;, \qquad (1.2.6)$$

where the subscript $_+$ denotes the differential part of a pseudodifferential operator. The Adler trace would be later extended by Wodzicki to pseudodifferential operators in arbitrary manifolds—a result of deep importance in many aspects of noncommutative geometry.

Two other important results were also contained in [**22**] concerning the hamiltonian structures of the KdV-type equations. On the one hand, the first Gel'fand–Dickey bracket was recognized as the Kirillov-Kostant bracket in a coadjoint orbit of the formal group of pseudodifferential operators of the form $1 + \sum_i a_i \partial^{-i}$—the Volterra group. This fact was independently observed by Lebedev and Manin [**23**], and puts the first Gel'fand–Dickey bracket on a solid conceptual framework. But perhaps more importantly, Adler conjectured a generalization of the Magri bracket (1.1.7), which has had a wide area of applicability outside the confines of the KdV-type equations. The Adler map was proven to be hamiltonian by Gel'fand and Dickey in [**24**] and the resulting Poisson bracket is known as the second Gel'fand–Dickey bracket. It lies at the heart of many results in W-algebras and conformal field theory. Most of the work in this thesis is framed in one way or another in the formalism developed by Adler and Gel'fand–Dickey. We will therefore spend considerable time developing it in Chapters Three and Four.

## 1.3. SOME IMPORTANT GENERALIZATIONS

### THE KP HIERARCHY

It follows from the Lax representation for the KdV equation that one can study the dynamical system defined by the isospectral deformations of more general differential operators. One of the virtues of the Adler–Gel'fand–Dickey scheme is that all these systems can be treated in parallel. The pivotal role played by the fractional powers of the Lax operator $L = \partial^n + \cdots$, and by its $n$th power $L^{1/n} = \partial + \cdots$ in particular, suggests the existence of a universal hierarchy containing all the other generalized KdV hierarchies. The idea is the following: the space of differential operators of the form $L = \partial^n + \cdots$ is in one-to-one correspondence with the space of pseudodifferential operators $\Lambda = \partial + \sum_i w_i \partial^{-i}$ whose $n$th power is differential. Since the spaces are isomorphic, they must have the same number of degrees of freedom and indeed the condition on $\Lambda$ singles out the first $n$ $w_i$ as independent. Clearly, the limit $n \to \infty$, which is not well-defined for $L$, makes perfect sense for $\Lambda$ and corresponds to the general pseudodifferential operator with all $w_i$ independent. The hierarchy of isospectral deformations of such an operator was introduced by the Kyoto school in the early 1980s **[25]** and is named the KP hierarchy after its first nontrivial equation—the equation introduced in the early 1970s by Kadomtsev and Petviashvili **[26]** as the simplest integrable extension of the KdV equation to $2 + 1$ dimensions.

The method of Hirota to solve the KdV equation can be extended to the KP hierarchy. It is here that the true depth of the idea reveals itself. We summarize the main result. The $\tau$-functions for the KP hierarchy are the points in an infinite-dimensional Grassmannian which has a very natural description in terms of two-dimensional quantum field theory. If one considers a complex fermion in two-dimensions, there is a canonical way to associate a $\tau$-function to the orbit through the vacuum of the infinite-dimensional group $GL(\infty)$ of invertible matrices. This group has a natural representation in terms of exponentials of fermion bilinears. Now bosonization will map the fermionic theory to a theory in one free boson in a way that commutes with the action of $GL(\infty)$. Thus we have a one-to-one correspondence between the $\tau$-functions of the KP hierarchy and the orbit of the vacuum in a bosonic Fock space. This correspondence can be made explicit as follows. The bosonic Fock space can be modeled as the space of polynomials in variables $t_1, t_2, \ldots$ corresponding to the creation modes of the boson—the vacuum being sent to the constant polynomial 1. To obtain other elements in the orbit of the vacuum we must simply bosonize exponentials of fermionic bilinears: but these are simply the vertex operators. Therefore we can obtain solitonic solutions of KP by acting with vertex operators on the bosonic vacuum. This accounts for the form of the two-soliton solution (1.1.11). The Hirota bilinear form now becomes the infinite-dimensional analog of the Plücker embedding of a Grassmannian in complex projective space. This circle of ideas is still very much under investigation and one can find a thorough discussion in [**27**].

Finally let us mention that the KP hierarchy shares many of the properties of the KdV-type hierarchies: integrability, bihamiltonian structure,... and lies at the heart of the study of (infinitely-generated) W-algebras of the $W_\infty$-type, whose role in string theory, quantum gravity, and even condensed matter physics is beginning to unfold.

A SUPERSYMMETRIC KDV EQUATION

One of the most remarkable symmetry principles to have appeared in recent times is that of supersymmetry: originally, the symmetry between bosons and fermions. For us it would be therefore interesting to see whether the KdV-type hierarchies admit supersymmetric extensions. A first step in this direction was taken by Kupershmidt [**28**] when he proposed a fermionic extension of the KdV equation. Nevertheless his equation—its bihamiltonian integrability notwithstanding—was not actually invariant under any supersymmetry. The first supersymmetric extension of the KdV hierarchy appeared in the seminal paper of Manin and Radul [**29**] on the supersymmetric KP hierarchy. This is a nonlinear partial differential equation for variables $u(x,t)$ and $\xi(x,t)$, where $u(x,t)$ (respectively $\xi(x,t)$) is a function taking values in the even (respectively odd) sector of an *a priori* infinitely-generated Grassmann algebra. The supersymmetric KdV (sKdV) equation of Manin and Radul now reads:

$$\begin{aligned}
\dot{u} &= 6uu' + u''' - 3\xi\xi'' \\
\dot{\xi} &= 3\xi u' + 3\xi' u + \xi''' ;
\end{aligned} \tag{1.3.1}$$

which reduces to the KdV equation upon putting $\xi = 0$. The sKdV equation is invariant under the following supersymmetry:

$$\delta\xi = u \text{ and } \delta u = \xi' , \tag{1.3.2}$$

which squares to an infinitesimal translation $\delta^2 = \partial$.

We can write the sKdV equation in a manifestly supersymmetric fashion by going over to superspace. Superspace is the natural arena for supersymmetry: it realizes supersymmetry geometrically in much the same way that Minkowski spacetime realizes Poincaré transformations. The relevant superspace in this case is $(1|1)$-dimensional with coordinates $(x, \theta)$: $\theta$ being the 'fermionic' partner of $x$. Because $\theta^2 = 0$ we can expand functions in superspace—traditionally known as superfields—as follows: $U(x, \theta) = \xi(x) + \theta u(x)$. Supersymmetry transformations are generated by the fermionic derivative $Q = \partial_\theta - \theta\partial$, so that $\delta U = \delta\xi - \theta\delta u = QU$. There is another fermionic derivative $D = \partial_\theta + \theta\partial$ which is supersymmetrically covariant—that is, anticommutes with the generator $Q$ of supersymmetry—and is hence called the supercovariant derivative. If we adopt the convention that on a superfield $U(x, \theta)$, $U' = DU$ and $U^{[i]} = D^i U$, then the sKdV equation (1.3.1) can be rewritten as follows

$$\dot{U} = 3(UU')'' + U^{[6]} \ . \tag{1.3.3}$$

This equation is manifestly supersymmetric precisely because $D$ is covariant, whence if $U$ transforms like a superfield, then so does $U'$.

The sKdV equation was shown by Mathieu [**30**] to be hamiltonian relative to a supersymmetric analog to the Magri bracket (1.1.7), which incidentally reproduces the $N = 1$ superVirasoro algebra on the modes. Its bihamiltonian structure was found by Oevel and Popowicz [**31**] and independently by Figueroa-O'Farrill, Mas, and Ramos [**32**], who proved that the sKdV analog of the Gardner bracket (1.1.5) is actually nonlocal. This is one of the many idiosyncrasies of supersymmetric integrable systems. Another idiosyncrasy is that contrary to the nonsupersymetric case, there is no natural supersymmetric KP hierarchy which is universal in the sense that the KP hierarchy is. There are many supersymmetric extensions of the KP hierarchy: the SKP hierarchy of Manin–Radul [**29**], the Jacobian SKP hierarchy [**33**] [**34**], and the even order SKP hierarchy or SKP$_2$ [**32**], among others. They are defined in Chapter Four and their study shall be a major theme of this thesis.

## 1.4. INTEGRABLE HIERARCHIES IN STRING THEORY

One hundred and fifty years after the discovery of the soliton, a different kind of wave would roll in to the shores of theoretical physics. In 1984, Green and Schwarz [**35**] discovered the now famous cancellation of anomalies in type II superstrings. It would take just five more years until both waves would come into contact. Indeed, the latest incarnation of the KdV equation is in string theory. This unexpected relation was brought about by the discovery that many of the properties of certain kinds of string theories are governed by the equations of the KdV-like hierarchies.

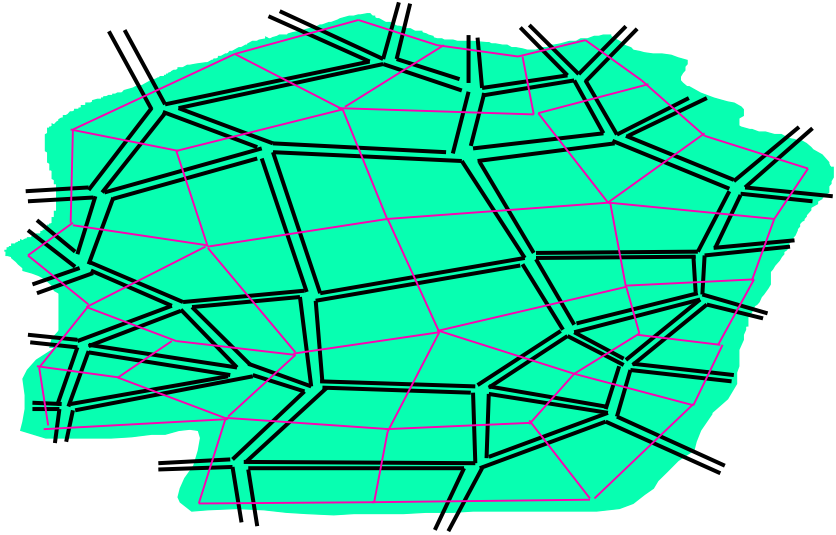

**Figure 1.3**  *A 'quadrangulated' surface and its dual graph*

A major problem in the standard perturbative approach to string theory (see [**36**] for a review) is that the topological expansion does not converge. Indeed, it was proven in [**37**] that the genus expansion for the bosonic string partition function $Z$ behaves as $Z \sim \sum_h (2h)!$, whence it is not even Borel summable. Matrix models were proposed to overcome this problem. The basic idea is the following. One first discretizes the worldsheet and substitutes the topological expansion and the integral over the metrics $g$ by a sum over all possible triangulations:

$$\sum_{\text{genera}} \int \mathcal{D}g \to \sum_{\text{triangulations}} \quad . \tag{1.4.1}$$

Each triangulation has a dual graph (see Fig. 1.3), whence the sum over triangulations can be substituted by a sum over the dual graphs. Remarkably the sum over graphs can be modeled by a finite-dimensional integral over the space of hermitian matrices. What makes this approach feasible is the discovery that this integral (or rather its free energy) can be computed recursively and that in the continuum limit the recursion relations are identical to those which the bihamiltonian structure imposes on the conserved charges of the KdV hierarchy.

More precisely, the partition function of the hermitian one-matrix model agrees in the continuum limit with a $\tau$-function of the KdV hierarchy satisfying an extra property known as the string equation. The string equation can be interpreted to say that the $\tau$-function is invariant under one of the additional symmetries of the KP hierarchy. These conditions translate into an infinite set of constraints on the partition function, enabling one to solve the theory completely—or at least up to a finite number of normalization constants. This remarkable correspondence persists between the hermitian $N$-matrix model and the $(N + 1)$-KdV hierarchy. This makes the study of additional symmetries of integrable hierarchies an important problem. We will devote Chapter Five to this theme.

Despite their success, matrix models pose two major theoretical challenges which have remained unmet. On the one hand, they seem to describe strings propagating in less than 1 dimension (the so-called $c$=1 barrier); and on the other hand, their extension to superstrings remains elusive despite many attempts to extend them. Nevertheless some progress has been made and at least one supersymmetric hierarchy has already made its appearance. Part of the work of this thesis is based in the identification of this hierarchy and in proving its integrability. This is done in Chapter Seven.

# Chapter Two

# HAMILTONIAN DYNAMICS AND INTEGRABILITY

The nature of the infinite-dimensional systems on which we shall focus our attention in this thesis is such, that they only afford some of the structure that we have come to expect from finite-dimensional hamiltonian dynamics. The remaining structure is nevertheless more than adequate to study these systems and the purpose of the present chapter is to motivate the formalism in a simple context. Departing from the familiar case of Hamilton's equations in $\mathbb{R}^{2n}$, we will arrive at an abstract definition of a hamiltonian dynamical system. To do so we must look closely at what is essential and what is superfluous in the usual formulation of hamiltonian dynamics. In the end, we will reach a formalism that is perfectly suited to the infinite-dimensional dynamical systems that we will study in the following chapters. Also in this chapter we briefly examine the notion of an integrable hierarchy and discuss the basics of hamiltonian reduction.

## 2.1. DYNAMICAL SYSTEMS ON POISSON MANIFOLDS

The usual arena of hamiltonian dynamics is symplectic geometry. Whereas this setting usually suffices, it is by no means necessary; and for the dynamical systems that will be the focus of this thesis, it is in fact too strong a requirement on our 'phase space'. The point of this section is then to extract the essential ingredients that make up a hamiltonian dynamical system.

Roughly speaking, there are two fundamental ingredients in the hamiltonian formulation of dynamics: one kinematical and one dynamical. The kinematical ingredient is the Poisson bracket, which allows us to associate a vector field with any function; and the dynamical ingredient is a choice of function (the hamiltonian) which via the Poisson bracket defines the time evolution. Manifolds admitting a Poisson bracket are called **Poisson manifolds**, and among them symplectic manifolds play a privileged role: they correspond to those Poisson manifolds with a nondegenerate Poisson bracket. The nature of the dynamical systems we will be discussing will force us to deviate from the customary in two important aspects. Firstly, we will abandon the symplectic category and settle for 'phase spaces' which are just Poisson manifolds. But also, we will have to trade our traditional geometric tools for others more algebraic which will be better suited to the dynamics on the infinite-dimensional spaces we will be considering. Poisson geometry (even of infinite-dimensional manifolds) is relatively well-understood, but for our purposes it is actually much more convenient to simply algebraize the relevant geometric notions. In doing so it may seem that we are leaving the 'substantial' and going into the formal; but this not the case at all. Indeed, the formalism is sufficiently general to allow one to specialize many of the results to the particular concrete situation.

Let us start with the familiar. Let us take as our phase space $M = \mathbb{R}^{2n}$ with coordinates $(q^i, p_i)$. Suppose we are given a function $H$ on $M$. Call it the hamiltonian. We can define a dynamical system starting from this data by imposing that the time evolution be governed by Hamilton's equations:

$$\dot{q}^i = \frac{\partial H}{\partial p_i} \qquad \dot{p}_i = -\frac{\partial H}{\partial q^i} \ . \tag{2.1.1}$$

Introducing the Poisson bracket of any two functions $f$ and $g$:

$$\{f\,,\,g\} = \sum_i \left( \frac{\partial f}{\partial q^i} \frac{\partial g}{\partial p_i} - \frac{\partial f}{\partial p_i} \frac{\partial g}{\partial q^i} \right) \ , \tag{2.1.2}$$

we can write Hamilton's equations simply as

$$\dot{f} = \{f\,,\,H\} \ , \tag{2.1.3}$$

for any function $f$ and, in particular, for the coordinates $(q^i, p_i)$. The Poisson bracket satisfies the following two properties. It is antisymmetric: $\{f\,,\,g\} = -\{g\,,\,f\}$; and it obeys the Jacobi identity: $\{f\,,\,\{g\,,\,h\}\} = \{\{f\,,\,g\}\,,\,h\} + \{g\,,\,\{f\,,\,h\}\}$. Antisymmetry is obvious from (2.1.2), whereas the Jacobi identity follows after a simple computation. Therefore the Poisson bracket defines the structure of a Lie algebra on the functions of $M$. More is true, however. Functions form a ring, and the Poisson bracket associates with every function on $M$ a derivation; that is, if $f$, $g$, and $h$ are functions on $M$, then

$$\{f\,,\,gh\} = g\,\{f\,,\,h\} + \{f\,,\,g\}\,h \ . \tag{2.1.4}$$

These facts make the functions on $M$ into a **Poisson algebra**. The derivation property allows one to give a more conceptual proof of the Jacobi identity. Simply notice that on the linear functions the Jacobi identity is obvious since the Poisson bracket of any pair of linear functions is a constant. Then one simply uses (2.1.4) to propagate the Jacobi identity to arbitrary functions.

Suppose that we now change coordinates to $x^i(q,p)$. The fundamental Poisson bracket of these coordinates is given by

$$\mho^{ij}(x) \equiv \left\{ x^i\,,\,x^j \right\} = \sum_k \left( \frac{\partial x^i}{\partial q^k} \frac{\partial x^j}{\partial p_k} - \frac{\partial x^i}{\partial p_k} \frac{\partial x^j}{\partial q^k} \right) \ . \tag{2.1.5}$$

It is easy to check that $\mho^{ij}$ transforms tensorially under a change of coordinates $x^i \to y^i(x)$ and thus defines an antisymmetric bivector—that is, a rank 2 antisymmetric covariant tensor: $\mho = \frac{1}{2}\mho^{ij}\partial_i \wedge \partial_j$. Furthermore, one can easily check that $\mho^{ij}$ is nondegenerate so that its inverse $\Omega_{ij}$ exists and defines a nondegenerate 2-form $\Omega = \frac{1}{2}\Omega_{ij}dx^i \wedge dx^j$ on $M$, called the symplectic form. The Jacobi identities of the Poisson bracket become a differential relation on $\mho$ which, when inverted, implies that the symplectic form is closed: $d\Omega = 0$. The differential relation on the bivector $\mho$ can be written very simply in terms of the Nijenhuis bracket, in terms of which $d\Omega = 0$ is equivalent to $[\mho, \mho] = 0$.

To summarize, starting with the usual coordinates $(q^i, p_i)$ and the usual Poisson brackets, we have uncovered an underlying geometric structure: the bivector $\mho$ obeying $[\mho, \mho] = 0$. This may seem overkill for $\mathbb{R}^{2n}$ but it allows us to do hamiltonian mechanics covariantly on any Poisson manifold $M$. A theorem going back to Lie and generalized recently by Weinstein [38] says that around each point of a Poisson manifold, once can find local coordinates $(q^i, p_i, c^\alpha)$ such that $\mho$ has the following form:

$$
\begin{array}{c}
\begin{array}{ccc} \phantom{q^i} & q^j \quad\quad p_j \quad\quad c^\beta \end{array} \\
\begin{array}{c} q^i \\ p_i \\ c^\alpha \end{array}
\begin{pmatrix} 0 & -\delta^i_j & 0 \\ \delta^j_i & 0 & 0 \\ 0 & 0 & 0 \end{pmatrix}
\end{array} . \tag{2.1.6}
$$

The $c^\alpha$ are called casimir functions and have vanishing Poisson brackets with everything. Clearly, if no casimirs exist, $\mho$ is nondegenerate, and we are in the symplectic case—the coordinates $(q^i, p_i)$ being in this case the familiar canonical coordinates. The symplectic instance of the Lie-Weinstein theorem is known as Darboux's theorem. It says that symplectic manifolds of the same dimension are locally isomorphic. Comparing with riemannian geometry, it basically comes to say that there is no symplectic curvature. The Lie-Weinstein theorem characterizes the local geometry of a Poisson manifold: locally a Poisson manifold is foliated by symplectic submanifolds; each symplectic leaf being specified uniquely by the values that the casimirs take on it. Because the casimirs have vanishing Poisson bracket with any function, they are constants of the motion relative to any hamiltonian and therefore the time evolution preserves each symplectic leaf. It may therefore seem that nothing is gained by considering dynamics on Poisson manifolds which are not symplectic. But there is a catch: global problems aside, the transformation necessary to bring the coordinates of a given Poisson manifold to the form $(q^i, p_i, c^\alpha)$ may be very cumbersome. This will be especially true in the integrable hierarchies with which we will be working.

There is another way to understand the Poisson structure $\mho$ that will better suit our needs. The derivation property (2.1.4) says that the to every function

$f$ there corresponds a vector field $X_f$

$$X_f \cdot g = \{f, g\} \ , \tag{2.1.7}$$

whose components in local coordinates are given by $X_f^i = -\mho^{ij}\partial_j f$. $X_f$ is called the hamiltonian vector field associated with $f$. In other words, $\mho$ gives rise to a tensorial map $J : \{\text{1-forms}\} \rightarrow \{\text{vector fields}\}$ defined by $X_f = -J(df)$. This is enough to specify $J$ completely since every one-form is locally a linear combination of one-forms of the form $g\,df$. In the non-symplectic case, this map will fail to be an isomorphism, but nevertheless its image will be a subalgebra of the vector fields. This important fact follows from the Jacobi identity of the Poisson bracket and the fact that one can always construct a local basis for the 1-forms out of gradients of functions. Therefore it is enough to show that for any two functions $f$ and $g$,

$$\left[X_f, X_g\right] = X_{\{f, g\}} \ . \tag{2.1.8}$$

But this follows trivially from the Jacobi identity. Indeed, acting on any function $h$,

$$
\begin{aligned}
\left[X_f, X_g\right] \cdot h &= X_f \cdot X_g \cdot h - X_g \cdot X_f \cdot h \\
&= \{f, \{g, h\}\} - \{g, \{f, h\}\} \qquad \text{(by (2.1.7))} \\
&= \{\{f, g\}, h\} \qquad\qquad\quad \text{(Jacobi identity)} \\
&= X_{\{f, g\}} \cdot h \ .
\end{aligned}
$$

The fact that the image of $J$ is a Lie subalgebra of the vector fields allows us to define a 2-form there as follows. If $\alpha$ and $\beta$ are 1-forms on $M$, then

$$\omega(J(\alpha), J(\beta)) = \langle J(\alpha), \beta \rangle \ , \tag{2.1.9}$$

where $\langle -, - \rangle$ is the dual pairing between vector fields and 1-forms. Applied to gradients $df$ and $dg$, we find that $\omega(J(df), J(dg)) = \omega(X_f, X_g) = \{f, g\}$. One can show that the Jacobi identity of the Poisson bracket implies that $\omega$ is closed; that is, that for any three 1-forms $\alpha$, $\beta$, and $\gamma$,

$$d\omega(J(\alpha), J(\beta), J(\gamma)) = 0 \ . \tag{2.1.10}$$

Apart from the degeneracy of the Poisson brackets, there is another aspect in which the formalism we will be using deviates from the usual one. The class of functions that we will be working with will turn out not to form a ring. In other words, the product of two functions will fall outside the class of functions we consider. This may seem at first problematic, but it turns out not to hinder the formalism at all. Let us then summarize the necessary ingredients in the formulation of hamiltonian dynamics. We will at the same time translate the relevant geometric data into algebraic terms. The following kinematical ingredients will be needed:

(1) a Lie algebra $\mathfrak{X}$ corresponding to the **vector fields**;

(2) a representation $\Omega^0$ of $\mathfrak{X}$ corresponding to the **functions**;

(3) a vector space $\Omega^1$ nondegenerately paired with $\mathfrak{X}$ via $\langle -, - \rangle$, and a linear map $d : \Omega^0 \to \Omega^1$ (we call $\Omega^1$ the **one-forms** and those one-forms in the image of $d$ **gradients**); and

(4) a linear map $J : \Omega^1 \to \mathfrak{X}$ satisfying the following properties (such a map will be called **hamiltonian**):

  (a) that the image of $J$ be a Lie subalgebra of $\mathfrak{X}$;

  (b) that $J$ be skewsymmetric: for all one-forms $\alpha, \beta \in \Omega^1$, $\langle J(\alpha), \beta \rangle = -\langle J(\beta), \alpha \rangle$; and

  (c) that the bracket $\{f, g\} = \langle J(df), dg \rangle$ of two functions $f, g \in \Omega^0$ satisfy the Jacobi identity; or, in other words, that the map $J \circ d : \Omega^0 \to \Omega^1 \to \mathfrak{X}$ be a Lie algebra morphism.

On such a structure we will then be able to define dynamics by choosing a function $H \in \Omega^0$ and defining the time evolution as the flow of the vector field $J(dH)$. Formally, we call the quintuple $(\mathfrak{X}, \Omega^0, \Omega^1, J, H)$ satisfying the above properties a **hamiltonian dynamical system**.

## 2.2. INTEGRABILITY AND DYNAMICAL HIERARCHIES

The dynamical systems that we will focus on are rather special in that they are completely integrable. The notion of complete integrability goes back to Jacobi and Liouville and has been substantially generalized in recent times. Let us start with a trivial example: the ubiquitous harmonic oscillator. Take $M = \mathbb{R}^2$ with coordinates $(q, p)$ and the standard Poisson bracket $\{q, p\} = 1$. We take as hamiltonian the function $H = \frac{1}{2}p^2 + \frac{1}{2}q^2$. The equations of motion are well known: $\dot{p} = -q$ and $\dot{q} = p$ which have as solutions:

$$q(t) = \bar{q} \cos t + \bar{p} \sin t \qquad \text{and} \qquad p(t) = \bar{p} \cos t - \bar{q} \sin t \ . \qquad (2.2.1)$$

The physical trajectories are circles centered at the origin and with radius $R = \sqrt{\bar{q}^2 + \bar{p}^2}$. Antisymmetry of the Poisson bracket implies that the Hamiltonian is conserved, and in fact we notice that on the trajectory at radius $R$, the hamiltonian obtains the value $\frac{1}{2}R^2$. Let us introduce polar coordinates (which are valid away from the origin) $q = r \cos \theta$ and $p = r \sin \theta$. In terms of these coordinates the physical trajectories take a very simple form

$$r(t) = R \qquad \text{and} \qquad \theta(t) = \theta(0) + t \ , \qquad (2.2.2)$$

while the symplectic form becomes $\Omega = dp \wedge dq = r\,dr \wedge d\theta = dH \wedge d\theta$. In other words, the change of variables $(q, p) \mapsto (\theta, H)$ is a canonical transformation which linearizes the dynamics. These coordinates are called **action-angle variables**. A hamiltonian dynamical system is called **(completely) integrable** if it admits action-angle variables; that is, if we can find coordinates $(H_i, \theta_i)$ relative to which the symplectic form becomes $\Omega = \sum_i dH_i \wedge d\theta_i$ and the time evolution is linear: $H_i(t) = H_i(0)$ and $\theta_i(t) = \theta_i(0) + t$.

This definition seems to suggest that integrability is an *a posteriori* consequence of solving the dynamics; but in fact a theorem due originally to Liouville gives us a necessary and sufficient condition for a hamiltonian dynamical system to be completely integrable. Liouville's theorem states the following. Let $(M, \Omega)$ be a $2n$-dimensional symplectic manifold and let us define some dynamics on $M$ by specifying a hamiltonian function $H$. Then the dynamical system is completely integrable if and only if there exist $n$ functions $H = H_1, H_2, \ldots, H_n$ whose gradients $dH_i$ are linearly independent almost everywhere[2] and such that they are in involution $\{H_i\,,\,H_j\} = 0$; in particular, the $H_i$ are all conserved quantities.

Liouville's theorem brings us naturally to the concept of an integrable hierarchy. From a purely formal point of view—that is, disregarding for a moment the physics we are describing—any one of the functions $\{H_i\}$ can be used as a hamiltonian and each of these hamiltonians defines an integrable dynamical system with the same functions $\{H_i\}$ as conserved quantities. Naturally, the physics will choose one particular hamiltonian that can be sensibly interpreted as generating *the* time evolution of the system, but from a structural point of view, there is no reason to prefer one over any other. The democratic thing to do is then to introduce $n$ 'times' $t_1, \ldots, t_n$ and define a hierarchy of flows $\frac{\partial f}{\partial t_i} = \{f\,,\,H_i\}$ for any function $f$. The involutivity of the hamiltonians imply that the flows commute:

$$\frac{\partial^2 f}{\partial t_i \partial t_j} - \frac{\partial^2 f}{\partial t_j \partial t_i} = \{\{f\,,\,H_i\}\,,\,H_j\} - \{\{f\,,\,H_j\}\,,\,H_i\}$$
$$= \{f\,,\,\{H_i\,,\,H_j\}\} \qquad \text{(Jacobi identity)}$$
$$= 0 \ . \qquad \text{(involutivity)}$$

It is meaningful to describe integrable hierarchies in the Poisson setting. Suppose that $(P, \eth)$ is a Poisson manifold of rank $2n$, by which we mean that the generic symplectic leaf is of dimension $2n$. By an integrable hierarchy we will understand a collection of $n$ functions in involution $H_1, H_2, \ldots, H_n$ such that their associated hamiltonian vector fields span an $n$-dimensional distribution (almost everywhere). Since hamiltonian vector fields are tangent to the symplectic leaves, this means that on a given symplectic leaf they give rise to an integrable hierarchy in the sense of Liouville.

---

[2]  This is unavoidable: already in the harmonic oscillator, $dH = 0$ at the origin.

For infinite-dimensional dynamical systems the existence of action-angle variables is problematic (although for the KdV hierarchy they do exist and are given in terms of scattering data [**8**]!) and one relaxes the notion of integrability by requiring an infinite number of conserved quantities in involution or sometimes even just an infinite number of commuting flows. This does not imply complete integrability in the strict sense, but it is a convenient working definition in the absence of a stronger yet still practical criterion.

A more rigorous notion of integrability for infinite-dimensional systems can in principle be defined by analogy with the KdV hierarchy. It is proven in [**13**] that every solution of the KdV hierarchy is 'close' (in some appropriate sense) to an $N$-soliton solution, for some $N$. Moreover, as mentioned briefly in the first chapter, any $N$-soliton solution of the KdV equation (indeed, hierarchy) can be effectively described by a completely integrable system in a $2N$-dimensional phase space. Therefore the union of all these finite-dimensional phase spaces is 'dense' in the space of solutions of the KdV hierarchy. This means that the phase space of the KdV hierarchy is the closure (in some appropriate topology) of an inductive limit of phase spaces of finite-dimensional completely integrable systems.

## 2.3. HAMILTONIAN REDUCTION

Finally we discuss in some detail the basic notions of hamiltonian reduction. In fact, we will only need in what follows a very particular case: the reduction induced by constraints of the second-class. The modern formulation of the theory of constraints goes back to Dirac [**39**]. A dynamical system on a phase space $M$, may actually only depend—via a mixture of kinematical and dynamical constraints—on some subspace. The way constraints arise normally is as follows. One usually describes a physical theory by specifying the configuration space and the action, which is a function on the tangent bundle. It may be, however, that this description is redundant and in fact the true physical degrees of freedom—that is, the physical configurations—comprise only a subspace of the full configuration space. This is always the case in gauge theories, but this phenomenon is not restricted to them. Dirac's treatment of constraints is purely hamiltonian. Given a Poisson manifold $M$ and some functions $\{\phi_i\}$ on $M$ with zero locus $M_o = \{m \in M | \phi_i(m) = 0 \; \forall i\}$, Dirac distinguishes two kinds of constraints: first and second class. Constraints are said to be of the first class, if their Poisson bracket is identically zero on $M_o$. At the other extreme we have second-class constraints: for which the matrix $\{\phi_i, \phi_j\}$ of Poisson brackets is nondegenerate on $M_o$. Of course, constraints will generally come mixed and it is something of an art to disentangle them. Fortunately, for our purposes we will only need to talk about second-class constraints. Notice that from their definition it follows that second-class constraints always come in pairs.

Dirac proved that if $(M, \Omega)$ is a Poisson manifold, then the zero locus $M_o$ of $2k$ second-class constraints $\{\phi_i\}$ inherits a symplectic structure from that of $M$. Moreover he gave an explicit formula for the Poisson bracket on $M_o$ in terms of that on $M$. A Poisson structure on $M_o$ is the same as a Poisson algebra structure on its ring of functions $\Omega^0(M_o)$. Any function on $M$ restricts to a function on $M_o$ and quite trivially any two functions on $M$ agree on $M_o$ if their difference vanishes there. On the other hand, and glossing over some regularity issues[3], every function on $M_o$ extends to a function on $M$. In other words, $\Omega^0(M_0) \cong \Omega^0(M)/I$, where $I$ is the ideal of those functions vanishing at $M_o$. Clearly the constraints belong to $I$ and in the regular case alluded to above, they generate it. That is, the typical element of $I$ is a linear combination $\sum_i f_i \phi_i$ where $f_i$ are arbitrary functions on $M$. Notice, however that $I$ is not a Poisson ideal. That is, the Poisson bracket of two elements of $I$ does not lie back in $I$. In fact, far from it: $\{\phi_i, \phi_j\}$ cannot all be zero on $M_o$, since the matrix is nondegenerate. Therefore we cannot expect to compute brackets on $M_o$ simply by extending functions to $M$, computing their bracket there, and restricting back to $M_o$. For this to be well-defined, the end result could not depend on the extension, but in fact it does precisely because $I$ is not a Poisson ideal. The idea of Dirac was to deform the Poisson bracket on $M$ in such a way that the constraints would commute with everything (at least on $M_o$). Thus one introduces the modified bracket

$$\{f, g\}_D \equiv \{f, g\} - \sum_{i,j} \{f, \phi_i\} C^{ij} \{\phi_j, g\} \;, \qquad (2.3.1)$$

where $C^{ij}$ is the matrix inverse of $C_{ij} \equiv \{\phi_i, \phi_j\}$.[4] Notice that the Dirac bracket obeys that for any function $f$, $\{f, \phi_i\} = 0$ for all $i$. Therefore if we change $f$ and $g$ by adding to them any arbitrary function which vanishes on $M_o$, their bracket on $M_o$ is not altered. Therefore equation (2.3.1) induces a Poisson structure on $M_o$, known as the **Dirac bracket**.

---

[3] Technically, we assume that $M_o$ is a regular embedded submanifold of $M$ or equivalently that $\mathbf{0}$ is a regular value of the function $\Phi : M \to \mathbb{R}^{2k}$ whose components are the constraints $\phi_i$.

[4] This bracket is really only defined on $M_o$ since there is no guarantee that the matrix $C_{ij}$ is invertible away from $M_o$; although in practice this subtlety seldom arises.

Now, in infinite dimensions we are sometimes forced to impose an infinite number of second-class constraints. In those cases it is usually impracticable to write the matrix $\{\phi_i\,,\,\phi_j\}$ explicitly—let alone to compute its inverse; so we must resort to other methods to perform the reduction. It is here that the understanding of the geometry underpinning Dirac's theory of constraints becomes essential.

We will see that the Dirac bracket corresponds to a particular choice of extension of functions from $M_o$ to $M$; or more precisely, to a choice of writing down gradients of functions on $M_o$ as one-forms on $M$. To understand this point it we too must go back to basics.

The tangent vectors to $M_o$ are naturally embedded in the tangent space to $M$ restricted to $M_o$. In other words, for every $m \in M_o$, $T_m M_o \subset T_m M$ in a natural way. Indeed, since $M_o$ is the zero locus of $\{\phi_i\}$, $X \in TM$ is tangent to $M_o$ if and only if $X \cdot \phi_i = 0$ for all $i$. But unlike tangent vectors, there is no natural way to embed $T_m^* M_o$ in $T_m^* M$. This is because $T_m^* M_o$ is defined as the dual space of $T_m M_o$ and the dual of a subspace is not naturally a subspace of the dual. In other words, if $U \subset V$ are vector spaces, then a choice of $U^* \subset V^*$ corresponds exactly to a choice of complement to $U$ in $V$; that is, if $V = U \oplus W$ then $U^* \cong W^o$ canonically, where $^o$ denotes the annihilator $W^o \equiv \{v^* \in V^* \mid \langle v^*, w \rangle = 0 \ \forall w \in W\}$.

We can illustrate this with an example. Consider the standard two-sphere $S^2$ embedded in $\mathbb{R}^3$ by the 'constraint' $x^2 + y^2 + z^2 = 1$. In terms of spherical coordinates $(r, \theta, \varphi)$ we can consider a function on the sphere as simply a function $f(\theta, \varphi)$ of the angular coordinates. Its gradient is well-defined as a one-form on the sphere, but it is not well-defined as a one-form on $\mathbb{R}^3$. In fact, to write it as a one-form on $\mathbb{R}^3$ we must first extend the function to a function $F(r, \theta, \varphi)$ such that $F(r = 1, \theta, \varphi) = f(\theta, \varphi)$, and then we simply take $dF$ (restricted to $r = 1$) as the one-form corresponding to $df$. But this clearly depends on $F$ since $\partial F / \partial r$ is not fixed at $r = 1$.

In other words, a choice of complement to $T_m M_o$ in $T_m M$ is exactly a choice of normal vectors to $M_o$. In the absence of any additional structure on a manifold $M$, there is no preferred choice. Suppose, however, that $M$ is given a riemannian metric. Then we can choose as normal vectors at $m \in M_o$ the orthocomplement $(T_m M_o)^\perp$ relative to the metric. This is clearly the natural choice in riemannian geometry and it has the following nice property: $T_m^* M_o \cong ((T_m M_o)^\perp)^o$ is mapped isomorphically onto $TM_o$ under the isomorphism $T^* M \to TM$ induced by the metric ('raising the index').

In symplectic geometry—which is a closer analog to the case we are interested in—there is no metric, but we have the next best thing: a non-degenerate 2-form $\Omega$. Suppose then that $(M, \Omega)$ is a symplectic manifold and that $M_o$ is a submanifold. We can try to mimic the same construction: namely, define the symplectic complement $(T_m M_o)^\perp$ as those $X \in T_m M$ such that $\Omega(X, Y) = 0$ for all $Y \in T_m M_o$. We would then like to have a direct sum decomposition for all $m \in M_o$

$$T_m M = T_m M_o \oplus (T_m M_o)^\perp \ . \tag{2.3.2}$$

Unfortunately this is not always possible and this condition defines a special kind of submanifolds $M_o$ known as **symplectic submanifolds**. As a trivial counterexample, suppose that $M_o$ is given by the zero locus of one constraint $\phi$. Then the hamiltonian vector field associated to the constraint is both tangent and symplectically normal to $M_o$, due to the antisymmetry of the symplectic form.

When $M_o$ is defined as the zero locus of independent constraints $\{\phi_i\}$, it is easy to give an alternative characterization of condition (2.3.2). Let $X_i$ denote the hamiltonian vector field associated to the the constraint $\phi_i$. Then for any vector $X \in TM$,

$$\Omega(X, X_i) = \langle X, d\phi_i \rangle = X \cdot \phi_i \ , \tag{2.3.3}$$

whence $X \in TM_o$ if and only if it is symplectically perpendicular to the $X_i$. In other words, $TM_o = \langle X_i \rangle^\perp$. Now condition (2.3.2) says that there is no linear combination of the $X_i$ which is symplectically perpendicular to the $X_i$; that is, that the restriction of the symplectic form to the subspace spanned by the $X_i$ is nondegenerate. In other words, that the matrix $\Omega(X_i, X_j) = \{\phi_i, \phi_j\}$ is invertible everywhere on $M_o$. In other words, we recover precisely Dirac's definition of second-class constraints.

Under the assumption that $M_o$ is a symplectic submanifold of $M$, we can then mimic the riemannian case and embed $T_m^* M_o$ in $T_m^* M$ as $\langle X_i \rangle^o$. That is, $\alpha \in T_m^* M$ belongs to $T_m^* M_o$ if and only if $\langle \alpha, X_i \rangle = 0$ for all $i$; or, equivalently, that $\Omega(J(\alpha), X_i) = 0$ for all $i$, where $J : T^* M \to TM$ is the isomorphism induced by the symplectic form. In other words, $T_m^* M_o$ consists precisely of those 1-forms on $M$ which map under $J$ to vectors tangent to $M_o$ (recall $(T_m M_o) = \langle X_i \rangle^\perp$). Notice that this last definition also makes sense in the Poisson case, since the Poisson structure $\mho$ defines a map $J$ from one-forms to vector fields. Of course, in this case, due to the degeneracy of $J$, $\alpha$ is not uniquely defined by this condition, since we can always add to it any one-form in the kernel of $J$.

Back to the symplectic case, the way to compute Poisson brackets is now clear: you take any two functions on $M_o$, you extend them to functions on $M$, you project their gradients to $T_m^* M_o$ as defined above and then use these projections to compute their Poisson bracket. The resulting expression should then be independent of the extension. Let us see this in a bit more detail. Let $f$ and $g$ be functions in $M_o$ thought of as restrictions of functions $f$ and $g$ on $M$. Let $X_f$ and $X_g$ be their respective hamiltonian vector fields. Under the decomposition (2.3.2) we can write them as

$$X_f = (X_f)_o + (X_f)_\perp \; , \qquad (2.3.4)$$

and the same for $g$. Since the $X_i$ span $TM_o^\perp$ we can expand $(X_f)_\perp$ in a linear combination $\sum_i \lambda_i X_i$ for some functions $\lambda_i$ which we now determine. On the one hand,

$$\Omega(X_j, X_f) = \{\phi_j \,, f\} \; , \qquad (2.3.5)$$

but also

$$\begin{aligned}
\Omega(X_j, X_f) &= \Omega(X_j, (X_f)_\perp) \\
&= \sum_i \lambda_i \Omega(X_j, X_i) \\
&= \sum_i \lambda_i \{\phi_j \,, \phi_i\} \; . \qquad (2.3.6)
\end{aligned}$$

Comparing the two expressions and letting $C^{ij}$ denote the inverse of the matrix $\{\phi_i \,, \phi_j\}$, we can solve for the $\lambda_i$ and write

$$(X_f)_o = X_f - \sum_{i,j} C^{ij} \{\phi_j \,, f\} X_i \; . \qquad (2.3.7)$$

Doing the same for $g$, the resulting Poisson bracket can be written as

$$\begin{aligned}
\{f \,, g\}_D &= \Omega((X_f)_o, (X_g)_o) \\
&= \Omega(X_f, X_g) - \sum_{i,j} \Omega(X_f, X_i) C^{ij} \Omega(X_j, X_g) \\
&= \{f \,, g\} - \sum_{i,j} \{f \,, \phi_i\} C^{ij} \{\phi_j \,, g\} \; ; \qquad (2.3.8)
\end{aligned}$$

that is, they are simply the Dirac brackets of equation (2.3.1). Following this geometric line of thought it is easy to show that the bracket does not depend on the choice of constraints $\{\phi_i\}$ used to describe $M_o$ and that it indeed obeys the Jacobi identity: this last remark following trivially from the fact that the Poisson bracket is defined with respect to the the pull-back of the symplectic form via the embedding $i : M_o \to M$ and the fact that the exterior differentiation commutes with pull-backs.

Let us summarize our geometric strategy to deal with second-class constraints. To compute the Dirac bracket of two functions in the constrained submanifold $M_o$, we will write them as functions on the ambient space and we will fix the ambiguity in their gradients in such a way, that when mapped by $J$ into vector fields on $M$, the vector fields turn out to be tangent to $M_o$. We then simply compute their Poisson bracket on $M$ but with these gradients. In some simple examples we will be able to compare with the Dirac prescription and check that they are in agreement, but in some others this more geometric approach will be the only way to proceed.

# Chapter Three

# LAX FORMALISM AND THE KP HIERARCHY

In this chapter we briefly review the Lax formalism for KdV-type equations. This represents a nontrivial example of a dynamical system of the type defined at the end of Section 2.1. The underlying space will be the space of formal differential operators of the form $\partial^n + \cdots$ generalizing the Sturm-Liouville operator $\partial^2 + u$ of the KdV equation. In Section 1 we begin to set up the basic differential calculus in this space: functions and vector fields. To define the one-forms it will be necessary to introduce the ring of formal pseudodifferential operators, which is done in Section 2. Also in Section 2 we turn our space into a formal Poisson manifold via the Adler map and the associated Gel'fand–Dickey brackets. Section 3 discusses the generalized $n$-KdV hierarchies and Section 4 is devoted to the KP hierarchy. The influence of Dickey in this section is evident. A fuller account of this topic can be found in his book [**40**].

## 3.1.  CALCULUS IN THE SPACE OF LAX OPERATORS

FUNCTIONS

By a Lax operator of order $n$ we mean a (one-dimensional) differential operator of the form

$$L = \partial^n + u_1 \partial^{n-1} + \cdots + u_n \, , \tag{3.1.1}$$

where the $u_i$ are to be thought of as either rapidly decreasing smooth functions on the real line, smooth functions on the circle, or simply as generators of a differential ring $R_L$ (or simply $R$ if no confusion can result). Respectively, $\partial$ is to be thought of as the derivative with respect to the coordinate on the real line, on the circle, or as the derivation on the ring $R$.

We let $\mathfrak{M}^n$ denote the space of Lax operators of a fixed order $n$. It is clearly an infinite-dimensional affine space modeled on the vector space of differential operators of order $n-1$ whose coefficients are differential polynomials of the $u_i$. We let $\mathcal{R}^{n-1}$ denote this space. Then $\mathfrak{M}^n = \partial^n + \mathcal{R}^{n-1}$. We shall drop the superscript from $\mathfrak{M}$ whenever no confusion can arise.

As functions on $\mathfrak{M}$ we shall take the integrals of differential polynomials of the $u_i$. So a typical function on $\mathfrak{M}$ will be of the form

$$F[L] = \int f(u) \; , \qquad (3.1.2)$$

where $f(u)$ is a polynomial of the $u_i$ and their derivatives, and where $\int$ means different things depending on the context: it means integration over the real line, integration over the circle, or simply the canonical projection $R \to R/\partial R$ in the more abstract algebraic setting. This last point may seem at first a bit confusing, but it is really very simple. Think of the case on the circle. It is clear that the integral of a total derivative vanishes. Furthermore, if an integral vanishes, its integrand is a total derivative. Therefore there is a one-to-one correspondence between {functions on the circle}/{total derivatives} and integrals. In the algebraic setup, $R$ plays the role of the functions on the circle and so the integrals are in one-to-one correspondence with $R/\partial R$. On a more pragmatic level, identifying the functions with $R/\partial R$ means that integration is linear and that we can integrate by parts.

VECTOR FIELDS

Vector fields in $\mathfrak{M}$ are first order deformations of the points in $\mathfrak{M}$. Since $\mathfrak{M}$ is an affine space, the tangent space at each point can be identified with the vector space it is modeled on, namely $\mathcal{R}^{n-1}$. Given a differential operator $A = \sum_{j=1}^{n} a_j \partial^{n-j} \in \mathcal{R}^{n-1}$ one can define a vector field $\partial_A$ on $\mathfrak{M}$, whose action on a function is given by the usual

$$\partial_A F[L] = \frac{d}{d\epsilon} F[L + \epsilon A]\bigg|_{\epsilon=0} \; . \qquad (3.1.3)$$

If $F$ is given by (3.1.2) then

$$\partial_A F = \int \sum_{i=1}^{n} \sum_{k=0}^{\infty} \frac{\partial f}{\partial u_i^{(k)}} a_i^{(k)} \; , \qquad (3.1.4)$$

where the superscript $(k)$ means $k^{\text{th}}$ derivative. Integrating by parts we can rewrite (3.1.4) as

$$\partial_A F = \int \sum_{i=1}^{n} \sum_{k=0}^{\infty} (-)^k \left( \frac{\partial f}{\partial u_i^{(k)}} \right)^{(k)} a_i \; , \qquad (3.1.5)$$

where we recognize the Euler operators

$$\mathcal{E}_i = \sum_{k=0}^{\infty} (-\partial)^k \cdot \frac{\partial}{\partial u_i^{(k)}} \; . \qquad (3.1.6)$$

Notice that acting on differential polynomials of the $u_i$ the above sums over $k$ are actually finite. It is easy to see that the Euler operator annihilates total derivatives, so that $\mathcal{E}_i \cdot f' = 0$. This implies that $\partial_A$ in (3.1.4) is well defined because if $F = \int (f + g')$ then the expression (3.1.5) is unaltered. In other words, $\partial_A$ descends to a map in $R/\partial R$. A more precise statement is that the derivation—also denoted $\partial_A$, with a little abuse of notation—of $R$ given by

$$\sum_{i=1}^{n} \sum_{k=0}^{\infty} a_i^{(k)} \frac{\partial}{\partial u_i^{(k)}} \tag{3.1.7}$$

commutes with $\partial$ and thus induces a map in the quotient $R/\partial R$.

A remark is in order. Our choice of functions does not form a ring: there is no natural multiplication on $R/\partial R$ induced from the one on $R$ because $\partial R$ is not a multiplicative ideal. Therefore one cannot even begin to wonder whether the vector fields act as derivations. It is possible to extend the functions in such a way that they do form a ring, but this is an unnecessary complication which does not enhance the formalism.

Vector fields, however, do form a Lie algebra in such a way that the functions form a representation. Let $A, B \in \mathcal{R}^{n-1}$ be tangent vectors and let $\partial_A$ and $\partial_B$ denote the vector fields that they define. Then one has

$$[\partial_A , \partial_B] = \partial_{[\![A,B]\!]} , \tag{3.1.8}$$

where $[\![A, B]\!]$ is defined by

$$\begin{aligned} [\![A, B]\!] &= \partial_A B - \partial_B A \\ &= \sum_{i=1}^{n} \sum_{j=1}^{n} \sum_{k=0}^{\infty} \left( \frac{\partial b_i}{\partial u_j^{(k)}} a_j^{(k)} - \frac{\partial a_i}{\partial u_j^{(k)}} b_j^{(k)} \right) \partial^i . \end{aligned} \tag{3.1.9}$$

## 3.2. ONE-FORMS AND PSEUDODIFFERENTIAL OPERATORS

### FORMAL PSEUDODIFFERENTIAL OPERATORS

To define the one-forms we have to have to introduce pseudodifferential operators ($\Psi$DO's). We first introduce a formal inverse to $\partial$, satisfying $\partial \partial^{-1} = \partial^{-1} \partial = 1$. This and the Leibniz rule imply the following multiplication law for $\partial^{-1}$:

$$\partial^{-1} a = a \partial^{-1} - a' \partial^{-2} + a'' \partial^{-3} - \cdots , \tag{3.2.1}$$

for $a \in R$ any differential polynomial. That this law is correct can be easily seen by applying $\partial$ on both sides of the equation. Repeated application of this relation yields the generalized Leibniz rule for any integer $m$

$$\partial^m a = \sum_{i=0}^{\infty} \frac{m(m-1)\cdots(m-i+1)}{i!} a^{(i)} \partial^{m-i} \ . \tag{3.2.2}$$

Notice that when $m > 0$ the sum truncates (at $m$, actually) and we have the usual Leibniz rule. Notice also that we are forced to consider formal Laurent series in $\partial^{-1}$. Therefore we define pseudodifferential operators as $R((\partial^{-1}))$ with multiplication given by the generalized Leibniz rule. One can check that this multiplication is associative. We will use $\mathcal{R}$ as shorthand for the ring of $\Psi$DO's. $\mathcal{R}$ splits as a direct sum $\mathcal{R} = \mathcal{R}_+ \oplus \mathcal{R}_-$ where $\mathcal{R}_+ = R[\partial]$ denotes the subring of differential operators and $\mathcal{R}_- = \partial^{-1}R[[\partial^{-1}]]$ the subring of 'integral' operators. Given any $\Psi$DO $P$ we will let $P_+$ denote its projection onto $\mathcal{R}_+$ and $P_- = P - P_+$ its projection onto $\mathcal{R}_-$.

It turns out that $\mathcal{R}_+$ and $\mathcal{R}_-$ are nondegenerately paired under a symmetric bilinear form defined on $\mathcal{R}$ using the **Adler trace** [**22**]. Given a $\Psi$DO $P = \sum_{i \leq N} a_i \partial^i$ we define its **residue** by $\operatorname{res} P = a_{-1}$ and its Adler trace by $\operatorname{Tr} P = \int \operatorname{res} P$. To be allowed to call this a trace, we actually have to show that it annihilates commutators. This follows from the following fact: if $P = a\partial^p$ and $Q = b\partial^q$ are two $\Psi$DO's, then the residue of their commutator is total derivative. More to the point,

$$\operatorname{res}[P,\,Q] = \left( \frac{p(p-1)\cdots(1-q)(-q)}{(p+q+1)!} \sum_{i=0}^{p+q} (-)^i a^{(i)} b^{(p+q-i)} \right)' \ . \tag{3.2.3}$$

Notice that the Adler trace of a differential operator is zero, thus it does not coincide with the standard operatorial trace. In fact, the Adler trace is the logarithmically divergent part of the operatorial trace. That this quantity should define a trace is not trivial: in the general case it is a result due to Dixmier [**41**]. With the Adler trace we can define a symmetric bilinear form on $\mathcal{R}$ by

$$\langle P, Q \rangle = \operatorname{Tr} PQ \ . \tag{3.2.4}$$

It is clear that this bilinear form pairs $\mathcal{R}_\pm$ nondegenerately with $\mathcal{R}_\mp$. This fact makes $(\mathcal{R}, \mathcal{R}_+, \mathcal{R}_-)$ into a Manin triple and, in particular, $\mathcal{R}_-$ into a Lie bialgebra. The corresponding Lie-Poisson group is called the Volterra group and will play an important role in what follows.

GRADIENTS AND ONE-FORMS

Vector fields are parametrized by $\mathcal{R}^{n-1}$, which is nondegenerately paired with $\mathcal{R}_-/\partial^{-n}\mathcal{R}_-$. In turn, this space is in one-to-one correspondence with $\Psi$DO's of the form

$$X = \partial^{-1}x_n + \partial^{-2}x_{n-1} + \cdots + \partial^{-n}x_1 \ . \tag{3.2.5}$$

Indeed, if $A = \sum_{i=1}^n a_i\partial^{n-i} \in \mathcal{R}^{n-1}$,

$$\operatorname{Tr} XA = \int \sum_{i=1}^n x_i a_i \ . \tag{3.2.6}$$

Therefore we let the one-forms be parametrized by $\mathcal{R}_-/\partial^{-n}\mathcal{R}_-$.

Given a function $F$ on $\mathfrak{M}$ its gradient is defined as the unique one-form $dF$ such that for any vector field $\partial_A$,

$$\partial_A F = \langle dF, A \rangle = \operatorname{Tr} A dF \ . \tag{3.2.7}$$

Comparing with (3.1.5) it follows that the gradient is given by

$$dF = \sum_{i=1}^n \partial^{-n+i-1}\frac{\delta F}{\delta u_i} = \sum_{i=1}^n \partial^{-n+i-1}\mathcal{E}_i \cdot f \ , \tag{3.2.8}$$

for $F = \int f$.

Summarizing, we have the following geometric setup in the space $\mathfrak{M}^n$ of Lax operators of order $n$:

- Functions $\Omega^0 = R/\partial R$,
- Vector fields $\mathfrak{X} = \{\partial_A | A \in \mathcal{R}^{n-1}\}$,
- One-forms $\Omega^1 = \mathcal{R}_-/\partial^{-n}\mathcal{R}_-$, and
- a dual pairing between $\Omega^1$ and $\mathfrak{X}$ given by the Adler trace of their product.

THE ADLER MAP AND GEL'FAND–DICKEY BRACKETS

According to the setup of Section 2.1, the next ingredient is the hamiltonian map $J : \Omega^1 \to \mathfrak{X}$ defining the Poisson bracket. Based on known results about the KdV and Boussinesq equations, Adler [**22**] proposed the following map (called the **Adler map**)

$$J(X) \equiv (LX)_+ L - L(XL)_+ \ , \tag{3.2.9}$$

where $X$ is a one-form. A priori $J$ maps $\mathcal{R}_- \to \mathcal{R}_+$ but a closer look at the expression reveals that $\partial^{-n}\mathcal{R}_-$ lies in its kernel, so that it induces a map $\Omega^1 \to \mathcal{R}_+$. Moreover, the image lies in $\mathcal{R}^{n-1}$. To see this notice that the RHS of (3.2.9) can be written alternatively as $L(XL)_- - (LX)_- L$ which explicitly shows that it has order less than $n$. Hence the Adler map defines a map from one-forms to tangent vectors. It is easy to prove that the above map is actually skew-symmetric. This only uses the fact that since the Adler trace only pairs up $\mathcal{R}_\pm$ with $\mathcal{R}_\mp$, we have that $\operatorname{Tr} PQ = \operatorname{Tr} P_+Q_- + \operatorname{Tr} P_-Q_+$.

The bracket

$$\{F\,,\,G\}_0 = \operatorname{Tr} J(dF)dG = \operatorname{Tr}\left[(LdF)_+LdG - L(dFL)_+dG\right] \qquad (3.2.10)$$

is therefore antisymmetric. Adler conjectured that it was Poisson and this was proven by Gel'fand and Dickey in [**24**], hence its name: the **second Gel'fand–Dickey bracket**. The reason for the name will be clear in a moment. We will not reproduce here the proof of this fact. A somewhat simplified version of the original combinatorial proof is reproduced in [**40**]. At least two other proofs are known. First of all Kupershmidt and Wilson [**15**] noticed that this bracket is induced by the Miura transformation as follows. Suppose that we formally factorize the Lax operator $L = (\partial - v_1)(\partial - v_2)\cdots(\partial - v_n)$. This formal factorization embeds the differential rings $R$ into the differential ring $S$ generated by the $v_i$. Let us define the following bracket on this ring. Let $\tilde{F} = \int f(v)$ and $\tilde{G} = \int g(v)$, and put

$$\left\{\tilde{F}\,,\,\tilde{G}\right\}_M = \sum_i \int \frac{\delta \tilde{F}}{\delta v_i}\left(\frac{\delta \tilde{G}}{\delta v_i}\right)' . \qquad (3.2.11)$$

This bracket is clearly Poisson since the Poisson operator is simply $\partial$ which is constant and hence trivially satisfies the Jacobi identity. The embedding $R \to S$ allows us to pull this back the above bracket. By construction, the induced bracket is Poisson; the remarkable fact is that it actually closes back into $R$; and in fact, that it agrees with the second Gel'fand–Dickey bracket (3.2.10). This fact is known as the Kupershmidt–Wilson theorem. The original proof of this theorem is very cumbersome and Dickey [**16**] gave a very simple and elementary proof. A third proof of this fact is due to Drinfel'd and Sokolov who obtain it by hamiltonian reduction from the Kirillov–Kostant Poisson structure (which we define later, in the supersymmetric case) in the dual of the affine Lie algebra $A_{n-1}^{(1)}$. Despite all these different proofs, the fact that the induced bracket should close back into $R$ has remained elusive until Wilson re-examined the result in the light of differential Galois theory [**42**]. In a nutshell, Wilson introduces yet another differential ring in which the $\{v_i\}$, and hence the $\{u_i\}$, are embedded and in which there is a natural (yet nonlocal) Poisson structure. It turns out that the structure is invariant under the action of $SL(N)$ and that the $\{v_i\}$ generate the differential subring of $B_+$-invariants and $\{u_i\}$ generate the differential subring of $SL(N)$-invariants. Here $B_+ \subset SL(N)$ is the Borel subgroup of upper triangular matrices. Therefore the $\{u_i\}$ inherit a Poisson structure which is easily seen to coincide with the second Gel'fand–Dickey bracket, whereas no such explanation exists for the Kupershmidt–Wilson theorem simply because the the flag space $SL(N)/B_+$ is not a Lie group due to $B_+$ not being a normal subgroup.

Kupershmidt and Wilson also noticed an amusing fact about this bracket. Suppose that we shift $L \to \hat{L} \equiv L + \lambda$, for $\lambda$ some constant parameter and let $J_\lambda(X) \equiv (\hat{L}X)_+\hat{L} - \hat{L}(X\hat{L})_+$. Expanding in powers of $\lambda$ we see that the quadratic terms drop and we have

$$J_\lambda(X) = J(X) + \lambda\,[L\,,\,X]_+\ .\qquad(3.2.12)$$

This induces a bracket

$$\{F\,,\,G\}_\lambda = \{F\,,\,G\}_0 + \lambda\,\{F\,,\,G\}_\infty\ ,\qquad(3.2.13)$$

where

$$\{F\,,\,G\}_\infty = \langle\,[L\,,\,dF]_+\,,dG\rangle = \mathrm{Tr}\,[L\,,\,dF]_+\,dG\qquad(3.2.14)$$

is nothing but the first Gel'fand–Dickey bracket. Since for any $L$, $J$ is a hamiltonian map, it follows that for all $\lambda$, $J_\lambda$ is hamiltonian and that $\{-\,,\,-\}_\lambda$ will satisfy the Jacobi identity. In particular, writing the Jacobi identity as a polynomial in $\lambda$, all coefficients must vanish separately. The free coefficient is just the Jacobi identity for the second Gel'fand–Dickey bracket and the coefficient in $\lambda^2$ is similarly the Jacobi identity for the first Gel'fand–Dickey bracket. The vanishing of the linear coefficient implies that the two Gel'fand–Dickey brackets are coordinated; that is, any linear combination is again a Poisson bracket. As mentioned in Chapter One, this is nontrivial since the Jacobi identity is quadratic and hence contains mixed terms that are only zero under very special circumstances: usually a symptom of integrability.

We finish this section with several interesting remarks about these brackets. The first remark is that the first bracket can be identified with a natural bracket on the coadjoint orbit of a formal Lie group. Notice that we can rewrite (3.2.14) as follows:

$$\begin{aligned}
\{F\,,\,G\}_\infty &= \mathrm{Tr}\,[L\,,\,dF]_+\,dG\\
&= \mathrm{Tr}\,[L\,,\,dF]\,dG\\
&= -\,\mathrm{Tr}\,L\,[dF\,,\,dG]\ .
\end{aligned}\qquad(3.2.15)$$

It was observed by Adler [**22**] and, independently, by Lebedev and Manin [**23**] that this is nothing but the Kirillov–Kostant bracket on a coadjoint orbit of the Volterra group. The Volterra group is the formal Lie group whose Lie algebra is given by the integral $\Psi\mathrm{DO}$'s $\mathcal{R}_-$ under the commutator. In other words, the Volterra group is just $G \equiv 1 + \mathcal{R}_-$, and the product is induced from the associative product in the ring of $\Psi\mathsf{DO}$'s. Under the Adler trace, $\mathcal{R}_+$ becomes identified with the coalgebra and the Volterra group acts in $\mathcal{R}_+$ by the coadjoint representation. Pick a Lax operator $L \in \mathfrak{M}^n$ and consider its orbit $\mathcal{O}_L = \{\phi L \phi^{-1} | \phi \in G\}$ under the Volterra group. Then $\mathcal{O}_L$ is given precisely by the affine subspace of $\mathfrak{M}^n$ defined by those $L$ with the same $u_1$ as $L$, and the Kirillov–Kostant bracket on it coincides with the first Gel'fand–Dickey bracket.

A second remark is that $\mathcal{R}_+ \cong \mathcal{R}_-^*$ is also a Lie algebra under the commutator and that makes $\mathcal{R}_-$ a Lie bialgebra and the Volterra group into a (formal) Lie-Poisson group. This approach has been fruitfully exploited by Semenov-Tyan-Shanskiĭ [**43**] to generalize the Gel'fand–Dickey brackets to the case of an associative algebra $A = A_+ \oplus A_-$ with $A_\pm$ subalgebras which are isotropic relative to the symmetric bilinear form associated to a trace. If we let $\pi_\pm$ denote the projections onto $A_\pm$ respectively, the operator $r = \pi_+ - \pi_-$ solves the classical Yang–Baxter equation and allows one to define two coordinated Poisson brackets on $A$ which reduce to the Gel'fand–Dickey ones when $A$ is the ring of $\Psi\mathsf{DO}$'s. We will prove a supersymmetric version of this theorem in Chapter Four.

Finally, we mention that under the second-class constraint $u_1 = 0$, the Gel'fand–Dickey brackets yield classical realizations of $\mathsf{W}$-algebras. In fact, the correspondence is as follows. The reduction of the Gel'fand–Dickey bracket associated to the Lax operator of order $n$ yields the $\mathsf{W}_n$ algebra. This is the generalization of the fact that the Magri bracket realizes Virasoro. This correspondence, first noticed by Khovanova [**44**], has been exploited by Fateev and Lukyanov [**45**] in order to define and quantize the $\mathsf{W}_n$ algebras by quantizing a deformation of the Miura transformation. Prior to the work of Fateev and Lukyanov, the only quantum $\mathsf{W}_n$ algebras known were the Virasoro algebra and the $\mathsf{W}_3$ algebra of Zamolodchikov, which was constructed as a solution to the conformal bootstrap.

### 3.3.  GENERALIZED KDV HIERARCHIES

Finally we introduce some dynamics in the space of Lax operators, thus fulfilling the setup of Section 2.1. In this section we will discuss the hierarchy of isospectral flows of a differential operator $L = \partial^n + \sum_i u_i \partial^{n-i}$. The isospectral problem associated to a differential operator consists in determining the flows which leave its spectrum invariant. We will be able to determine all isospectral flows which are of Lax type. As we will see, these flows commute and are bihamiltonian with respect to the Gel'fand–Dickey brackets. Moreover the hamiltonians generating these flows provide us with an infinite number of nontrivial, independent, polynomial, conserved charges in involution, rendering the hierarchy (formally) integrable.

We start by considering an arbitrary Lax operator

$$L = \partial^n + \sum_{i=1}^{n} u_i \partial^{n-i} \ . \tag{3.3.1}$$

By arbitrary we mean that the coefficients are differentially independent or, in other words, that they freely generate the differential ring $R$. By a flow we mean a derivation $\partial_t$ which commutes with $\partial$ and such that

$$\partial_t L = \sum_{i=1}^{n} (\partial_t u_i) \partial^{n-i} \ . \tag{3.3.2}$$

Suppose now that we realize the Lax operator as a differential operator acting on smooth functions somewhere, and that $\psi$ is a (formal) eigenfunction with eigenvalue $\lambda$:

$$L \cdot \psi = \lambda \psi \ . \tag{3.3.3}$$

Then the isospectral problem associated to $L$ consists of determining all flows $\partial_t$ such that $\partial_t \lambda = 0$. In other words, applying $\partial_t$ to (3.3.3), we find that an isospectral flow obeys

$$\partial_t L \cdot \psi + L \cdot \partial_t \psi - \lambda \partial_t \psi = 0 \ . \tag{3.3.4}$$

We can obtain isospectral flows by the Lax method. Let $P$ be another differential operator, and define

$$\partial_t L \equiv [P \, , \, L] \ . \tag{3.3.5}$$

Then defining $\partial_t \psi \equiv P \cdot \psi$, we find that $\partial_t \lambda = 0$. However, not every differential operator $P$ gives rise to an isospectral flow in this fashion. From (3.3.2) we see that $\partial_t L$ is a differential operator of degree $n - 1$, whence we must impose that $[P \, , L]$ have at most that order. Suppose that $P$ has order $m$, then in general $[P \, , L]$ has order at most $m + n - 1$. Demanding that it have order at most $n - 1$ imposes $m$ conditions on $P$, which has a priori $m + 1$ independent components. Thus we expect that $P$ will not be uniquely determined by its order alone. However we notice that there is always the possibility of adding to any such $P$ a zeroth order operator since for any function $f$, $[L \, , f]$ is a possible isospectral flow. We can take care of these 'trivial' flows in the following fashion. For any function $\varphi$, the operators $L$ and $e^{-\varphi} L e^{\varphi}$ have the same spectrum. Furthermore, choosing $\varphi$ judiciously[5] we can gauge away the coefficient in $L$ of order $n-1$. Therefore we can restrict ourselves to operators $L$ of the form

$$L = \partial^n + \sum_{i=2}^{n} u_i \partial^{n-i} \ . \tag{3.3.6}$$

It is then easy to see that a differential operator $P$ or order $m$ gives a consistent flow if $[P\,,L]$ has order at most $n-2$ which implies $m+1$ conditions on the $m+1$ components of $P$. We will see, in fact, that there is precisely one such $P$ of a given order.

FRACTIONAL POWERS AND THE COMMUTANT

Let us define the subset $\Omega_L$ of differential operators to be the set of those differential operators $P$ such that $[P\,,L]$ is a consistent evolution equation for the coefficients of $L$—in other words, $[P\,,L]$ has order at most $n-1$ for the generic $L$ or order $n-2$ if $u_1 = 0$. The isospectral problem associated to $L$ consists in determining this set. Towards this goal it is convenient to consider the commutant $Z_L$ of $L$ defined as those pseudodifferential operators commuting with $L$. The next results links the commutant intimately to $\Omega_L$.

LEMMA 3.3.7. *If $A \in Z_L$, then $A_+ \in \Omega_L$. In fact, $[A_+\,,L]$ has order at most $n-2$.*

PROOF. Break up $A$ as $A = A_+ + A_-$. Then since $A \in Z_L$ it follows that $[A_+\,,L] = [L\,,A_-]$. The right-hand side is a differential operator; whereas the left-hand side has order[6] at most $n-2$. □

Therefore it behooves us to study the commutant $Z_L$ of $L$. It does not cost anything extra to study this in a bit more generality; so we will let $L$ be a pseudodifferential operator. The following result is crucial:

PROPOSITION 3.3.8. *Let $L = \partial^n + \cdots$ be an otherwise arbitrary $\Psi\mathsf{DO}$. Then there exists a unique $\Psi\mathsf{DO}$ $L^{1/n} = \partial + \cdots$ which obeys $(L^{1/n})^n = L$; and, moreover, its coefficients are differential polynomials in the coefficients of $L$.*

---

[5] Of course, this choice of $\varphi$ involves integrating $u_1$ and hence does not live in $R$. Nevertheless, the modifications to the other $u_j$ always involve derivatives of $\varphi$ and hence belong to $R$.

[6] We use here the following fact: if $P$ and $Q$ are $\Psi\mathsf{DO}$'s of orders $p$ and $q$ respectively, then their product $PQ$ has order at most $p+q$, and their commutator $[P\,,Q]$ has order $p+q-1$.

PROOF. Indeed, suppose that $L$ is given by $L = \partial^n + \sum_{i \geq 1} u_i \partial^{n-i}$ and let $\Lambda = \partial + \sum_{i \geq 1} a_i \partial^{1-i}$. Computing one finds that

$$\Lambda^n = \partial^n + n a_1 \partial^{n-1} + O(\partial^{n-2}) \ , \qquad (3.3.9)$$

whence, if we take $a_1 = \frac{1}{n} u_1$, $\Lambda^n - L = O(\partial^{n-2})$. Suppose now that $a_1, \ldots, a_{k-1}$, differential polynomials in (the coefficients of) $L$, have been found so that $\Lambda^n - L = O(\partial^{n-k-2})$. A brief calculation shows that the term of order $\partial^{n-k-2}$ is given by $n a_k - X - u_k$, where $X$ is some differential polynomial in the $u_i < k$ and in the $a_{i<k}$—hence in the $u_{i<k}$. Therefore setting $a_k = \frac{1}{k}(u_k + X)$, allows to extend the induction hypothesis one step further. It is clear that in the limit, $L^{1/n} = \Lambda$ is the desired $n^{\text{th}}$ root. $\qquad \square$

Notice that it follows from the proof that if $L$ is a differential operator, then the first $n$ coefficients $a_1, a_2, \ldots, a_n$ of $L^{1/n}$ are differentially independent. In fact, the map sending $\{u_1, u_2, \ldots, u_n\}$ to $\{a_1, a_2, \ldots, a_n\}$ is a differential ring isomorphism.

The existence of the $n$-th root allows us to define fractional powers $L^{k/n}$ of $L$ for any $k \in \mathbb{Z}$, by

$$L^{k/n} = \begin{cases} (L^{1/n})^k, & \text{for } k \geq 0 \ , \\ (L^{-1/n})^{-k}, & \text{for } k < 0 \ ; \end{cases}$$

where $L^{-1/n} = \partial^{-1} + \cdots$ is the unique inverse to $L^{1/n}$ whose existence is proven in exactly the same way as the existence of the $n^{\text{th}}$ root. It is clear that the fractional powers commute with $L$ since they are both integer powers of $L^{1/n}$. But, in fact, it turns out that the fractional powers precisely generate the commutant.

PROPOSITION 3.3.10. *As a vector space over the constants, $Z_L$ is generated by the fractional powers $L^{k/n}$, for $k \in \mathbb{Z}$.*

PROOF. Let $A = \sum_{i \leq m} a_i \partial^i$ be a $\Psi\mathsf{DO}$ of order $m$ which commutes with $L$. We compute $[A \, , L]$ and we set each coefficient equal to zero. The term of highest order, $O(\partial^{n+m-1})$, is proportional to $a'_m$, whence we deduce that $a_m$ is a constant. Therefore $A - a_m L^{m/n}$ is a $\Psi\mathsf{DO}$ of order at most $m-1$ which commutes with $L$. By induction we are done. $\qquad \square$

It follows from Lemma 3.3.7 that for all non-negative integers $r$, $L_+^{r/n} \in \Omega_L$. In fact we have the following

PROPOSITION 3.3.11. *For $L = \partial^n + \cdots$ an otherwise arbitrary $\Psi\mathsf{DO}$, $\Omega_L$ is generated by the $L_+^{r/n}$ for $r \in \mathbb{N}$ and by $R$. When we impose that $u_1 = 0$, then $\Omega_L$ is generated only by the $L_+^{r/n}$, for $r = 0, 1, \ldots$.*

PROOF. Suppose that $P = \sum_{i=0}^{m} a_i \partial^i \in \Omega_L$, for some $m > 0$. Then the highest order term in $[P, L]$ is of order $\partial^{n+m-1}$ which, for $m > 0$, must vanish for $P \in \Omega_L$. The coefficient is proportional to $a'_m$, whence we find that $a_m$ is a constant. Therefore $P - a_m L_+^{m/n} \in \Omega_L$ is a differential operator of order $m - 1$. Continuing in this fashion we arrive at a zeroth order operator—*i.e.*, an element of $R$—which from a previous remark is trivially in $\Omega_L$ as long as $u_1$ is different from zero. Otherwise, we must demand that $[P, L]$ have order at most $n - 2$, whence only the constants in $R$ survive.                                          □

FLOWS AND CONSERVED CHARGES

Each differential operator in $\Omega_L$ defines a flow on the space of Lax operators as follows:

$$\partial_i L \equiv \left[ L_+^{i/n} , L \right] . \tag{3.3.12}$$

If $L$ is a differential operator then the flows $\partial_{rn}$ are trivial since $L_+^r = L^r$ commutes with $L$. The above hierarchy of flows determines the $n$-KdV hierarchy. The first important property of the Lax flows is that they commute. The proof follows by a routine calculation.

PROPOSITION 3.3.13. *For all $i, j \in \mathbb{N}$, $\partial_i \partial_j L = \partial_j \partial_i L$.*                    □

This means that we can introduce an infinite number of times $\{t_1, t_2, \ldots\}$ such that $\partial_i = \partial/\partial t_i$.

Because the Lax flows are given in terms of commutators, we can immediately write down conservation laws using the Adler trace. Indeed, for $r \in \mathbb{N}$, let

$$H_r \equiv \frac{n}{r} \operatorname{Tr} L^{r/n} . \tag{3.3.14}$$

Again, if $L$ is a differential operator, $H_{jn} = 0$ since $L^j$ is differential and therefore traceless. It is clear that these functions are polynomial and moreover that they are conserved, since

$$\frac{\partial H_r}{\partial t_i} = \frac{n}{r} \operatorname{Tr} \left[ L_+^{i/n} , L^{r/n} \right] = 0 , \tag{3.3.15}$$

where we have used that $\partial_t L = [P, L]$ if and only if $\partial_t L^q = [P, L^q]$, for any fractional power $q$.

We will now show that they are nontrivial and independent. Independence follows simply from a grading argument. Let us define the following grading: $[\partial] = 1$, and let us define the grading on $R$ by demanding that $[L] = n$, so that $[u_i] = n - i$, and that $[f'] = [f] + 1$ for any homogeneous element $f$ in $R$. If we further define $[\partial^{-1}] = -1$, then the ring of $\Psi$DO's becomes a graded ring and hence $[L^{r/n}] = r$. Therefore $[\operatorname{Tr} L^{r/n}] = r$ and its gradient—which will be computed shortly—has grade $[dH_r] = r - n$, whence the gradients are linearly independent.

From now on we will restrict $L$ to be a differential operator. This way we can apply the differential calculus we set up in the space $\mathfrak{M}$ of Lax operators.

Let us first compute the gradient of $H_r$. By definition, if $A$ is any vector field, the gradient $dH_r$ is defined by

$$\operatorname{Tr} dH_r A = \frac{d}{d\epsilon} H_r[L + \epsilon A]\Big|_{\epsilon=0} \; . \tag{3.3.16}$$

A simple exercise in the calculus developed in Section 1 yields,

$$dH_r = L_-^{r/n-1} \bmod \partial^{-n}\mathcal{R}_- \; . \tag{3.3.17}$$

### BIHAMILTONIAN STRUCTURE

We are now in a position to prove that the $H_r$ generate the Lax flows relative to the Gel'fand–Dickey brackets.

Recall that the Gel'fand–Dickey brackets are induced from the following hamiltonian maps:

$$J_0(X) = (LX)_+ L - L(XL)_+ = L(XL)_- - (LX)_- L \; ,$$
$$J_\infty(X) = [L\,,\,X]_+ \; ,$$

for any 1-form $X$. In particular, let us compute

$$\begin{aligned}
J_0(dH_r) &= J_0(L_-^{r/n-1}) \\
&= L(L_-^{r/n-1}L)_- - (LL_-^{r/n-1})_- L \\
&= LL_-^{r/n} - L_-^{r/n}L \\
&= \left[L_+^{r/n}\,,\,L\right] = \frac{\partial L}{\partial t_r} \; , 
\end{aligned} \tag{3.3.18}$$

which implies that for any function $F$, its evolution in the $t_r$ direction is given by

$$\frac{\partial F}{\partial t_r} = \{H_r\,,\,F\}_0 \; . \tag{3.3.19}$$

Similarly,

$$J_\infty(dH_{r+n}) = \left[L\,,\,L_-^{r/n}\right] = \left[L_+^{r/n}\,,\,L\right] = \frac{\partial L}{\partial t_r} \; , \tag{3.3.20}$$

whence

$$\frac{\partial F}{\partial t_r} = \{H_{r+n}\,,\,F\}_\infty \; . \tag{3.3.21}$$

In other words, the Lax flows are bihamiltonian with respect to the two Gel'fand–Dickey brackets—the $r^{\text{th}}$ flow $\partial_{t_r}$ being generated by $H_{r+n}$ and $H_r$ relative to the first and second brackets, respectively. The relations

$$J_\infty(dH_{r+n}) = J_0(dH_r) \tag{3.3.22}$$

are known as the Lenard recursive relations for the conserved charges. In principle, knowing $H_r$ for $r = 1, 2, \ldots, n-1$, one can determine all the others using the recursion relations.

As a result of the conservation of the $H_r$, we find that they are in involution with respect to both Gel'fand–Dickey brackets. In other words, for all $i, j$

$$0 = \frac{\partial H_i}{\partial t_j} = \{H_j\,,\, H_i\}_0 = \{H_{j+n}\,,\, H_i\}_\infty \ . \tag{3.3.23}$$

It is not difficult to show, using the bihamiltonian structure, that the conserved charges are in fact nontrivial, for if $H_i$ were trivial, so would the flows $\partial_i$ and $\partial_{i-n}$, and this can in turn be seen to imply that $H_{i-n}$ is trivial. Continuing in this fashion, we arrive at a contradiction since it is easy to see explicitly that the first $n-1$ conserved quantities $H_1, H_2, \ldots, H_{n-1}$ are nontrivial.

In summary, we have exhibited an infinite number of nontrivial, independent, polynomial, conserved quantities in involution relative to the bihamiltonian structure. This proves that $n$-KdV is (formally) completely integrable.

## 3.4. THE KP HIERARCHY: A UNIVERSAL KDV HIERARCHY

We saw in the previous section that, in the systematic treatment of the generalized KdV hierarchies, a crucial role was played by the $n$-th root $L^{1/n}$ of the Lax operator. This is a $\Psi\mathsf{DO}$ of the form

$$\Lambda = \partial + \sum_{i=1}^\infty a_i \partial^{1-i} \ , \tag{3.4.1}$$

obeying the constraint

$$\Lambda^n_- = 0 \ , \tag{3.4.2}$$

and such that $n$ is the smallest natural number for which this is true. We also saw that this constraint meant that the first $n$ coefficients in $\Lambda$ are differentially independent and that, in fact, they freely generated the differential ring of the coefficients of $L$. The Kadomtsev-Petviashvili (KP) hierarchy is defined as the hierarchy of isospectral flows of a $\Psi\mathsf{DO}$ $\Lambda$ of the form (3.4.1) without imposing the constraint (3.4.2); that is, a $\Psi\mathsf{DO}$ $\Lambda$ all whose coefficients are independent. Similar arguments to the ones given in the last section justify the reduction $a_1 = 0$.

LAX FLOWS AND CONSERVED CHARGES

We are again interested in flows of Lax type; that is, generated by differential operators $\Pi$ such that

$$\partial_t \Lambda = [\Pi, \Lambda] \tag{3.4.3}$$

is a consistent evolution equation. This means that the right-hand side has order at most 0 for the unreduced operator or $-1$ when $a_1 = 0$. Let us call the set of such $\Pi$, $\Omega_\Lambda$. We let $Z_\Lambda$ denote the commutant of $\Lambda$ in the set of $\Psi\mathsf{DO}$'s. As in the KdV hierarchies, there is an intimate connection between $Z_\Lambda$ and $\Omega_\Lambda$.

LEMMA 3.4.4. *If $A \in Z_\Lambda$, then $A_+ \in \Omega_\Lambda$ and, in fact, $[A_+, \Lambda]$ has order at most $-1$.*

PROOF. If $[A, \Lambda] = 0$, then $[A_+, \Lambda] = [\Lambda, A_-]$, which has order at most $-1$. $\square$

Analogous to Proposition 3.3.10 and Proposition 3.3.11 we have the following two results:

PROPOSITION 3.4.5. *As a vector space over the constants, $Z_\Lambda$ is generated by the powers $\Lambda^k$ of $\Lambda$, for $k \in \mathbb{Z}$.* $\square$

PROPOSITION 3.4.6. *$\Omega_\Lambda$ is generated by $\Lambda_+^r$ for $r \in \mathbb{N}$ and by the differential ring $R_\Lambda$ generated by the coefficients of $\Lambda$. When we impose that $a_1 = 0$, then $\Omega_\Lambda$ is simply generated by $\Lambda_+^r$, for $r = 0, 1, \ldots$.* $\square$

Until further notice we will work with the reduced KP operator where $a_1 = 0$. Each $\Lambda_+^i \in \Omega_\Lambda$ defines an isospectral flow

$$\partial_i \Lambda = \left[ \Lambda_+^i, \Lambda \right] ; \tag{3.4.7}$$

and, again, one can show that, analogously to Proposition 3.3.13, they commute so that we can introduce an infinite number of times $\{t_1, t_2, \ldots, t_n\}$ so that $\partial_i = \partial/\partial t_i$:

PROPOSITION 3.4.8. *For all $i, j \in \mathbb{N}$, $\partial_i \partial_j \Lambda = \partial_j \partial_i \Lambda$.* $\square$

For each $i \in \mathbb{N}$ let us define the function $h_i \equiv \frac{1}{i} \operatorname{Tr} \Lambda^i$. Using the fact that the Adler trace annihilates commutators, it follows that these obviously polynomial quantities are conserved. They are nontrivial because $\operatorname{res} \Lambda^i = a_{i+1} + \cdots$—where $\cdots$ is some differential polynomial of the $a_{j \leq i}$—and if it were a total derivative, there would exist a differential relation between the $a_i$, which violates the hypothesis of their independence. Independence of the $h_i$ is proven again via the introduction of a grading such that $[\partial] = 1$ and $[a_i] = i$ such that $[\Lambda] = 1$. Then $[\operatorname{Tr} \Lambda^i] = i + 1$ and their gradients are thus linearly independent.

REDUCTION TO THE $n$-KDV HIERARCHY

The phase space of the $n$-KdV hierarchy is the space of Lax operators $L$ of order $n$ with $u_1 = 0$. Since every such operator has a unique $n^{\text{th}}$ root of the form $L^{1/n} = \partial + O(\partial^{-1})$, this space is isomorphic to the space of $\Psi$DO's $\Lambda$ of the form $\Lambda = \partial + O(\partial^{-1})$ such that $\Lambda^n_- = 0$, which is clearly a subspace of the phase space of the KP hierarchy. Moreover, this subspace is preserved by the KP flows. Indeed, if $\Lambda$ obeys the evolution equation

$$\frac{\partial \Lambda}{\partial t_i} = \left[ \Lambda^i_+ \, , \, \Lambda \right] \; , \tag{3.4.9}$$

so does $\Lambda^n$–that is,

$$\frac{\partial \Lambda^n}{\partial t_i} = \left[ \Lambda^i_+ \, , \, \Lambda^n \right] \; . \tag{3.4.10}$$

Therefore, if $\Lambda^n_- = 0$, then

$$\left( \frac{\partial \Lambda^n}{\partial t_i} \right)_- = \left[ \Lambda^i_+ \, , \, \Lambda^n \right]_- = 0 \; , \tag{3.4.11}$$

since $\Lambda^n = \Lambda^n_+$. Under the map $\Lambda^n = L$, these flows are precisely the flows

$$\frac{\partial L}{\partial t_i} = \left[ L^{i/n}_+ \, , \, L \right] \tag{3.4.12}$$

of the $n$-KdV hierarchy. Moreover, and up to a factor $n$, the conservation laws are precisely the ones of the KdV hierarchy, the ones for $i$ a multiple of $n$ being trivial.

Therefore we conclude that the $n$-KdV hierarchy is a reduction of the KP hierarchy; or, in other words, that the KP hierarchy is a universal hierarchy for the KdV series.

The calculus developed in Section 1 can be extended to handle the space of pseudodifferential operators $\Lambda$. This is done, for example, in [**40**]. It turns out that the KP hierarchy is again bihamiltonian with respect to a version of the Gel'fand–Dickey brackets induced by an Adler-type map. In fact, in view of the result of Semenov-Tyan-Shanskiĭ quoted at the end of Section 1, this is not surprising. What may be surprising is that the KP hierarchy is hamiltonian relative to a one-parameter family of such hamiltonian structures analogous to the second Gel'fand–Dickey bracket. This result was obtained independently by Figueroa-O'Farrill, Mas and Ramos in [**46**] and by Khesin and Zakharevich [**47**]. It extends previous work of Radul [**48**] who found an infinite discrete family of such structures. The original hamiltonian structure for the KP hierarchy was discovered by Dickey [**49**].

The Poisson structures of the KP hierarchy provide examples of infinitely-generated W-algebras of the $W_\infty$-type. These are deformations of the algebra $w_\infty$ of area preserving diffeomorphisms on a two-dimensional phase space. These algebras are not yet classified, but all known such algebras can be related [**50**] via contractions or reductions to the one-parameter family of hamiltonian structures for the KP hierarchy discovered in [**46**] and [**47**]. $W_\infty$-type algebras have started to appear naturally in physical problems like the quantum Hall effect.

DRESSING TRANSFORMATIONS

The KP hierarchy can be understood as a dynamical system on a formal Lie group. Let $G$ denote the formal Lie group of $\Psi$DO's of the form $1 + \sum_i w_i \partial^i$. We don't take the $w_i$ to be elements of $R$ for reasons that will become obvious in a moment, but rather in some extension. Let $\Lambda$ be a $\Psi$DO of the form $\partial + \sum_{i \geq 1} a_i \partial^{1-i}$. It is not hard to prove that provided that we allow ourselves to formally integrate the $a_i$, we can undress $\Lambda$ as follows:

$$\Lambda = \phi \partial \phi^{-1} \qquad \text{for some } \phi \in G \ . \tag{3.4.13}$$

In fact, $\phi$ is unique up to multiplication on the right by an element of $G$ with constant coefficients. We can fix this ambiguity by demanding homogeneity of $\phi$ and demanding that the only constants have degree zero. We can now lift the KP flows via (3.4.13) to flows on the Volterra group $G$. The following results after a trivial computation:

PROPOSITION 3.4.14. *The flows*

$$\partial_n \phi = \frac{\partial \phi}{\partial t_n} \equiv -(\phi \partial^n \phi^{-1})_- \phi$$

*induce via (3.4.13) the flows (3.4.7).* □

It is amusing and moreover practical when we consider the additional symmetries in Chapter Five, to understand these flows in a different way. Let us extend the ring of $\Psi$DO's by the derivations $\partial_n$ of (3.4.7) in such a way that $\partial \partial_n = \partial_n \partial$. Then consider the trivial relation $[\partial_n - \partial^n, \partial] = 0$. If we dress this relation with $\phi$ we recover the flows in Proposition 3.4.14. We will exploit this approach to the KP hierarchy and to its supersymmetric extensions in Chapter Five when we discuss the algebra of additional symmetries.

# Chapter Four

# SUPERSYMMETRIC INTEGRABLE HIERARCHIES

We now start the discussion of supersymmetric integrable hierarchies. Just as in the nonsupersymmetric case we need to develop some formalism to handle the infinite-dimensional spaces of Lax operators. As before the arena will be the ring of formal superpseudodifferential operators $\mathsf{S\Psi DO}$'s. We discuss them in Section 1. After defining and reviewing the basic properties of this ring we discuss the formal geometry in the space of Lax operators and we prove that this space can be given the structure of a Poisson manifold relative to the supersymmetric version of the Adler map. The proof of this fact is done in somewhat more generality. It is the supersymmetric analogue of a theorem of Semenov-Tyan-Shanskiĭ, and we believe that it appears here for the first time. In Section 2 we discuss the various supersymmetric extensions of the KP hierarchy and we set the stage for the study of their additional symmetries in the following chapter. Most of the material in this chapter follows the series of papers [**29**], [**33**], [**51**], [**32**], and [**52**].

## 4.1. SUPERSYMMETRIC LAX FORMALISM

### PSEUDODIFFERENTIAL OPERATORS IN SUPERSPACE

Let $k$ be an arbitrary field of characteristic zero and let $S = S_0 \oplus S_1$ denote a $\mathbb{Z}_2$-graded ring over $k$. Let $S$ be moreover endowed with an odd derivation $D$. Then $D^2$ is an even derivation which we will call $\partial$. We will think of $S$ as our function space. As an example we can take $S$ to be the ring

$$k[[x, \theta]] = k[[x]] \oplus k[[x]]\theta \qquad (4.1.1)$$

of formal power series in an even variable $x$ and an odd variable $\theta$, satisfying $x\theta = \theta x$ and $\theta^2 = 0$. The $\mathbb{Z}_2$-grading is defined by putting $|x| = 0$ and $|\theta| = 1$, and the odd derivation is given by $D = \partial_\theta + \theta\partial$, where $\partial = \partial/\partial x$. When we come to discuss supersymmetric hierarchies, we will take $S$ to be the differential ring generated by some superfields $U_i \in k[[x, \theta]]$.

In any case, the odd derivation $D$ obeys the supersymmetric analog of the

Leibniz rule

$$D(ab) = D(a)b + (-)^{|a|}aD(b) \ , \tag{4.1.2}$$

where $a$ is a homogeneous element of $S$ of $\mathbb{Z}_2$-degree $|a|$ and $|D| = 1$. We further define the ring of supersymmetric pseudodifferential operators (S$\Psi$DO) with coefficients in $S$

$$\mathcal{S} \equiv S((D^{-1})) = \left\{ P = \sum_{i \gg -\infty} a_i D^{-i} \,\middle|\, a_i \in S \right\} \ . \tag{4.1.3}$$

The ring of S$\Psi$DO's can be given the structure of a superalgebra using the generalized Leibniz rule

$$D^k a = \sum_{i=0}^{\infty} \begin{bmatrix} k \\ k-i \end{bmatrix} (-)^{|a|(k-i)} a^{[i]} D^{k-i} \ , \tag{4.1.4}$$

where $a$ is a homogeneous element of $S$ and $\begin{bmatrix} k \\ k-i \end{bmatrix}$ are the so-called superbinomial coefficients given by

$$\begin{bmatrix} k \\ k-i \end{bmatrix} = \begin{cases} 0 & \text{for } i < 0 \text{ or } (k,i) \equiv (0,1) \bmod 2; \\ \begin{pmatrix} \left[\frac{k}{2}\right] \\ \left[\frac{k-i}{2}\right] \end{pmatrix} & \text{for } i \geq 0 \text{ and } (k,i) \not\equiv (0,1) \bmod 2. \end{cases}$$

The generalized Leibniz rule follows from the one for $\partial$ (given in (3.2.2)) by taking $D^{2k} = \partial^k$ and $D^{2k+1} = \partial^k D$. Since the $\mathbb{Z}_2$-grading gets induced here we have that $\mathcal{S} = \mathcal{S}_0 \oplus \mathcal{S}_1$ where

$$\mathcal{S}_0 = S_0((D^{-1})) = \left\{ \sum_{i \gg -\infty} a_i D^{-i} \,\middle|\, |a_{2i}| = 0 \ , |a_{2i+1}| = 1 \right\} \ , \tag{4.1.5}$$

and

$$\mathcal{S}_1 = S_1((D^{-1})) = \left\{ \sum_{i \gg -\infty} a_i D^{-i} \,\middle|\, |a_{2i}| = 1 \ , |a_{2i+1}| = 0 \right\} \ , \tag{4.1.6}$$

and we have thus defined the notion of an even (respectively odd) S$\Psi$DO.

In the case $S = k[[x, \theta]]$ or some freely-generated subring, let us remark the following fact:

$$S((D^{-1})) = S((\partial^{-1})) \oplus S((\partial^{-1}))\partial_\theta \ . \tag{4.1.7}$$

Indeed, on the one hand we clearly have $S((\partial^{-1})) \oplus S((\partial^{-1}))\partial_\theta \subset S((D^{-1}))$ since $D^2 = \partial$ and $\partial_\theta = D - \theta D^2$. On the other hand any S$\Psi$DO can be written in the

following manner:

$$\sum_i a_i D^i = \sum_i a_{2i} D^{2i} + \sum_i a_{2i+1} D^{2i+1}$$

$$= \sum_i a_{2i} \partial^i + \sum_i a_{2i+1} \partial^i (\partial_\theta + \theta \partial)$$

$$= \sum_i \left( a_{2i} + a_{2i+1} \theta \right) \partial^i + \sum_i a_{2i-1} \partial^i \partial_\theta \; , \qquad (4.1.8)$$

so that we have also $S((D^{-1})) \subset S((\partial^{-1})) \oplus S((\partial^{-1})) \partial_\theta$.

In general it is important to distinguish in the ring of $\mathsf{S\Psi DO}$'s the subring of supersymmetric differential operators ($\mathsf{SDOP}$'s)

$$\mathcal{S}_+ \equiv S[D] = \left\{ \sum_{0 \le i \ll \infty} a_i D^i \middle| a_i \in S \right\} \; , \qquad (4.1.9)$$

with respect to which we have the splitting

$$\mathcal{S} = \mathcal{S}_+ \oplus \mathcal{S}_- \; , \qquad (4.1.10)$$

where

$$\mathcal{S}_- \equiv D^{-1} S[[D^{-1}]] = \left\{ \sum_{i=1}^\infty a_i D^{-i} \middle| a_i \in S \right\} \qquad (4.1.11)$$

denotes the integral $\mathsf{S\Psi DO}$'s. If $P \in \mathcal{S}$ is any $\mathsf{S\Psi DO}$ we shall denote by $P_\pm$ its projection onto $\mathcal{S}_\pm$ along $\mathcal{S}_\mp$.

The ring $\mathcal{S}$ of $\mathsf{S\Psi DO}$'s can be made into a filtered associative $k$-algebra if we define the space supersymmetric pseudodifferential operators of order $n$ by

$$\mathcal{S}^n = \left\{ P = \sum_i a_i D^i \in S[D] \middle| \operatorname{ord} P \le n \right\} \; , \qquad (4.1.12)$$

where we say that $\operatorname{ord} P = n$ if $a_i = 0$ for all $i > n$ and $a_n \ne 0$. We have then

$$\mathcal{S}^n \subset \mathcal{S}^{n+1} \;\; \text{and} \;\; \mathcal{S} = \cup_{n \in \mathbb{Z}} \mathcal{S}^n \qquad (4.1.13)$$

and under the multiplication $\mathcal{S}^p \times \mathcal{S}^q \to \mathcal{S}^{p+q}$ . Moreover defining a bracket

$$[ \;\; ] : \mathcal{S}^p \times \mathcal{S}^q \to \mathcal{S}^{p+q} \qquad (4.1.14)$$

via the graded commutator $[AB] \equiv AB - (-)^{|A||B|} BA$, $\mathcal{S}$ becomes a Lie superalgebra.

Let us end this discussion on the ring of $\mathsf{S\Psi DO}$'s by introducing the super-symmetric analog of the Adler trace. If $P = \sum_i a_i D^i$, we define its noncommutative super-residue by $\operatorname{sres} P = a_{-1}$. In order to define the supertrace we need to introduce a notion of integration. As in Section 3.1 this can be defined abstractly as the canonical projection $\int : S \to S/DS$, which simply means dropping the perfect derivatives. In the presence of a more explicit realization for $S$, say as a subring of $k[[x, \theta]]$, we can make the notion of integration more concrete by considering the Berezinian. For any homogeneous differential polynomial of $\{U_i = u_i + \theta v_i\}$, $f(U) = a(u, v) + \theta b(u, v)$, the Berezinian is defined by $\int_B f(U) = \int b(u, v)$. For such rings $S$, the only difference between the two notions of integration consists in the fact that the abstract integration is an even operation, whereas the Berezinian has a $\mathbb{Z}_2$-degree of one. In any case the Adler supertrace is defined by

$$\operatorname{Str} = \int \circ \operatorname{sres} \ . \tag{4.1.15}$$

. It is a straightforward computation to show that, analogously to (3.2.3), the super-residue of a graded commutator is a perfect $D$ derivative, so that $\operatorname{Str}[A, B] = 0$, for two $\mathsf{S\Psi DO}$'s $A$ and $B$.

GEOMETRY OF THE SPACE OF LAX OPERATORS

One of the central objects in our formalism is the space of supersymmetric Lax operators of degree $n$, defined by

$$\mathfrak{M}_n \equiv \left\{ L = D^n + \sum_{i=1}^{\infty} U_i D^{n-i} \, \middle| \, U_i \in S, |U_i| \equiv i \bmod 2 \right\} \ . \tag{4.1.16}$$

(We shall drop the subscript $n$ whenever no confusion can arise.) Given any $L \in \mathfrak{M} \subset \mathcal{S}$ one can define $S_L$ the differential subring generated by the $U_i$'s which will obviously induce the corresponding subrings $S_L[D] \subset S[D]$ and $S_L[[D^{-1}]] \subset S[[D^{-1}]]$. $\mathfrak{M}$ is an infinite-dimensional affine space modeled on the vector space $\mathcal{S}^{n-1}$ of $\mathsf{S\Psi DO}$'s of order $n-1$. Our aim is to endow this space with a Poisson structure. Using the formalism in Sections 3.1-2 as a guide, we will define Poisson brackets on functions of the form:

$$F[L] = \int_B f(U) \ , \tag{4.1.17}$$

where $f(U)$ is a homogeneous differential polynomial of the $U$'s and $\int_B$ is the previously defined Berezinian. It is worth remarking that whereas $S_L$ is a graded supercommutative algebra, $DS_L$ is not an ideal and hence the multiplication in $S_L$ does not get induced in the quotient. This means, in particular, that it will not make sense to demand of our Poisson brackets to satisfy the usual derivation property but this, fortunately, will not affect the formalism.

The tangent space $T_L\mathfrak{M}$ to $\mathfrak{M}$ at the point $L$ is isomorphic to the infinitesimal deformations of $L$, which are pseudodifferential operators of order at most $n-1$. In other words the tangent space is isomorphic to $\mathcal{S}^{n-1}$ itself, namely

$$T_L\mathfrak{M} \equiv \left\{ A = \sum_{i=1}^{\infty} A_i D^{n-i} \,\middle|\, A_i \in S, |A_i| \equiv |A| + n - i \bmod 2 \right\} . \qquad (4.1.18)$$

Then to every tangent vector $A \in T_L\mathfrak{M}$ one can associate a vector field $D_A$ whose action on any $f \in S_L$ is defined by

$$\begin{aligned} D_A f &\equiv \frac{d}{d\epsilon} f\left(U_i + \epsilon A_i\right)\bigg|_{\epsilon=0} \\ &= \sum_{i=1}^{\infty}\sum_{k=0}^{\infty} (-)^{(|A|+n)k} A_i^{[k]} \frac{\partial f}{\partial U_i^{[k]}} , \end{aligned} \qquad (4.1.19)$$

where we do not impose a priori that $\epsilon$ be even; in other words, $|L|$ and $|A|$ need not agree. A straightforward computation shows us that

$$D_A D = (-)^{(|A|+n)} D D_A . \qquad (4.1.20)$$

Notice that $D_A : S_L \to S_L$ induces a map—also denoted $D_A$ with some abuse of notation—$D_A : S_L/DS_L \to S_L/DS_L$ given by $D_A \int f = \int D_A f$. More explicitly if we denote $F[L] = \int_B f$, then the vector field $D_A$ defined by $A$ is given by

$$D_A F = (-)^{|A|+n} \int_B \sum_{k=0}^{n-1} A_k \mathcal{E}_k \cdot f, \qquad (4.1.21)$$

where we have introduced the Euler operators

$$\mathcal{E}_k = \sum_{i=0}^{\infty} (-)^{|U_k|i + i(i+1)/2} D^i \frac{\partial}{\partial U_k^{[i]}} , \qquad (4.1.22)$$

with $U_k^{[i]} = D^i U_k$. One can check that $D_A$ is well defined. But in this case since $S_L/DS_L$ is no longer a superalgebra, $D_A$ is no longer a derivation.

We expect the cotangent space $T_L^*\mathfrak{M}$ to $\mathfrak{M}$ at $L$ to be defined as the dual space of $T_L\mathfrak{M}$. In order to see this we introduce a dual pairing in $\mathcal{S}$ given by

$$< A, B > \equiv \mathrm{Str}\, AB , \qquad (4.1.23)$$

under which $S[D]$ and $D^{-1}S[[D^{-1}]]$ are maximally isotropic spaces and nondegenerately paired with each other. Indeed, if we take $X = \sum_{k=0}^{\infty} D^{-k-1} X_k \in \mathcal{S}_-$ and $A = \sum_{k=0}^{n-1} A_k D^k \in \mathcal{S}_+$, their pairing is given by

$$\mathrm{Str}\, AX = (-)^{|X|+1} \int_B \sum_{k=0}^{n-1} (-)^k A_k\, X_k \ . \tag{4.1.24}$$

Hence the tangent space $T_L\mathfrak{M}$ is nondegenerately paired (via the pairing defined in (4.1.23)) with the quotient space $\mathcal{S}/D^{-n}\mathcal{S}_-$ and therefore we have

$$T_L^*\mathfrak{M} \cong \mathcal{S}/D^{-n}\mathcal{S}_- \ . \tag{4.1.25}$$

A generic element here will be then an integral operator $X \in \mathcal{S}_-$ of the form $X = \sum_{k=0}^{n-1} D^{-k-1}X_k$ and, with a little abuse of notation, we also let $X$ denote the one-form it gives rise to at $L$. Thus if $X$ and $A$ are as above, the pairing between the vector field $D_A$ and the one-form $X$ is given by

$$\langle D_A, X \rangle \equiv (-)^{|A|+|X|+n+1}\, \mathrm{Str}\, AX = (-)^{|A|+n} \int_B \sum_{k=0}^{n-1} (-)^k A_k\, X_k \ . \tag{4.1.26}$$

The strange choice of signs has been made to avoid undesirable signs later on. Given a function $F = \int_B f$ we define its gradient $dF$ by $\langle D_A, dF \rangle = D_A F$ whence, comparing with (4.1.21), yields

$$dF = \sum_{k=0}^{n-1} (-)^k D^{-k-1} \mathcal{E}_k \cdot f \ . \tag{4.1.27}$$

Thus, the gradient of a function is a one-form as expected.

To define a Poisson bracket we need a hamiltonian map $J : T_L^*\mathfrak{M} \to T_L\mathfrak{M}$. Given any two functions $F$ and $G$, their Poisson bracket $\{F, G\}$ is defined by

$$\{F, G\} \equiv D_{J(dF)}G = \langle D_{J(dF)}, dG \rangle = (-)^{|J|+|F|+|G|+n+1}\, \mathrm{Str}\, J(dF)dG \ . \tag{4.1.28}$$

$J$ is hamiltonian if and only if this obeys the appropriate (anti)symmetry properties and the Jacobi identity.

If $L$ is a superdifferential operator, the supersymmetric analogue of the Adler map $J : T_L^*\mathfrak{M} \to T_L\mathfrak{M}$ given by

$$J(X) = (LX)_+ L - L(XL)_+ = L(XL)_- - (LX)_- L \tag{4.1.29}$$

was shown in [**18**] to be hamiltonian as a corollary of a supersymmetric version of the Kupershmidt–Wilson theorem. In other words, what was shown was that $J$ is induced from a much simpler hamiltonian map in a different set of variables $\Phi_i$ defined by the formal factorization $L = (D - \Phi_n)(D - \Phi_{n-1}) \cdots (D - \Phi_1)$. The fundamental Poisson bracket in these variables is given by

$$\{\Phi_i(X)\,,\,\Phi_j(Y)\} = (-)^i \delta_{ij} D \delta(X - Y)\,, \qquad (4.1.30)$$

where, if $X = (x, \theta)$ and $Y = (y, \omega)$, then $\delta(X - Y) = \delta(x - y)(\theta - \omega)$. The change of variables from the $U_j$ to the $\Phi_i$ is a supersymmetric version of the Miura transformation. This factorization depends crucially on $L$ being differential. When $L$ is a pseudodifferential operator, there is no known analogue of the Miura transformation even in the nonsupersymmetric case. Many of the hierarchies we will consider, however, will be defined as isospectral deformations of a suitable class of $\mathsf{S\Psi DO}$'s for which we will need to prove the hamiltonian nature of the corresponding Adler map. Therefore it prompts us to find a proof that does not rely on the Kupershmidt–Wilson theorem. Again for $L$ differential, this was proven in [**51**] in a combinatorial fashion. In the next section we will give a more general proof which is valid for a variety of algebraic situations and not just for superpseudodifferential operators. This is the supersymmetric analogue of a theorem due originally to Semenov-Tyan-Shanskiĭ.

## 4.2.  THE HAMILTONIAN PROPERTY OF THE ALGEBRAIC ADLER MAP

Let $\mathfrak{g}$ be an associative superalgebra over a field $k$ of zero characteristic and suppose that it decomposes as the vector space direct sum of two subsuperalgebras $\mathfrak{g} = \mathfrak{g}_+ \oplus \mathfrak{g}_-$. In other words, $\mathfrak{g}$ is a $\mathbb{Z}_2$-graded vector space such that there is an associative multiplication which respects the grading and $\mathfrak{g}_\pm$ is a graded subspaces which is closed under the multiplication. From now on we will simply call $\mathfrak{g}$ an algebra and speak of $\mathfrak{g}_\pm$ as a subalgebra. Given any element $X \in \mathfrak{g}$ we denote by $X_\pm$ its projection to $\mathfrak{g}_\pm$ along $\mathfrak{g}_\mp$. Suppose further than we have a nondegenerate supertrace form $\mathrm{Str} : \mathfrak{g} \to k$ inducing a supersymmetric bilinear form $\langle X, Y \rangle = \mathrm{Str}\, XY$ which is maximally split; that is, such that the subalgebras $\mathfrak{g}_\pm$ are maximally isotropic; in other words, $\mathrm{Str}\, X_\pm Y_\pm = 0$.

Choose a homogeneous element $L \in \mathfrak{g}$ and define the generalized Adler map $J : \mathfrak{g} \to \mathfrak{g}$ as

$$J(X) = (LX)_+ L - L(XL)_+ = L(XL)_- - (LX)_- L\,. \qquad (4.2.1)$$

We can view this as an infinitesimal deformation $\delta_X L = J(X)$. More geometrically, however, we can view $J(X)$ as a vector field tangent to $\mathfrak{g}$ at $L$ as follows. Since $\mathfrak{g}$ is a linear space, we can identify its tangent and cotangent spaces with the algebra itself and, moreover, the dual pairing between the tangent and cotangent spaces is given by the trace form. Then $J$ can be interpreted as a way to assign vector fields to one-forms. Notice, that the parity of the vector field $\delta_X$ is not necessarily the parity of $X$, but rather it is the sum of the parities of $X$ and $L$. Let us then introduce the following notation: we will denote by $|X|$ the parity of $X$ and by $|\bar{X}| = |X| + |L|$, where we of course sum modulo 2. We furthermore use the abbreviation $\varepsilon = |L|$. We continue introducing conventions. If $B$ is a bilinear form on $\mathfrak{g}$ we shall say that it is **s-skewsymmetric** if

$$B(X, Y) = -(-)^{|\bar{X}||\bar{Y}|} B(Y, X) \ , \tag{4.2.2}$$

for any two homogeneous elements $X$ and $Y$. This will turn out to be the natural notion of skewsymmetry in the $\mathbb{Z}_2$-graded case. If $X, Y, Z$ are homogeneous elements of $\mathfrak{g}$ and $f : (X, Y, Z) \mapsto f(X, Y, Z)$ is any function (perhaps $\mathfrak{g}$-valued), we define the $\mathbb{Z}_2$ graded version of cyclic and signed permutations:

$$\mathop{\mathbf{C}}_{X,Y,Z} f(X, Y, Z) \equiv f(X, Y, Z) + (-)^{|\bar{X}|(|\bar{Y}|+|\bar{Z}|)} f(Y, Z, X)$$
$$+ (-)^{|\bar{Z}|(|\bar{X}|+|\bar{Y}|)} f(Z, X, Y)$$
$$\mathop{\mathbf{S}}_{X,Y,Z} f(X, Y, Z) \equiv \mathop{\mathbf{C}}_{X,Y,Z} \left( f(X, Y, Z) - (-)^{|\bar{Y}||\bar{Z}|} f(X, Z, Y) \right) \ .$$

The first thing we prove is that the vector fields obtained by $J$ close under the Lie bracket or, in terms of the infinitesimal deformations $\delta_X$, that they too form a closed algebra under the (graded)commutator. It is clear that by linearity we can restrict ourselves to homogeneous $X$.

LEMMA 4.2.3. *For all $X, Y \in \mathfrak{g}$, independent of $L$,*

$$[\delta_X \, , \, \delta_Y] = \delta_{[X \, , \, Y]_L^*} \ ,$$

*where $[X \, , \, Y]_L^*$ is given, modulo the kernel of $J$, by*

$$[X \, , \, Y]_L^* = X(LY)_- - (XL)_+ Y - (-)^{|\bar{X}||\bar{Y}|}(X \leftrightarrow Y) \ . \tag{4.2.4}$$

PROOF. Using that $\delta_X L = J(X)$, what we want to show is equivalent to

$$\delta_X J(Y) - (-)^{|\bar{X}||\bar{Y}|} \delta_Y J(X) = J([X \, , \, Y]_L^*) \ . \tag{4.2.5}$$

The LHS of this equation can be expanded to

$$(J(X)Y)_+ L + (-)^{|\bar{X}||\bar{Y}|}(LY)_+ J(X) - J(X)(YL)_+$$
$$\qquad - (-)^{|\bar{X}||\bar{Y}|}L(YJ(X))_+ - (-)^{|\bar{X}||\bar{Y}|}(X \leftrightarrow Y)$$
$$= ((LX)_+ LY)_+ L - (L(XL)_+ Y)_+ L + (-)^{|\bar{X}||\bar{Y}|}(LY)_+(LX)_+ L$$
$$\quad - (-)^{|\bar{X}||\bar{Y}|}\underbrace{(LY)_+ L(XL)_+ - (LX)_+ L(YL)_+}_{} - (-)^{|\bar{X}||\bar{Y}|}L(Y(LX)_+ L)_+$$
$$\quad + (-)^{|\bar{X}||\bar{Y}|}L(YL(XL)_+)_+ + L(XL)_+(YL)_+ - (-)^{|\bar{X}||\bar{Y}|}(X \leftrightarrow Y)$$
$$= (LX(LY)_-)_+ L - (L(XL)_+ Y)_+ L + L(X(LY)_+ L)_+$$
$$\quad - L((XL)_- YL)_+ - (-)^{|\bar{X}||\bar{Y}|}(X \leftrightarrow Y) \ ,$$

where the underbraced terms cancel under $X \leftrightarrow Y$ and we notice that

$$((LX)_+ LY)_+ L - (LX)_+(LY)_+ L = (LX(LY)_-)_+ L$$
$$L(XL)_+(YL)_+ - L(XL(YL)_+)_+ = -L((XL)_- YL)_+ \ .$$

Moreover since

$$(LX(LY)_-)_+ L + L(X(LY)_+ L)_+ = J(X(LY)_-) + L(XLYL)_+$$
$$(L(XL)_+ Y)_+ L + L((XL)_- YL)_+ = J((XL)_+ Y) + L(XLYL)_+ \ ,$$

we can rewrite the LHS of (4.2.5) as

$$J(X(LY)_- - (XL)_+ Y) - (-)^{|\bar{X}||\bar{Y}|}(X \leftrightarrow Y) \ ,$$

proving thus the lemma.                                                    $\square$

   If $X$ and $Y$ are allowed to depend on the point $L$—as they very well could, under their identification with one-forms on $\mathfrak{g}$—we would incur in terms of the form $(-)^{|\bar{X}|\varepsilon}J(\delta_X Y) - (-)^{|\bar{X}||\bar{Y}|}(X \leftrightarrow Y)$ in the LHS of (4.2.5). These terms clearly go along for the ride and all they do is modify the bracket $[X\,,\,Y]_L^*$. This modification, though, is important; for suppose that one asks oneself whether $[-\,,\,-]_L^*$ is a Lie bracket. It is clearly s-skewsymmetric, so that all we need to check is the Jacobi identity. From the definition, $[\delta_X\,,\,\delta_Y] = \delta_{[X\,,\,Y]_L^*}$ and the fact that the commutator on the LHS satisfies the Jacobi identity trivially, we expect that

$$\mathrm{Jacobi}_L(X, Y, Z) \equiv \underset{X,Y,Z}{\mathbf{C}}\, \big[X\,,\,[Y\,,\,Z]_L^*\big]_L^* = 0 \ . \tag{4.2.6}$$

But in computing the nested $[-\,,\,-]_L^*$ brackets, we notice that even if $X$, $Y$, and $Z$ do not depend on $L$, $[Y\,,\,Z]_L^*$ does, so we need to include the terms $\delta_X [Y\,,\,Z]_L^*$. These terms are indeed crucial to check the Jacobi identity for $[-\,,\,-]_L^*$, as we now show.

PROPOSITION 4.2.7. *The bracket*

$$[X\,,\,Y]_L^* \equiv (-)^{|\bar{X}|\varepsilon}\delta_X Y + X(LY)_- - (XL)_+Y - (-)^{|\bar{X}||\bar{Y}|}(X \leftrightarrow Y)$$

*satisfies the Jacobi identity.*

PROOF. Let $X, Y, Z \in \mathfrak{g}$ be homogeneous. We will show that (4.2.6) holds. For simplicity we work under the assumption that $X, Y, Z$ are $L$-independent so that we have no terms of the form $\delta_X Y$. The general case is no harder to prove. By definition,

$$\begin{aligned}
\big[X\,,\,[Y\,,\,Z]_L^*\big]_L^* = (-)^{|\bar{X}|\varepsilon}\delta_X\,[Y\,,\,Z]_L^* &+ X(LY(LZ)_-)_- - X(L(YL)+Z)_- \\
&-(XL)_+Y(LZ)_- + (XL)_+(YL)_+Z - (-)^{|\bar{X}|(|\bar{Y}|+|\bar{Z}|)}Y(LZ)_-(LX)_- \\
&+(-)^{|\bar{X}|(|\bar{Y}|+|\bar{Z}|)}(YL)_+Z(LX)_- + (-)^{|\bar{X}|(|\bar{Y}|+|\bar{Z}|)}(Y(LZ)_-L)_+X \\
&-(-)^{|\bar{X}|(|\bar{Y}|+|\bar{Z}|)}((YL)_+ZL)_+X - (-)^{|\bar{Y}||\bar{Z}|}(Y \leftrightarrow Z)\,.
\end{aligned}$$

Also by definition,

$$\delta_X\,[Y\,,\,Z]_L^* = (-)^{|\bar{X}||Y|}Y(J(X)Z)_- - (-)^{|\bar{X}||Y|}(YJ(X))_+Z - (-)^{|\bar{Y}||\bar{Z}|}(Y \leftrightarrow Z)\,,$$

which we choose to write as $(-)^{|\bar{X}||Y|}$ times

$$\begin{aligned}
Y((LX)_+LZ)_- - Y(L(XL)_+Z)_- &- (Y(LX)_+L)_+Z \\
&+(YL(XL)_+)_+Z - (-)^{|\bar{Y}||\bar{Z}|}(Y \leftrightarrow Z)\,.
\end{aligned}$$

Therefore we can write

$$\begin{aligned}
\mathrm{Jacobi}_L(X,Y,Z) = \mathop{\text{\Large S}}_{X,Y,Z} \Big[ &(-)^{|\bar{X}||\bar{Y}|}Y((LX)_+LZ)_- - (-)^{|\bar{X}||\bar{Y}|}Y(L(XL)_+Z)_- \\
&-(-)^{|\bar{X}||\bar{Y}|}(Y(LX)_+L)_+Z + (-)^{|\bar{X}||\bar{Y}|}(YL(XL)_+)_+Z + X(LY(LZ)_-)_- \\
&-X(L(YL)_+Z)_- - \underbrace{(XL)_+Y(LZ)_-} + (XL)_+(YL)_+Z - \\
&(-)^{|\bar{X}|(|\bar{Y}|+|\bar{Z}|)}Y(LZ)_-(LX)_- + (-)^{|\bar{X}|(|\bar{Y}|+|\bar{Z}|)}\underbrace{(YL)_+Z(LX)_-} \\
&+(-)^{|\bar{X}|(|\bar{Y}|+|\bar{Z}|)}(Y(LZ)_-L)_+X - (-)^{|\bar{X}|(|\bar{Y}|+|\bar{Z}|)}((YL)_+ZL)_+X \Big]\,,
\end{aligned}$$

where the underbraced terms cancel after taking into account the signed permutations. Now we notice that

$$\mathop{\text{\Large S}}_{X,Y,Z} -(XL)_+Y(LZ)_- + (-)^{|\bar{X}|(|\bar{Y}|+|\bar{Z}|)}(YL)_+Z(LX)_- = 0\,,$$

$$\mathop{\text{\Large S}}_{X,Y,Z} X(LY(LZ)_-)_- - (-)^{|\bar{X}|(|\bar{Y}|+|\bar{Z}|)}Y(LZ)_-(LX)_- = \mathop{\text{\Large S}}_{X,Y,Z} X((LY)_+LZ)_-\,,$$

and

$$\mathop{\text{\Large S}}_{X,Y,Z}(XL)_+(YL)_+Z + (-)^{|\bar{X}||\bar{Y}|}(YL(XL)_+)_+Z = \mathop{\text{\Large S}}_{X,Y,Z} -((XL)_-YL)_+Z\,.$$

We obtain in this fashion that

$$\text{Jacobi}_L(X, Y, Z)$$
$$= \mathop{\mathsf{S}}_{X,Y,Z} (X(LY)_+ L)_+ Z - ((XL)_- YL)_+ Z + (X(LY)_- L)_+ Z - ((XL)_+ YL)_+ Z \ ,$$

which after a minor rearrangement is easily seen to vanish.          □

We now define a bilinear form on the image of $J$ as follows:

$$\omega(J(X), J(Y)) = (-)^{|\bar{X}| + |\bar{Y}| + \varepsilon + 1} \, \text{Str} \, J(X)Y \ . \tag{4.2.8}$$

LEMMA 4.2.9. *$\omega$ is s-skewsymmetric.*

PROOF. This is a simple computation using the definition (4.2.1) of the Adler map and the isotropy of $\mathfrak{g}_\pm$.          □

Hence under the identification of $J(X)$ as a vector fields, $\omega$ is to be interpreted as a two-form; only that it is only defined on those vector fields in the image of $J$. In fact, if $X$ were the gradient of a function $F : \mathfrak{g} \to k$, then $J(dF)$ would be the hamiltonian vector field associated to the function and $\omega(J(dF), J(dG)) = \{F, G\}$ would be the Poisson bracket associated to $J$. By Lemma 4.2.9 this bracket is s-skewsymmetric, hence to prove that $J$ is a hamiltonian map, it is equivalent to prove that this bracket obeys the Jacobi identity. This is in turn equivalent to the two-form $\omega$ defined on $\text{Im} \, J$ being closed. This condition makes sense precisely because of Lemma 4.2.3.

For $X, Y, Z \in \mathfrak{g}$ homogeneous, we define the exterior derivative of $\omega$ as the following three-form on the image of $J$:

$$d\omega(J(X), J(Y), J(Z)) = \mathop{\mathsf{C}}_{X,Y,Z} \left( \delta_X \omega(J(Y), J(Z)) - \omega(J([X, Y]_L^*), J(Z)) \right) \ . \tag{4.2.10}$$

We now have the following:

LEMMA 4.2.11. *For any homogeneous functions $F, G, H$,*

$$d\omega(J(dF), J(dG), J(dH)) = - \mathop{\mathsf{C}}_{dF,dG,dH} \{F, \{G, H\}\}$$

PROOF. Because of (4.2.10), (4.2.8) and the definition of the bracket we have

$$d\omega(J(dF), J(dG), J(dH))$$

$$= \mathop{\boldsymbol{\mathsf{C}}}_{dF,dG,dH} \delta_{dF} \{G\,,\,H\} - \delta_{[dF\,,\,dG]_L^*} H$$

$$= \mathop{\boldsymbol{\mathsf{C}}}_{dF,dG,dH} \{F\,,\,\{G\,,\,H\}\} - [\delta_{dF}\,,\,\delta_{dG}]\; H$$

$$= \mathop{\boldsymbol{\mathsf{C}}}_{dF,dG,dH} \{F\,,\,\{G\,,\,H\}\} - \delta_{dF} \{G\,,\,H\} + (-)^{|F||G|}\delta_{dF} \{G\,,\,H\}$$

$$= \mathop{\boldsymbol{\mathsf{C}}}_{dF,dG,dH} \{F\,,\,\{G\,,\,H\}\} - \{F\,,\,\{G\,,\,H\}\} + (-)^{|F||G|} \{G\,,\,\{F\,,\,H\}\}$$

$$= - \mathop{\boldsymbol{\mathsf{C}}}_{dF,dG,dH} \{F\,,\,\{G\,,\,H\}\}\;,$$

where we have used that $|F| = |\bar{dF}|$ and the definition of the gradient. $\qquad\square$

Before getting into the proof of $d\omega = 0$, let us pause for a moment to mention a curious fact. Suppose that $X$ and $Y$ were independent of $L$. Then,

$$\omega(J(X), J(Y)) = (-)^{|\bar{X}|+|\bar{Y}|+\varepsilon+1} \operatorname{Str} J(X)Y$$
$$= (-)^{|\bar{X}|+|\bar{Y}|+\varepsilon+1} \tfrac{1}{2} \operatorname{Str} L\,[X\,,\,Y]_L^*\;, \qquad (4.2.12)$$

which is reminiscent of the Kirillov–Kostant Poisson structure on the dual of a Lie algebra.

Let us discuss this briefly. Suppose that $\mathfrak{h}$ is a Lie algebra and $\mathfrak{h}^*$ its dual. This is a linear space whose tangent space can be identified naturally with itself and its cotangent space with the algebra $\mathfrak{h}$. Therefore one can identify one-forms with elements of the algebra. In particular, if $F$ is any function on $\mathfrak{h}^*$, its gradient $dF(L)$ at a point $L \in \mathfrak{h}^*$ can be identified with an element of the algebra. The Kirillov–Kostant bracket of two functions $F$ and $G$ at the point $L$ is defined by

$$\{F\,,\,G\}\,(L) \equiv \langle L, [dF\,,\,dG] \rangle\;. \qquad (4.2.13)$$

If, as in our case, $\mathfrak{h}$ has a nondegenerate trace form, we can identify $\mathfrak{h}$ and $\mathfrak{h}^*$, and then the Kirillov–Kostant bracket (4.2.13) becomes simply

$$\{F\,,\,G\}\,(L) = (-)^{|F|+|G|+\varepsilon+1} \operatorname{Str} L\,[dF\,,\,dG]\;, \qquad (4.2.14)$$

which is to be compared with (4.2.12). The Jacobi identity of the bracket (4.2.13) follows from the Jacobi identity for the Lie bracket in $\mathfrak{h}$. Of course, we do not quite have the Lie bracket in $\mathfrak{h}$ but rather a modified bracket which nevertheless does obey Jacobi. In fact, we can think of our modified bracket as the linearized Poisson bracket at the point $L$. Either from this analogy or from our experience with Poisson manifolds, we should expect that the following holds.

THEOREM 4.2.15. *For all $X, Y, Z$, independent of $L$,*

$$d\omega(J(X), J(Y), J(Z)) = 0 \ . \tag{4.2.16}$$

PROOF. By definition

$$
\begin{aligned}
d\omega(J(X), J(Y), J(Z)) &= \underset{X,Y,Z}{\mathbf{C}}(-)^{|\bar{X}|+|\bar{Y}|+|\bar{Z}|+\varepsilon}(\mathrm{Str}\, J([X,Y]_L^*)Z - \mathrm{Str}\, \delta_X J(Y)Z) \\
&= \underset{X,Y,Z}{\mathbf{C}}(-)^{|\bar{X}|+|\bar{Z}|+\varepsilon+1+|\bar{X}||\bar{Y}|}\delta_Y \, \mathrm{Str}\, J(X)Z \\
&= -\underset{X,Y,Z}{\mathbf{C}}(-)^{|\bar{X}|+|\bar{Z}|+\varepsilon+1+|\bar{X}|(|\bar{Y}|+|\bar{Z}|)}\delta_Y \, \mathrm{Str}\, J(Z)X \\
&= -\underset{X,Y,Z}{\mathbf{C}}(-)^{|\bar{X}|+|\bar{Y}|+|\bar{Z}|+\varepsilon+1}\, \mathrm{Str}\, \delta_X J(Y)Z \ ,
\end{aligned}
$$

where we have used the antisymmetry of $\omega$. Comparing the first and the last line we find that

$$
\begin{aligned}
d\omega(J(X), J(Y), J(Z)) &= \underset{X,Y,Z}{\mathbf{C}}(-)^{|\bar{X}|+|\bar{Y}|+|\bar{Z}|+\varepsilon+(|\bar{X}|+|\bar{Y}|)|\bar{Z}|}\tfrac{1}{2}\, \mathrm{Str}\, J(Z)[X,Y]_L^* \\
&= \underset{X,Y,Z}{\mathbf{C}}(-)^{(|\bar{X}|+|\bar{Y}|+|\bar{Z}|)(\varepsilon+1)}\tfrac{1}{2}\, \mathrm{Str}[X,Y]_L^* J(Z) \ .
\end{aligned}
$$

Therefore all we have to prove is that

$$\underset{X,Y,Z}{\mathbf{C}}\, \mathrm{Str}[X,Y]_L^* J(Z) = 0 \ . \tag{4.2.17}$$

We use now the fact that for all $W$, $Z$

$$
\begin{aligned}
\mathrm{Str}\, &W J(Z) \\
&= -\tfrac{1}{2}(-)^{\varepsilon(|\bar{W}|+|\bar{Z}|+\varepsilon)}\, \mathrm{Str}\, L([W,Z]_L^* - (-)^{|\bar{W}|\varepsilon}\delta_W Z + (-)^{|\bar{Z}|\varepsilon+|\bar{W}||\bar{Z}|}\delta_Z W) \ .
\end{aligned}
$$

In our case $W = [X,Y]_L^*$ so that $\delta_W Z = 0$ but not $\delta_Z W$ which must be taken into account. Using Proposition 4.2.7 we rewrite (4.2.17) simply as

$$\underset{X,Y,Z}{\mathbf{C}}(-)^{|\bar{Z}|(|\bar{X}|+|\bar{Y}|+\varepsilon)}\, \mathrm{Str}\, L\delta_Z[X,Y]_L^* = 0 \ , \tag{4.2.18}$$

which we proceed to prove.

$$\mathop{\mathbf{C}}_{X,Y,Z}(-)^{|\bar{Z}|(|\bar{X}|+|\bar{Y}|+\varepsilon)}\operatorname{Str} L\delta_Z[X,Y]_L^*$$

$$\begin{aligned}
&= \mathop{\mathbf{S}}_{X,Y,Z}(-)^{|\bar{Z}|(|\bar{X}|+|\bar{Y}|+\varepsilon)}\operatorname{Str}(-)^{|X||\bar{Z}|}(LX(J(Z)Y)_- - L(XJ(Z))_+Y)\\
&= \mathop{\mathbf{S}}_{X,Y,Z}\operatorname{Str}\big[LX((LY)_+LZ)_- - LX(L(YL)_+Z)_- - L(X(LY)_+L)_+Z\\
&\qquad\qquad + L(XL(YL)_+Z)_+\big]\\
&= \mathop{\mathbf{S}}_{X,Y,Z}\operatorname{Str}[(LX)_+(LY)_+LZ - (LX)_+LYLZ + L(XL(YL)_+)_+Z]\\
&= \mathop{\mathbf{S}}_{X,Y,Z}\operatorname{Str}[-(LX)_+(LY)_-LZ + (-)^{\varepsilon(|X|+|Y|+|Z|)}XL(YL)_+(ZL)_-]\\
&= \operatorname{Str}(-LXLYLZ + (-)^{\varepsilon(|X|+|Y|+|Z|)}XLYLZL) - (-)^{|\bar{Y}||\bar{Z}|}(Y \leftrightarrow Z)\ ,
\end{aligned}$$

which clearly vanishes by cyclicity of the trace.                    □

We mention that in the last step of the above proof we have used the following:

LEMMA 4.2.19. *For any* $A, B, C \in \mathfrak{g}$, $\mathop{\mathbf{C}}_{A,B,C}\operatorname{Str}A_+B_-C = \operatorname{Str}ABC$.

PROOF. We expand $\operatorname{Str}ABC$ by writing each $A, B, C$ explicitly into its $+$ and $-$ projections. Using the fact that $\mathfrak{g}_\pm$ are isotropic subalgebras, we find that of the eight possible terms only six survive. Using the s-cyclicity of the supertrace, these can be easily seen to rearrange themselves into:

$$\mathop{\mathbf{C}}_{A,B,C}\operatorname{Str}A_+B_-C\ .\qquad\qquad\qquad\square$$

A remark is in order. We could restrict ourselves to $X$, $Y$, and $Z$ which do not depend on $L$ since the above result is a statement at a point. In terms of Poisson brackets this is simply the fact that the gradient of any function can be substituted at a point by the gradient of a linear function in the proof of the Jacobi identity. Had we taken general $X$, $Y$, and $Z$ we would have incurred in terms of the form $\delta_X Y$ which again would have been seen to cancel. The sceptical reader should verify this her/himself.

Finally we mention a corollary of the above theorem. Suppose that $\mathfrak{g}$ has an identity element—which represents no loss of generality since every associative algebra can be augmented to a unital algebra. Then we can deform the Adler map by shifting $L \mapsto L + \lambda$ for some scalar $\lambda$ assumed to be of the same parity as $L$. Then we find $J_\lambda(X) = J(X) + \lambda J_\infty(X)$, where

$$J_\infty(X) = [L\,,\,X]_+ - [L\,,\,X_+]\ ,\qquad\qquad (4.2.20)$$

where in the above graded commutators the parity of $X$ is $|\bar{X}|$. Then the bracket defined by $J_\infty$ obeys the Jacobi identity and is, moreover, coordinated to the bracket coming from the unperturbed Adler map. If we try to write this new bracket in Kirillov–Kostant form (*cf.* (4.2.13)) we find that the Lie bracket from which it arises is

$$[X\,,\,Y]_R \equiv [R(X)\,,\,Y] + [X\,,\,R(Y)]\ ,\qquad (4.2.21)$$

where $R : \mathfrak{g} \to \mathfrak{g}$ is defined by $R(X) = X_- - X_+$. It is an easy exercise to verify that this bracket does indeed satisfy the Jacobi identity. This makes $R$ into a classical $r$-matrix. Notice also that this new bracket $[-\,,\,-]_R$ is (up to a factor of 2) the one obtained by deforming $[-\,,\,-]_L^*$ by $L \mapsto L + \lambda$.

We can now descend from the abstract into the concrete applications we have in mind. In particular, we can take $\mathfrak{g}$ to be the algebra of $\mathsf{S\Psi DO}$'s with the usual split into differential and integral operators and the supertrace being given by the Adler supertrace (4.1.15). Then the Adler map defines a Poisson bracket at all points $L \in \mathfrak{g}$; that is, for *any* $\mathsf{S\Psi DO}$. Restricting ourselves to different submanifolds of $\mathfrak{g}$—for example, the affine subspaces $\mathfrak{M}_n$—the Adler map induces a Poisson bracket in each one. All these spaces have the property that they can be decomposed as orbits of the formal supergroup associated to the Lie superalgebra of integral operators. For example, if $L \in \mathfrak{M}_n$ is a superdifferential operator of the form $D^n + \sum_{i=1}^n U_i D^{n-i}$, and if $g = 1 + \sum_{i=1}^\infty B_i D^{-i}$ is an element of the formal superVolterra group, $gLg^{-1} \in \mathfrak{M}_n$. In fact, if $n$ is even, the orbit is determined by the invariants $U_1$ and $U_2$; whereas for $n$ odd, both $U_1$ and $U_2$ transform. As already mentioned, the first Gel'fand–Dickey bracket is simply the Kirillov–Kostant bracket on the orbit. As we have now seen the second Gel'fand–Dickey bracket is also of the Kirillov–Kostant type, but with a different Lie bracket. It is an interesting problem to elucidate this relationship further. In particular, can this also be understood as a coadjoint orbit of some (formal) group?

## 4.3.  SUPERSYMMETRIC HIERARCHIES OF THE KP TYPE

We saw in Chapter Three that the Lax formalism provides an ideal framework for studying integrable hierarchies of the KdV type, giving rise to a coherent and robust edifice. Indeed we saw in Section 3.4 that any generalized $n$-KdV hierarchy can be obtained as a reduction of the KP hierarchy by restricting to the subspace of KP operators whose $n$th power is purely differential. The dynamics is described by an infinite number of commuting flows. Each of these hierarchies is formally integrable in the sense that it possesses an infinite number of conserved charges in involution with respect to two coordinated Poisson brackets and a bihamiltonian structure defined via the Adler map. Moreover one can identify the second Gel'fand–Dickey brackets for every $n$-KdV hierarchy with the classical version of the $\mathsf{W}_n$-algebra.

It seems therefore natural to try follow the same path in our attempt to define and study supersymmetric generalizations of KdV-type hierarchies. More precisely, we should define our generic $n$-sKdV hierarchy starting from a Lax operator of the form

$$L = D^n + \sum_{i=1}^{n} U_i D^{n-i} \ . \tag{4.3.1}$$

In light of our previous discussion we would like to be able to characterize all these $n$-sKdV hierarchies as particular reductions of a universal supersymmetric KP hierarchy, whose Lax operator should be given by the $n$-th root of (4.3.1). But, already at this point, we encounter a problem. As shown in [29] every homogeneous SΨDO of the form $L = D^n + U_1 D^{n-1} + \cdots$ has a unique $n$th root for odd $n$, whereas for even $n$ the $n$th root need not exist or even if it does, it need not be unique. Nevertheless for homogeneous SΨDO's of even order, say $L = D^{2n} + U_1 D^{2n-1} + \cdots$, it was proven in [32] that there exists a unique $n$th root $L^{1/n} = D^2 + \cdots$, whose coefficients are differential polynomials in the coefficients of $L$. This means that one is forced to treat separately the odd and even order sKdV hierarchies and hence to consider two supersymmetric KP hierarchies. On the one hand, we have the supersymmetric KP hierarchy (MRSKP) defined by Manin and Radul in [29] which will reduce to the odd order sKdV hierarchies; and on the other hand the even supersymmetric KP hierarchy (SKP$_2$) defined by Figueroa-O'Farrill, Mas, and Ramos [32], which will reduce to those of even order. We should remark that the sKdV equation (1.3.3) is not obtained from the SKP$_2$ hierarchy simply by demanding that some power of the Lax operator be differential. In fact, the Lax operator corresponding to the sKdV hierarchy is $L = D^4 + UD$ which corresponds to the BSKP$_2$ hierarchy to be defined in Section 6.3.

One could of course ask whether these are the only supersymmetric extensions that the KP hierarchy admits. The answer to this question is clearly negative if we consider it in a slightly more general context. If we supersymmetrize the KP hierarchy (*cf.* Section 3.4) by considering flows analogous to those in Proposition 3.4.14 in the superVolterra group one can then construct a hierarchy which does not possess a standard Lax formulation. This hierarchy is called the Jacobian SKP hierarchy (JSKP) and it has been introduced by Mulase [33] and Rabin [34].

THE MRSKP HIERARCHY

The MRSKP hierarchy has been introduced by Manin and Radul in [29] as a supersymmetric extension of the KP hierarchy consisting in an infinite system of flows for an infinite set of even and odd fields, depending on the space variables $(x, \theta)$ of the $(1|1)$ superspace and on an infinite set of odd and even time variables $(t_1, t_2, \ldots)$ and having the KP hierarchy as a natural reduction. In this section we shall consider some of the basic properties that will prove to be of interest for us in the sequel.

The MRSKP hierarchy is defined as the universal family of isospectral flows deforming a S$\Psi$DO $\Lambda = D + \sum_{i \geq 1} U_i D^{1-i}$, with $U_i \in R$. But in contrast to the nonsupersymmetric case this infinite family of odd and even flows satisfy a *nonabelian* Lie superalgebra whose commutation relations are

$$[D_{2i}, D_{2j}] = 0 \ , \quad [D_{2i}, D_{2j-1}] = 0 \ , \quad [D_{2i-1}, D_{2j-1}] = -2D_{2i+2j-2} \ . \quad (4.3.2)$$

(We have adopted here the same sign conventions for the time parameters $t_i$ as in [**53**], [**34**].) A particular representation of (4.3.2) in terms of an infinite number of odd and even times $\{t_1, t_2, t_3, \ldots\}$ is given by

$$D_{2i} = \frac{\partial}{\partial t_{2i}}$$
$$D_{2i-1} = \frac{\partial}{\partial t_{2i-1}} - \sum_{j \geq 1} t_{2j-1} \frac{\partial}{\partial t_{2i+2j-2}} \ , \quad (4.3.3)$$

where the odd times are odd variables satisfying $t_{2i-1}t_{2j-1} = -t_{2j-1}t_{2i-1}$ and hence $t_{2i-1}^2 = 0$. These flows are initially defined on $R$ but one can extend them on the whole $\mathcal{R}$ as evolutionary derivations, that is,

$$[D_{2i}, D] = [D_{2i-1}, D] = 0 \ , \quad (4.3.4)$$

and one can thus write the Lax flows of the MRSKP hierarchy in the following manner:

$$D_{2i}\Lambda = -[\Lambda_-^{2i}, \Lambda] = [\Lambda_+^{2i}, \Lambda] \quad (4.3.5)$$
$$D_{2i-1}\Lambda = -[\Lambda_-^{2i-1}, \Lambda] = [\Lambda_+^{2i-1}, \Lambda] - 2\Lambda^{2i} \ . \quad (4.3.6)$$

As a concrete example, let us write down the first flow $D_1$ for a few fields:

$$D_1 U_1 = -2U_2$$
$$D_1 U_2 = -U_2' \quad (4.3.7)$$
$$D_1 U_3 = -U_3' - 2U_4 - 2U_2^2 - U_2 U_1' \ .$$

The fact that MRSKP admits a Lax formulation is doubtlessly a remarkable feature that proves to be very important in applications. This seems to be connected to what may be seen as a drawback, namely the fact that the algebra of flows is no longer abelian. Indeed, one can redefine the odd flows in such a way that they (anti)commute, but at the the price of the flows not being strictly of Lax form, but explicitly dependent on the time parameters.

The initial value problem associated to the MRSKP hierarchy has been proven [**53**] to be uniquely solvable. Nevertheless, its complete integrability as a dynamical system is far from being obvious. One can proceed by analogy with the nonsupersymmetric case and write down an infinite number of conserved charges

$$H_n = \frac{1}{n} \operatorname{Str} \Lambda^n \quad \text{for all } n \in \mathbb{N} ;\tag{4.3.8}$$

but as one can easily see, for $n = 2k$, $\Lambda^2 k = \frac{1}{2} \left[ \Lambda^k , \Lambda^k \right]$ and $H_{2k}$ is trivial since the supertrace annihilates (graded) commutators. Hence the first problem that arises is to find out whether there exist even conserved charges. To the best of my knowledge this is still an open problem, although as a systematic search at low degree seems to indicate that there are no *local* even conserved charges [**54**].

This brings us to the major open problem concerning MRSKP, namely the existence of a hamiltonian structure. Here too, since the original attempt of Manin and Radul of writing the even flows of the hierarchy in a form reminiscent of the first Gel'fand–Dickey bracket (by using exactly the nontrivial odd supercharges) not much progress has been made.

### THE $\mathrm{SKP}_2$ HIERARCHY

The $\mathrm{SKP}_2$ hierarchy was introduced in [**32**] as a supersymmetric generalization of the KP hierarchy, out of which all even order sKdV hierarchies could be obtained by reduction. As shown in that paper, $\mathrm{SKP}_2$ possesses an infinite set of commuting even flows and all its sKdV-like reductions are integrable and bihamiltonian. The bihamiltonian integrability of the unreduced hierarchy, although implicit in [**32**], was proven by Yu in [**55**].

Given a generic even order sKdV hierarchy, the introduction of $\mathrm{SKP}_2$ is prompted by the following simple fact. Any even order supersymmetric Lax operator $L = D^{2k} + \sum_{i=1}^{2k} U_i D^{2k-i}$, will satisfy a Lax-type evolution equation of the form

$$\frac{\partial L}{\partial t} = [P , L] ,\tag{4.3.9}$$

if and only if

$$\frac{\partial L^{1/k}}{\partial t} = \left[ P , L^{1/k} \right] .\tag{4.3.10}$$

Therefore we are led—analogously to the nonsupersymmetric case—to the study of the hierarchy based on the general supersymmetric Lax operator $\mathcal{L} = D^2 + \sum_{i=1}^{\infty} U_i D^{2-i}$. Then by imposing the constraint $(\mathcal{L}^k)_- = 0$ we will obtain—perhaps after imposing further constraints—all the even-order sKdV hierarchies.

The discussion of this even order SKP hierarchy can be made following closely the nonsupersymmetric case of the KP hierarchy. Indeed the space $\Omega_{\mathcal{L}}$ of SDOP's

$P$ for which the equation

$$\frac{\partial \mathcal{L}}{\partial t} = [P, \mathcal{L}] \tag{4.3.11}$$

is a consistent (local) evolution equation can be related to the space $Z_{\mathcal{L}}$ of $\mathsf{S\Psi DO}$'s commuting with $\mathcal{L}$, namely if $M \in Z_{\mathcal{L}}$ then $M_+ \in \Omega_{\mathcal{L}}$. $Z_{\mathcal{L}}$ on the other hand, as a vector space over the constants, is spanned by the powers $\mathcal{L}^n$, for $n \in \mathbb{Z}$. From this one can immediately characterize $\Omega_{\mathcal{L}}$ in the following fashion. The most general element of $\Omega_{\mathcal{L}}$ is given by a linear combination with constant coefficients of $\mathcal{L}^n_+$, for $n \in \mathbb{N}$ and by any superdifferential operator of the form $fD^2 + gD + h$, where $h$ is an arbitrary differential polynomial of $\mathcal{L}$ and $f$ and $g$ are differential polynomials of $\mathcal{L}$ subject to the condition

$$f'' + U_1 f' + 2g U_1 = 0 , \tag{4.3.12}$$

where $U_1$ is the coefficient of $D$ in $\mathcal{L}$.

Thus, we have an infinite number of even flows $\partial_n \mathcal{L} = [\mathcal{L}^n_+, \mathcal{L}]$ and one can easily check that they commute with each other. We can therefore introduce an infinite number of 'time' variables $t_n$ for $n \in \mathbb{N}$ and define the following flows associated to them:

$$\frac{\partial \mathcal{L}}{\partial t_n} = [\mathcal{L}^n_+, \mathcal{L}] . \tag{4.3.13}$$

The first few equations that one obtains by explicitly computing the first flow on the first, say, four fields read:

$$\begin{aligned}
\frac{\partial U_1}{\partial t_1} &= 0 \\
\frac{\partial U_2}{\partial t_1} &= 2U_3 U_1 \\
\frac{\partial U_3}{\partial t_1} &= U_3'' + U_3' U_1 - U_3 U_1' \\
\frac{\partial U_4}{\partial t_1} &= U_4'' + 2U_5 U_1 - U_4' U_1 - U_3 U_1'' - U_3 U_2' .
\end{aligned} \tag{4.3.14}$$

The $\mathrm{SKP}_2$ hierarchy has an infinite number of nontrivial independent polynomial conserved quantities. Indeed define

$$H_n = \frac{1}{n} \operatorname{Str} \mathcal{L}^n \quad \text{for } n \in \mathbb{N} . \tag{4.3.15}$$

They are obviously integrals of polynomial densities and moreover one can actually prove that they are nontrivial and independent. Thus they form an infinite set of even (under the $\mathbb{Z}_2$-grading) conserved charges for $\mathrm{SKP}_2$. Hence $\mathrm{SKP}_2$ is formally completely integrable.

THE JSKP HIERARCHY

Since there is no unique supersymmetric extension of the KP hierarchy one could of course ask what distinguishes the different supersymmetric KP hierarchies or which one of them is a more natural generalization of the KP hierarchy. We have previously argued that the MRSKP hierarchy has the advantage of possessing a standard Lax formulation. Nevertheless, from a geometrical point of view, it is not the MRSKP hierarchy the one that seems the most natural supersymmetric generalization of the KP hierarchy. The flows of these KP-type hierarchies have a fascinating geometric interpretation [**34**][**33**].

Morally speaking, one can understand this as follows. The spectrum of any reasonable operator on a compact manifold is discrete. By duality, one expects that if an operator is defined on a discrete set (say, a point) its spectrum would be a compact manifold. Now, if $L$ is a Lax operator for a KdV-type hierarchy, its components are taken to be formal power series and hence will generically have zero radius of convergence. In other words, we can think of it as being defined on a point. It turns out that its spectrum can be thought of some Riemann surface $\Sigma$. The Lax flows, being isospectral, preserve this Riemann surface and in fact can be understood as deformations of holomorphic line bundles over $\Sigma$. By continuity, the flows must preserve the topology of the bundle but not necessarily the holomorphic structure. In other words, the KdV-type equations can be understood as flows on the moduli space of (flat) holomorphic line bundles over $\Sigma$; in other words, the Jacobian variety of $\Sigma$. According to the geometric analysis of Rabin [**34**], one cannot understand the MRSKP flows in exactly this fashion. In other words, the MRSKP flows do not just deform line bundles over a fixed superRiemann surface, but actually deform the superanalytic structure of the superRiemman surface itself. The flows are not generic though: they are such that the hierarchy remains integrable. If we insist in having strictly Jacobian flows—that is, preserving the superRiemman surface—one is forced to introduce a different hierarchy: the Jacobian SKP hierarchy (JSKP) of Mulase [**33**] and Rabin [**34**]. This hierarchy seems to provide the the closest geometric analog of the KP hierarchy in the supersymmetric case since its flows are defined on the supersymmetric Jacobian variety of an algebraic supercurve.

To define the Jacobian SKP hierarchy it is necessary to abandon momentarily the Lax form for the evolution equations. Instead, it is convenient to mimic the treatment of the last subsection of Section 3.4 on dressing transformations and try to write a natural set of flows in the superVolterra group.

Let us first consider the MRSKP hierarchy in this light. As shown in [**29**], the necessary and sufficient condition for the existence of an even $\mathsf{S\Psi DO}$, $\phi = 1 + \sum_{i \geq 1} V_i D^{-i}$, with $V_i \in R$, such that $\Lambda = \phi D \phi^{-1}$ is $U_1^{[1]} + 2U_2 = 0$. If we restrict ourselves to such $\Lambda$'s then we can alternatively define the MRSKP hierarchy as the family of flows on the dressing operator $\phi$

$$D_i\phi = -\left(\phi D^i \phi^{-1}\right)_- \phi \, , \qquad (4.3.16)$$

or equivalently

$$\frac{\partial\phi}{\partial t_{2i}} = -\left(\phi D^{2i} \phi^{-1}\right)_- \phi$$
$$\frac{\partial\phi}{\partial t_{2i-1}} = -\left(\phi \left(D^{2i-1} + \sum_{j\geq 1} t_{2j-1} D^{2i+2j-2}\right)\phi^{-1}\right)_- \phi \, . \qquad (4.3.17)$$

One can prove that provided $\Lambda$ is dressable, the two definitions of the MRSKP hierarchy are indeed equivalent. First of all it is obvious that given the flows on the dressing operator (4.3.16) one obtains the Lax flows on $\Lambda$:

$$\begin{aligned} D_i\Lambda &= D_i\left(\phi D\phi^{-1}\right) \\ &= -\left(\phi D^i\phi^{-1}\right)_- \phi D\phi^{-1} + (-)^i \phi D\phi^{-1}\left(\phi D^i\phi^{-1}\right)_- \\ &= -[\Lambda^i_-, \Lambda] \, . \end{aligned} \qquad (4.3.18)$$

In order to prove the converse let us introduce the dressed version of $\Lambda$ in the Lax flows and rewrite them in the following form

$$\left(D_i\phi + \left(\phi D^i\phi^{-1}\right)_- \phi\right)D\phi^{-1} - (-)^i\phi D\phi^{-1}\left(D_i\phi + \left(\phi D^i\phi^{-1}\right)_- \phi\right)\phi^{-1} = 0 \, . \qquad (4.3.19)$$

Suppose now that $D_i\phi + \left(\phi D^i\phi^{-1}\right)_- \phi = A_N D^N + A_{N-1}D^{N-1} + \ldots$, for some arbitrary $N$. Then one obtains the following conditions for the leading coefficients:

$$A_N = 0 \quad \text{for } N \text{ odd,}$$

and

$$\left.\begin{aligned} 2A_{N-1} - (-)^n A'_N - (-)^n 2V_1 A_N &= 0 \\ A'_{N-1} + 2V_1 A_{N-1} - (-)^n V'_1 A_N &= 0 \end{aligned}\right\} \quad \text{for } N \text{ even.} \qquad (4.3.20)$$

That is, in both cases we obtain that—provided we drop the constants—the leading coefficient $A_N$ must vanish and hence (4.3.16) is satisfied.

Notice that one can dress the following obvious commutation relations

$$[D_{2i} - D^{2i}, D] = 0 \qquad (4.3.21)$$
$$[D_{2i-1} - D^{2i-1}, D] = -2D^{2i} \qquad (4.3.22)$$

and obtain the Lax flows (4.3.5) and (4.3.6).

We can now define the JSKP hierarchy as the infinite family of odd and even commuting flows on the superVolterra group given by

$$\frac{\partial \phi}{\partial t_{2i}} = -\left(\phi \partial^i \phi^{-1}\right)_{-} \phi$$

$$\frac{\partial \phi}{\partial t_{2i+1}} = -\left(\phi \partial^i \partial_\theta \phi^{-1}\right)_{-} \phi \ , \qquad (4.3.23)$$

where $\phi = 1 + \sum_{\geq 1} V_i D^{-i}$ and $\{t_1, t_2, t_3, \ldots\}$ is the same infinite set of odd and even times as in the case of the MRSKP hierarchy. One can easily work out the first few equations, for the first few fields:

$$\frac{\partial V_1}{\partial t_1} = V_{1,\theta}$$

$$\frac{\partial V_2}{\partial t_1} = V_{2,\theta} + V_1 V_2 - V_3 \qquad (4.3.24)$$

$$\frac{\partial V_3}{\partial t_1} = V_{3,\theta} + V_1 V_3 \ ,$$

where $V_{i,\theta} = (\partial_\theta V_i)$.

Clearly, the even flows of JSKP coincide with the even flows of MRSKP being actually nothing but the original KP system. On the other hand, since $\phi \partial_\theta \neq \partial_\theta \phi$, it seems there is no simple way of writing the odd flows in terms of a Lax operator $L = \phi D \phi^{-1}$; in other words, the JSKP hierarchy (or more precisely its odd part) does not have a Lax representation in terms of fractional powers of $L$. We can nevertheless write the JSKP flows in a Lax form (3.3.5) by defining $L \equiv \phi \partial \phi^{-1}$ and $M = \phi \partial_\theta \phi^{-1}$, in terms of which the flows can be written as follows:

$$\frac{\partial L}{\partial t_{2i}} = -\left[L_{-}^i \, , \, L\right] \quad \text{and} \quad \frac{\partial L}{\partial t_{2i+1}} = -\left[(L^i M)_{-} \, , \, L\right] \ . \qquad (4.3.25)$$

Then the trivial commutation relations which give upon dressing the flows of the hierarchy in the Lax form read

$$[D_{2i} - \partial^i, \partial] = 0$$

$$[D_{2i+1} - \partial^i \partial_\theta, \partial] = 0 \ . \qquad (4.3.26)$$

# Chapter Five

# ADDITIONAL SYMMETRIES

The notion of integrability is intimately linked to the notion of symmetry. The idea that a group of symmetries acting on phase space can be used to solve a dynamical system goes back to Jacobi and Laplace and the method of 'elimination of nodes.' We are all familiar with the fact that the two-body problem reduces down to a one-body problem relative to the center of mass. In general, when a group of symmetries acts on a phase space in such a way that the Poisson brackets are preserved, there is a well-defined procedure (called hamiltonian reduction) by which to construct a lower-dimensional phase space. If in addition the symmetries preserve the dynamics, then these can be effectively described in the reduced phase space. It is not hard to show that a system is completely integrable if and only if it can be reduced in this fashion down to a trivial phase space consisting of isolated points. Liouville's theorem on complete integrability can be understood in precisely this fashion. Given a set of conserved quantities in involution, their flows give rise to an action of (some quotient of) the affine group. The resulting hamiltonian reduction yields a phase space that in the best of cases consists of only one point. When the motion is quasi-periodic, this is the essence of the Kolmogorov-Arnold-Moser theorem on invariant tori. The coordinates on the tori are the angle variables and the action variables are canonically conjugate ones which are functions of the conserved charges and the hamiltonian.

Given an integrable evolution equation generated by some hamiltonian, the flows generated by the conserved charges in involution are dynamical symmetries. Since these symmetries are sufficient to reduce the phase space down to a discrete set of points, one may naively think that one cannot find other continuous symmetries. In particular, for the KdV-type hierarchies introduced in Chapter Three, we classified all the possible Lax flows and they turned out to generate the flows of the hierarchy. One would then not expect that there should exist any 'additional' symmetries of Lax type. It thus came as some surprise when Orlov and Schulman [**56**] in 1986 discovered an infinite set of additional symmetries for the KP equation which can be written in Lax form. The catch was that these symmetries are explicitly time-dependent.

In this chapter we review from a different perspective the additional symmetries of the KP hierarchy and then turn our attention to the determination of the additional symmetries for the supersymmetric extensions discussed in the previous Chapter. For the KP case we will follow the treatment in [**57**] which exploits a representation of the algebra of differential operators on the Volterra group. This action is analogous to the map defined by Radul from the algebra of differential operators on the space of Lax operators of the KdV-hierarchies, hence we discuss this map briefly below. In Section 1, we also determine the additional symmetries of the KP hierarchy and show that they are isomorphic to $W_\infty$ following [**57**]. In Section 2 we study the additional symmetries of three supersymmetric extensions of the KP hierarchy: MRSKP, SKP$_2$, and JSKP using a supersymmetric variant of the Radul map. The work in this second part is contained in my paper [**52**].

## 5.1. ADDITIONAL SYMMETRIES OF THE KP HIERARCHY

### THE RADUL MAP

The aim of this section is to introduce the Radul map. This will provide us with an elegant framework in which to define the additional flows of KP and make transparent $W_\infty$ as the algebra of additional symmetries of this hierarchy.

The context in which the Radul map appeared for the first time is nevertheless slightly different. Its original motivation lies in the general frame of attempts to understand the $W$-symmetry, in particular by trying to relate $W$-algebras to algebraic structures that are better understood. One method to investigate how a class of algebras fits within other algebraic structures is to try and establish maps (morphisms) between its objects and other well-known objects. In the case of $W$-algebras, examples of such maps are the Miura transformation [**15**], the (generalized) Drinfel'd–Sokolov reduction [**17**], and the Radul map [**58**]. This last one is a Lie algebra homomorphism from the differential operators on the circle to the algebra of vector fields on the space of Lax operators, some of which generate $W$-transformations.

Consider the subring $\mathcal{R}_+$ of differential operators on the circle and give it a Lie algebra structure by the commutator. We call the resulting Lie algebra DOP. We follow the notation in Chapter Three, and we let $\mathfrak{M}$ denote the space of $\Psi$DO's for the form $\partial + \sum_{i\geq 1} u_i \partial^{1-i}$. Taking $n = 1$ we are in the space of Lax operators for the KP hierarchy, whereas taking $u_{i>n} = 0$ we are in the space of Lax operators of the $n$-KdV hierarchy.

The Radul map $W : \mathsf{DOP} \to T_L\mathfrak{M}$ is defined by

$$W(E) = (LEL^{-1})_- L = LE - (LEL^{-1})_+ L \; . \tag{5.1.1}$$

On $T_L\mathfrak{M}$ we can define a Lie bracket as in (3.1.8). Let us recall this. Every

$A \in T_L\mathfrak{M}$ of the above form defines a vector field $\partial_A$ by

$$\partial_A = \sum_{i=1}^{\infty} \sum_{k=0}^{\infty} a_i^{(k)} \frac{\partial}{\partial u_i^{(k)}} \ . \tag{5.1.2}$$

In particular, $\partial_A L = A$. We then define the Lie bracket $[\![A, B]\!]$ of two vectors $A, B \in T_L\mathfrak{M}$ by

$$\partial_{[\![A,B]\!]} = [\partial_A \, , \, \partial_B] \ ; \tag{5.1.3}$$

or equivalently

$$[\![A, B]\!] = \partial_A B - \partial_B A \ . \tag{5.1.4}$$

Notice that this is not the ordinary commutator in $\mathsf{DOP}$.

THEOREM 5.1.5. *The Radul map is a Lie algebra isomorphism*

$$[\![W(E), W(F)]\!] = W([E, \, F]_c) \ , \tag{5.1.6}$$

*where the modified bracket on* $\mathsf{DOP}$ *is defined by*

$$[E, \, F]_c \equiv \partial_{W(E)}F - \partial_{W(F)}E + [E, \, F] \ . \tag{5.1.7}$$

(For a proof in the more general case of generalized pseudodifferential operators see [**57**].)

The image of the generalized Adler map is a subalgebra of $T_L\mathfrak{M}$, and this allows us to pull back the bracket $[\![-, -]\!]$ on $T_L\mathfrak{M}$ to a bracket $[-, \, -]_L^*$ on $T_L^*\mathfrak{M}$ defined by requiring that the Adler map be a homomorphism. Explicitly, for $X, Y \in T_L^*\mathfrak{M}$, we have

$$[\![J(X), J(Y)]\!] = J([X, \, Y]_L^*) \ , \tag{5.1.8}$$

where

$$[X, \, Y]_L^* = \partial_{J(X)}Y + X(LY)_- - (XL)_+Y - (X \leftrightarrow Y) \ . \tag{5.1.9}$$

Moreover there exists a Lie algebra homomorphism $R : \mathsf{DOP} \to T_L^*\mathfrak{M}$ defined by $R(E) = -EL^{-1} \bmod \partial^{-n}\mathcal{R}_-$, which means that

$$[R(E), \, R(F)]_L^* = R([E, \, F]_c) \ , \tag{5.1.10}$$

or, in other words, the following diagram is commutative

$$\begin{array}{ccc} & & T_L^*\mathfrak{M} \\ & \raisebox{0.5ex}{$R$} \nearrow & \downarrow {\scriptstyle J} \\ \mathsf{DOP} & \xrightarrow{W} & T_L\mathfrak{M} \end{array}$$

Let us now consider the immediate application of the homomorphism property of the Radul map to the identification originally due to Aoyama and Kodama in [**59**] of $\mathsf{W}_\infty$ as the algebra of additional symmetries of the KP hierarchy.

### ADDITIONAL SYMMETRIES

As we have seen in Section 3.4, the KP hierarchy is defined as the universal family of isospectral deformations of a pseudodifferential operator $\Lambda$ of the form (3.4.1). The evolution of $\Lambda$ is specified by the commuting family of flows $\partial_i$ given by (3.4.7). If one restricts oneself to operators satisfying $a_1 = 0$, then we saw that one can lift the KP flows to the Volterra group $G$. The Volterra group acts naturally via dressing transformations $L \mapsto \phi^{-1}L\phi$, where $\phi = 1 + \sum_{i \geq 1} w_i \partial^{-i} \in G$ is the dressing operator. In terms of the dressing operator, the flows of the KP hierarchy are given by Proposition 3.4.14.

One can write these flows in a different way by using an analogue of the Radul map [**58**]. The similarity between the expression for the Radul map (5.1.1) and the one of the KP flows given by Proposition 3.4.14 suggests us to define a map

$$W'(E) = (\phi E \phi^{-1})_- \phi \ , \tag{5.1.11}$$

from $\mathsf{DOP}$ to the Lie algebra $\mathcal{R}_-$ of the Volterra group. The KP flows become now $\partial_n \phi = -W'(\partial^n) = -\partial_{W'(\partial^n)}\phi$, where $\partial_{W'(\partial^n)}$ is then a flow on the Volterra group. The map (5.1.11) now translates the trivial fact $[\partial^n, \partial^m] = 0$ into the commutativity of the flows $[\partial_n, \partial_m] = 0$. This allows us to represent the flows in terms of an infinite set of times, $\partial_i = \frac{\partial}{\partial t_i}$, with $i = 1, 2, \ldots$. One interpretation of this feature is that every flow possesses an infinite number of symmetries given by the other flows. This interpretation begs the question whether these are all or, if on the contrary, there exist additional symmetries. Remarkably enough, it turns out that one can construct a larger family of times-dependent flows which contains as a subset the original KP flows and commute with them. This new family of flows satisfies a nonabelian algebra with respect to which the KP hierarchy forms its center. Thus we adopt here the following definition.

DEFINITION 5.1.12. By (additional) symmetries of an integrable hierarchy of flows, we mean its centralizer in the algebra of times-dependent vector fields.

The fact that these symmetries contain the original hierarchy, although largely taken for granted, is only true provided the flows of the hierarchy themselves satisfy an abelian algebra; and it is to these cases that the word 'additional' can be applied. We will see in fact that this is not generally the case for supersymmetric hierarchies.

Along with Definition 5.1.12, it is in practice convenient to have a 'working definition' that is more suitable for computation. Our working definition is motivated by the following fact. The flows $\partial_{W'(\Gamma)}$ generated via (5.1.11) by differential operators $\Gamma$ satisfying

$$[\partial_i - \partial^i, \Gamma] = 0 \tag{5.1.13}$$

commute with the KP flows. Indeed following [**57**] we have that

$$\begin{aligned}
[\partial_{W'(\Gamma)}, \partial_i] &= -[\partial_{W'(\Gamma)}, \partial_{W'(\partial^i)}] \\
&= -\partial_{W'([\Gamma, \partial^i]_\phi)} \ ,
\end{aligned} \tag{5.1.14}$$

where $[\Gamma, \partial^i]_\phi$ is a modified Lie bracket analogous to (5.1.7). This particular bracket is given by

$$\begin{aligned}
[\Gamma, \partial^i]_\phi &= \partial_{W'(\Gamma)} \partial^i - \partial_{W'(\partial^i)} \Gamma + [\Gamma, \partial^i] \\
&= [\partial_i, \Gamma] + [\Gamma, \partial^i] \\
&= [\partial_i - \partial^i, \Gamma] \\
&= 0 \ .
\end{aligned} \tag{5.1.15}$$

We therefore call 'additional symmetries' the flows generated by operators $\Gamma$ subject to (5.1.13). It is conceivable that (5.1.13) is also a necessary condition— that is, that all additional symmetries arise in this fashion; but we shall not attempt to prove it here.

This means that looking for the additional symmetries comes down to trying to find solutions for (5.1.13). An obvious solution to this equation is simply $\Gamma = \partial$, which, introduced in (5.1.13) and after applying a dressing transformation, gives precisely the KP flows in Proposition 3.4.14. This agrees with the fact that the KP flows commute with each other.

A more interesting solution can be obtained if we allow for an explicit dependence on the time parameters of the hierarchy; that is, if we extend our ring of functions by the infinite set of independent variables $\{t_1, t_2, \ldots\}$, in which case we have to extend the derivative operator $\partial$ as a derivation in this new ring.

A priori, since $x$ and all the $t_i$ are independent variables of our infinite set of partial differential equations, we can automatically conclude that $\partial$ has to be extended trivially to the new ring. Nevertheless in the case KP, since the first flow (for dressable $L$) reads $\partial_1 L = [L_+, L] = [\partial, L]$ and therefore gives $\partial = \partial_1$, one can identify $x$ with $t_1$. One can then define (see, for instance, [**40**]) a formal infinite-order differential operator

$$\Gamma = \sum_{j \geq 1} j t_j \partial^{j-1} \ , \tag{5.1.16}$$

which satisfies (5.1.13), and from it a two-parameter family of flows

$$\partial_{m,k}\phi = \left(\phi\Gamma^k\partial^m\phi^{-1}\right)_-\phi \ , \tag{5.1.17}$$

that satisfy $[\partial_{m,k}, \partial_n] = 0$. Notice that for $k = 0$ and $m > 0$ they agree with the KP flows. Moreover since $[\partial, \Gamma] = 1$ it follows that the Lie algebra generated by $\Gamma^k\partial^m$, $k \geq 0$ and $m \in \mathbb{Z}$ is isomorphic (as a Lie algebra) to $\mathsf{W}_{1+\infty}$ and hence the algebra of additional symmetries is nothing but $\mathsf{W}_\infty$. This was proven in [**59**] by a direct computation in modes, but the proof using the Radul-like map is more conceptual.

One can alternatively write the two-parameter family of flows in a Lax form

$$\partial_{m,k}L = -[(M^kL^m)_-, L] \ ,$$

where $M = \phi\Gamma\phi^{-1}$ is the dressed version of $\Gamma$.

Although realized here without it, $\mathsf{W}_\infty$ has a natural central extension given, as a subalgebra of $\mathsf{DOP}$, by the Khesin–Kravchenko [**60**] logarithmic cocycle. In the KP context, the central extension appears when acting on the $\tau$-functions— equivalently, when we realize $\mathsf{W}_\infty$ as free fermion bilinears in a two-dimensional conformal field theory.

## 5.2. ADDITIONAL SYMMETRIES OF SKP HIERARCHIES

### THE SUPERSYMMETRIC RADUL MAP

In this section we shall introduce a supersymmetric generalization of the Radul map and we shall see that it defines a Lie algebra homomorphism between the space of $\mathsf{SDOP}$'s and $T_L\mathfrak{M}$. In order to do this we have first of all to define a Lie (super)algebra structure on $T_L\mathfrak{M}$. Of course, since the elements of $T_L\mathfrak{M}$ are in particular $\mathsf{S\Psi DO}$'s of order at most $n-1$ one always has the obvious bracket given by the graded commutator. Still this is not the one that will allow us to exhibit the supersymmetric Radul map as a Lie algebra homomorphism. Instead let us consider the natural Lie bracket on vector fields on $\mathfrak{M}$, namely

$$[D_A, D_B] = D_A D_B - (-)^{|D_A||D_B|} D_B D_A \ . \tag{5.2.1}$$

This will induce in $T_L\mathfrak{M}$ a bracket $[\![-, -]\!]$ by

$$[D_A, D_B] = D_{[\![A,B]\!]} \ , \tag{5.2.2}$$

whose explicit form we shall obtain now.

LEMMA 5.2.3. *The Lie bracket $[\![-,-]\!]$ in $T_L \mathfrak{M}$ is given by*

$$[\![A, B]\!] = D_A B - (-)^{|D_A||D_B|} D_B A , \qquad (5.2.4)$$

*for any two* S$\Psi$DO's $A$ *and* $B$ *in* $T_L \mathfrak{M}$.

PROOF. Consider $f$ an arbitrary function in $S_L$. Then

$$[D_A, D_B] f = \left( D_A D_B - (-)^{|D_A||D_B|} D_B D_A \right) f , \qquad (5.2.5)$$

which using the fact that $D_A f^{[k]} = (-)^{k|D_A|} (D_A f)^{[k]}$ (which follows by repeated application of (4.1.20)) becomes

$$
\begin{aligned}
[D_A, D_B] f &= \sum_{i,j=1}^{\infty} \sum_{k,l=0}^{\infty} \left( (-)^{k|D_B|+(l+k)|D_A|} A_j^{[l]} \frac{\partial B_i}{\partial U_j^{[k]}} \right. \\
&\qquad\qquad \left. -(-)^{|D_A||D_B|+k|D_A|+(l+k)|D_B|} B_j^{[l]} \frac{\partial A_i}{\partial U_j^{[l]}} \right)^{[k]} \frac{\partial f}{\partial U_i^{[k]}} \\
&= \sum_{i=1}^{\infty} \sum_{k=0}^{\infty} (-)^{k(|D_A|+|D_B|)} [\![A, B]\!]_i^{[k]} \frac{\partial f}{\partial U_i^{[k]}} , \qquad (5.2.6)
\end{aligned}
$$

where

$$[\![A, B]\!]_i = \sum_{j=1}^{\infty} \sum_{l=0}^{\infty} (-)^{l|D_A|} A_j^{[l]} \frac{\partial B_i}{\partial U_j^{[k]}} - (-)^{|D_A||D_B|+l|D_B|} B_j^{[l]} \frac{\partial A_i}{\partial U_j^{[l]}} , \qquad (5.2.7)$$

and we get indeed (5.2.4). $\qquad\qquad \square$

As we saw in Chapter Four one can pull the Lie bracket $[\![-,-]\!]$ on $T_L \mathfrak{M}$ back to $T_L^* \mathfrak{M}$ via the Adler map $J(X) = (LX)_+ L - L(XL)_+$. In other words one can define a bracket $[-,-]_L^*$ on $T_L^* \mathfrak{M}$ such that

$$[\![J(X), J(Y)]\!] = J([X, Y]_L^*) . \qquad (5.2.8)$$

Computing $[X, Y]_L^*$ one finds **[51]**

$$
\begin{aligned}
[X, Y]_L^* = &(-)^{n(n+|X|)} D_{J(X)} Y + X(LY)_- \\
&- ((XL)_+ Y)_- - (-)^{(n+|X|)(n+|Y|)} (X \leftrightarrow Y) . \qquad (5.2.9)
\end{aligned}
$$

This already tells us that the Adler map is a Lie (super)algebra homomorphism mapping the cotangent space to the tangent space of $\mathfrak{M}$ at $L$, each of them being considered with the corresponding Lie algebra structure.

Now we can finally define the supersymmetric analog of the Radul map

$$W : \mathsf{SDOP} \to T_L\mathfrak{M} \tag{5.2.10}$$

sending any $E \in \mathsf{SDOP}$ to the tangent vector $W(E)$ defined by

$$W(E) \equiv LE - (LEL^{-1})_+L = (LEL^{-1})_-L \ . \tag{5.2.11}$$

THEOREM 5.2.12. *The supersymmetric Radul map is a Lie (super)algebra homomorphism, i.e.,*

$$[\![W(E), W(F)]\!] = W([E,F]_L) \ , \tag{5.2.13}$$

*where $[E,F]_L$ is the modified Lie bracket on* $\mathsf{SDOP}$ *given by*

$$[E,F]_L = [E,F] + (-)^{n|E|}D_{W(E)}F - (-)^{|E||F|+n|F|}D_{W(F)}E \ . \tag{5.2.14}$$

PROOF. By direct computation in the right hand side we have

$$
\begin{aligned}
[\![W(E), W(F)]\!] &\equiv D_{W(E)}W(F) - (-)^{|D_{W(E)}||D_{W(F)}|}D_{W(F)}W(E) \\
&= D_{W(E)}(LFL^{-1})_-L - (-)^{|D_{W(E)}||D_{W(F)}|}(E \leftrightarrow F) \\
&= (W(E)FL^{-1})_-L + (-)^{n|D_{W(E)}|}(LD_{W(E)}FL^{-1})_-L \\
&\quad - (-)^{|F||D_{W(E)}|}(LFL^{-1}W(E)L^{-1})_-L \\
&\quad + (-)^{|F||D_{W(E)}|}(LFL^{-1})_-W(E) - (-)^{|D_{W(E)}||D_{W(F)}|}(E \leftrightarrow F) \\
&= ((LEL^{-1})_-LFL^{-1})_-L + (-)^{|E||F|}((LFL^{-1})_-LEL^{-1})_-L \\
&\quad - (-)^{|E||F|}(LFEL^{-1})_-L + (-)^{n|E|}(LD_{W(E)}FL^{-1})_-L \\
&\quad - (-)^{|D_{W(E)}||D_{W(F)}|}(E \leftrightarrow F) \\
&= (L[E,F]L^{-1})_-L + (-)^{n|E|}(LD_{W(E)}FL^{-1})_-L \\
&\quad - (-)^{|F|(n+|E|)}(LD_{W(F)}EL^{-1})_-L \\
&= W([E,F]_L) \ , \tag{5.2.15}
\end{aligned}
$$

which proves the theorem.                                                    □

Notice that in the case where $E$ and $F$ are independent of $L$ we recover the usual Lie bracket on $\mathsf{SDOP}$.

We have in this moment the following diagram where both maps $W$ and $J$ have been proven to be Lie algebra homomorphisms:

$$
\begin{array}{ccc}
 & & T_L^*\mathfrak{M} \\
 & & \downarrow {\scriptstyle J} \\
\mathsf{SDOP} & \xrightarrow{\ W\ } & T_L\mathfrak{M}
\end{array}
$$

It would be thus interesting to see whether one can complete this diagram with a homomorphism $R$ such that $J \circ R = W$.

We consider therefore the map $R : \mathsf{SDOP} \to T_L^* \mathfrak{M}$ defined by

$$R(E) = -(EL^{-1})_- \bmod D^{-n} \mathcal{S}_- \;, \tag{5.2.16}$$

for any $E$ in $\mathsf{SDOP}$. Since $D^{-n} \mathcal{S}_- \subseteq \ker J$ we have that indeed

$$J \circ R(E) = W(E) \;, \tag{5.2.17}$$

for any $E$ in $\mathsf{SDOP}$ and therefore $J \circ R = W$.

THEOREM 5.2.18.  *$R$ is a Lie algebra homomorphism, with*

$$[R(E), R(F)]_L^* = R([E, F]_L) \;. \tag{5.2.19}$$

PROOF.  Using the fact that $J \circ R = W$ and that $|R(E)| = |E| + n$ we have

$$\begin{aligned}
[R(E), R(F)]_L^* = {}& -(-)^{n|E|} D_{W(E)}(FL^{-1})_- + (EL^{-1})_-(L(FL^{-1})_-)_- \\
& - (((EL^{-1})_- L)_+(FL^{-1})_-)_- - (-)^{|E||F|}(E \leftrightarrow F) \\
= {}& -(-)^{n|E|}(D_{W(E)}FL^{-1})_- + (-)^{|E||F|}(FL^{-1}W(E)L^{-1})_- \\
& + (EL^{-1})_-(LFL^{-1})_- - (E(FL^{-1})_-)_- \\
& + ((EL^{-1})_+ L(FL^{-1})_-)_- - (-)^{|E||F|}(E \leftrightarrow F) \\
= {}& -([E, F]L^{-1})_- - (-)^{n|E|}(D_{W(E)}FL^{-1})_- \\
& + (-)^{n|F| + |E||F|}(D_{W(F)}EL^{-1})_- \\
= {}& R([E, F]_L) \;. \tag{5.2.20}
\end{aligned}$$

$\square$

COROLLARY 5.2.21.  *We have the following commutative diagram of Lie algebras:*

$$\begin{array}{ccc}
& & T_L^* \mathfrak{M} \\
& {}^{R} \nearrow & \downarrow {\scriptstyle J} \\
\mathsf{SDOP} & \xrightarrow{\;W\;} & T_L \mathfrak{M}
\end{array}$$

$\square$

### THE MRSKP HIERARCHY

We shall start in this section the study of the additional symmetries of supersymmetric KP hierarchies by considering the supersymmetric extension of KP defined by Manin and Radul in [**29**], the MRSKP hierarchy. The additional symmetries of this particular hierarchy have been studied also in [**61**] and we find agreement with their results.

The flows (4.3.16) are reminiscent of the supersymmetric Radul map (where in this case $n = 1$) and suggest us to define a map $W' : \mathsf{SDOP} \to \mathcal{S}_-$ by

$$W'(E) = \left( \phi E \phi^{-1} \right)_- \phi \; , \qquad (5.2.22)$$

for any $\mathsf{SDOP}$ $E$ such that $D_n \phi = -W'(D^n) = -D_{W'(D^n)} \phi$, where $D_{W'(D^n)}$ is a flow on the superVolterra group. The algebra of flows of MRSKP becomes in light of this definition a simple consequence of Theorem 5.2.12.

PROPOSITION 5.2.23. *The MRSKP flows satisfy the Lie superalgebra given in* (4.3.2).

PROOF. Following step by step the proof of Theorem 5.2.12 and replacing $L$ with $\phi$ we have that

$$[D_{W'(E)}, D_{W'(F)}] = D_{W'([E,F]_\phi)} \; . \qquad (5.2.24)$$

Applying this to our MRSKP flows we get for instance

$$\begin{aligned} [D_{2i-1}, D_{2j-1}] &= [D_{W'(D^{2i-1})}, D_{W'(D^{2j-1})}] \\ &= D_{W'(2D^{2i+2j-2})} \\ &= -2D_{2i+2j-2} \; . \end{aligned} \qquad (5.2.25)$$

One can in a similar way check that all the other commutators in (4.3.2) do indeed vanish. $\qquad \square$

After these general considerations concerning the MRSKP hierarchy we are now prepared to tackle the problem of finding its (additional) symmetries. We have seen that in the case of KP one defines a larger family of flows (*i.e.*, containing the KP flows) which satisfy an algebra whose center is the KP hierarchy itself. Here the situation will turn out to be slightly different since the MRSKP flows themselves do not commute with each other but rather they obey the nonabelian (super)algebra (4.3.2).

DEFINITION 5.2.26. We call (additional) symmetries of the MRSKP hierarchy the centralizer of the algebra (4.3.2) of flows of MRSKP in the algebra of times-dependent vector fields on $\mathfrak{M}$.

Analogous to the nonsupersymmetric case, one way to look for additional symmetries is to look for operators $\Gamma$ satisfying

$$[D_i - D^i, \Gamma] = 0 \; . \qquad (5.2.27)$$

The additional flow associated to $\Gamma$ is then obtained via the supersymmetric Radul map and is given by $D_{W'(\Gamma)}$.

One obvious solution is $\Gamma = \partial$, the even derivation on the ring $\mathcal{S}$ and the generator (via the Radul map) of the even flows of the hierarchy. Notice nevertheless that the odd derivation $D$—the generator of the odd flows—does not obey (5.2.27) but for even $i$. One is therefore forced to conclude that only the even flows are actually symmetries of the hierarchy, this being the most striking distinction from the nonsupersymmetric case. Apart from this 'trivial' solution, one can of course ask whether there also exist times-dependent symmetries of the MRSKP hierarchy. The answer to this question is the object of the following lemma.

LEMMA 5.2.28. *Let $S[t_i]$ be the extension ring of $S$ by the time variables $\{t_i\}$ and let*

$$\Gamma_0 = x + \frac{1}{2}\sum_{j\geq 1} jt_{2j}D^{j-2} - \frac{1}{2}\sum_{j\geq 1} t_{2j-1}\partial^{j-2}Q + \frac{1}{2}\sum_{i,j\geq 1}(i-j)t_{2i-1}t_{2j-1}\partial^{i+j-2} ,$$

$$\Gamma_1 = \theta + \sum_{j\geq 1} t_{2j-1}\partial^{j-1}$$

*where $Q = \partial_\theta - \theta\partial$, be formal infinite order (super)differential operators in $S[t_i][[D]]$ of $\mathbb{Z}_2$-degrees $|\Gamma_0| = 0$, $|\Gamma_1| = 1$. These operators enjoy the following properties:*

a) $[D_i - D^i, \Gamma_0] = 0$, $[D_i - D^i, \Gamma_1] = 0$, *and* $[D_i - D^i, Q] = 0$ *for any $i \geq 1$;*
b) $[Q, \Gamma_1] = 1$, $[Q, \Gamma_0] = -\Gamma_1$, $[\partial, \Gamma_0] = 1$;
c) $[\Gamma_0, \Gamma_1] = 0$, $[\Gamma_1, \Gamma_1] = 0$, $[\Gamma_0, \Gamma_0] = 0$. $\qquad\qquad \square$

PROOF. There is one point that ought to be mentioned here, concerning the extension of the derivations $\partial$ and $D$ to the ring $S[t_i]$. We recall that in the case of the KP hierarchy the first even time could be identified with $x$ because of the first flow which read $\partial_1 = \partial$. Here, although the first even flow tells us again that $\partial_2 = \partial$, things turn out to be different. Indeed $D$ cannot be (analogously to $\partial$) identified with $D_1$, as one can easily convince oneself by writing down the first odd flow. We are therefore forced to proceed safely and do not identify $x$ with $t_2$, but rather keep them as independent variables and extend trivially the action of $\partial$ and $D$ to the ring $S[t_i]$. $\qquad\qquad \square$

We can now define the 'additional' flows of the MRSKP hierarchy as the following four-parameter family of odd and even flows

$$D_{m,k,l,p}\phi = W'(\Gamma_0^k\Gamma_1^l Q^p \partial^m) , \qquad\qquad (5.2.29)$$

with $k \geq 0$, $l = 0, 1$, $p = 0, 1$ and $m \in \mathbb{Z}$, where the even MRSKP flows can be obtained as a particular case, namely $D_{m,0,0,0} = -D_{2m}$ for $m > 0$.

**Theorem 5.2.30.** *The additional flows are symmetries of the MRSKP hierarchy, that is they commute with the MRSKP flows:*

$$[D_i, D_{m,k,l,p}] = 0 \ . \tag{5.2.31}$$

**Proof.** Using the expression of the flows in terms of the supersymmetric Radul map and Theorem 5.2.12 we have that $[D_i, D_{m,k,l,p}] = D_{W'(-[D^i, \Gamma_0^k \Gamma_1^l Q^p \partial^m]_\phi)}$, with

$$\begin{aligned}
[D^i, \Gamma_0^k \Gamma_1^l Q^p \partial^m]_\phi &= D_{W'(D^i)} \Gamma_0^k \Gamma_1^l Q^p \partial^m - (-)^{i(l+p)} D_{W'(\Gamma_0^k \Gamma_1^l Q^p \partial^m)} D^i \\
&\quad + [D^i, \Gamma_0^k \Gamma_1^l Q^p \partial^m] \\
&= -[D_i - D^i, \Gamma_0^k \Gamma_1^l Q^p \partial^m] \ ,
\end{aligned} \tag{5.2.32}$$

which using Lemma 5.2.28 gives us the announced result. $\qquad\square$

**Corollary 5.2.33.** *The algebra of additional symmetries of the MRSKP hierarchy given by* (5.2.29) *is isomorphic to the Lie algebra of SDOP, which is isomorphic (as a Lie algebra) to* $\mathsf{SW}_{1+\infty}$.

**Proof.** Indeed, the isomorphism is given by

$$\begin{aligned}
z \mapsto -\partial \ , &\qquad \xi \mapsto Q + \Gamma_1 \partial \\
\partial_z \mapsto \Gamma_0 \ , &\qquad \partial_\xi \mapsto \Gamma_1 \ .
\end{aligned} \tag{5.2.34}$$

The isomorphism between $\mathsf{SDOP}$ and $\mathsf{SW}_{1+\infty}$ is standard (see, *e.g.*, [**62**]). $\qquad\square$

The fact that we have introduced the generator $Q$ of supertranslations may seem unsatisfactory to the purist, given that the MRSKP hierarchy is only defined in terms of abstract derivations $D_i$ and $D$. One could therefore ask whether it is really necessary to break manifest supersymmetric covariance in this fashion instead of trying to construct another even generator $\widetilde{\Gamma}_0$ that would behave like $x$ and that would satisfy $[D_i - D^i, \widetilde{\Gamma}_0] = 0$, $[D, \widetilde{\Gamma}_0] = \Gamma_1$, and $[\partial, \widetilde{\Gamma}_0] = 1$. This turns out to be impossible, essentially because $D$ itself is not a symmetry of the hierarchy. Indeed, an explicit calculation shows that

$$\begin{aligned}
[D_{2i-1} - D^{2i-1}, \Gamma_1] &= [D_{2i-1} - D^{2i-1}, [D, \widetilde{\Gamma}_0]] \\
&= -2[\partial^i, \widetilde{\Gamma}_0] \\
&= -2i\partial^{i-1} \ ,
\end{aligned} \tag{5.2.35}$$

which is different from zero and which thus contradicts the theorem. Hence such an operator $\widetilde{\Gamma}_0$ cannot exist. One could nevertheless insist that the very definition of (additional) symmetry is not appropriate. Namely, one could argue that by the very nature of an integrable hierarchy, every flow of MRSKP should be thought of as a symmetry of all its other flows. In other words one should include $D$ too as a generator of the additional symmetries. This would of course require redefining the additional symmetries by adding to the previously found flows (5.2.29) the actual flows of the hierarchy. One could even go further and claim that once we allowed for the odd flows of the hierarchy to be part of the additional symmetries, what we have done is really to relax the condition (5.2.27) in order to include (4.3.22) as a particular case. But then consistency would force us to also look for possible times-dependent solutions of (5.2.27) where the right hand side would be proportional with an appropriate power of $\partial$. If one carries on this computation one finds for instance a whole family of odd operators $\Gamma_1 = \theta + \sum_{j \geq 1} a_j t_{2j-1} \partial^{j-1}$ satisfying

$$[D_{2i} - D^{2i}, \Gamma_1] = 0 \ , \tag{5.2.36}$$

$$[D_{2i-1} - D^{2i-1}, \Gamma_1] = (a_i - a_1)\partial^{i-1} \ . \tag{5.2.37}$$

This embarrassment of riches suggests that this more relaxed notion of additional 'symmetry' is of little interest.

A final remark on the additional symmetries on the MRSKP hierarchy is in order. Consider the additional flows $D_{m,0,0,1}$ generated by $Q\partial^m$. These flows obey an algebra isomorphic to the one obeyed by the odd MRSKP flows themselves. Therefore it seems that one could consider them as the odd flows of yet another supersymmetric extension of the KP hierarchy, having the odd flows of MRSKP as additional symmetries and in fact sharing the same additional symmetries as MRSKP. This hierarchy is in fact the object of a recent paper by Ramos [**63**]. It would be interesting to understand how this hierarchy fits in the geometric picture of Mulase and Rabin.

### the SKP$_2$ hierarchy

We recall from Section 4.2 that the SKP$_2$ hierarchy is defined as the universal family of isospectral deformations of a SΨDO of the form $\mathcal{L} = D^2 + \sum_{i \geq 1} U_i D^{2-i}$, with $U_i \in S$ and its evolution is described by a commuting family of flows $\partial_i \mathcal{L} = -[\mathcal{L}_-^i, \mathcal{L}] = [\mathcal{L}_+^i, \mathcal{L}]$, where all the flows are even and therefore can be represented in terms of an infinite set of even times $\{t_1, t_2, \ldots\}$ by $\partial_i = \frac{\partial}{\partial t_i}$. In the following we shall restrict ourselves to operators $\mathcal{L}$ which are dressable; that is, which satisfy the conditions $U_1 = U_2 = 0$.

Notice that one can dress the following obvious commutation relations

$$[\partial_n - \partial^n, \partial] = 0 \tag{5.2.38}$$

with an arbitrary $\phi = 1 + \sum_{i \geq 1} V_i D^{-i}$, $V_i \in S$, and obtain the SKP$_2$ flows.

Let us now consider the problem of finding the additional symmetries for this hierarchy. Fortunately we can use our previous experience with KP and MRSKP to write down the generators of additional symmetries for SKP$_2$. Indeed since the hierarchy has only even flows, it follows that the $x$-like generator for the additional flows of KP still commutes with the SKP$_2$ flows. Moreover, and because of the same reason, both $D$ and $Q$ can now be considered generators of additional symmetries. In fact we have the following result:

**LEMMA 5.2.39.** *Let $S[t_i]$ be the extension ring of $S$ by the even time variables $\{t_i\}$ and let*

$$\Gamma = x + \sum_{t \geq 1} jt_j \partial^{j-1}$$

*be a formal infinite order (super)differential operator in $S[t_i][[\partial]]$. This operator enjoys the following properties: $[\partial_i - \partial^i, \Gamma] = 0$ and $[\partial, \Gamma] = 1$. Moreover, the operators $D$ and $Q$ obey: $[\partial_i - \partial^i, D] = [\partial_i - \partial^i, Q] = 0$.* $\qquad\square$

We can now define the 'additional' flows of the SKP$_2$ hierarchy as the following four-parameter family of odd and even flows

$$D_{m,k,l,p}\phi = W'(\Gamma^k D^l Q^p \partial^m) \ , \qquad (5.2.40)$$

with $k \geq 0$, $l = 0, 1$, $p = 0, 1$ and $m \in \mathbb{Z}$. Again, the original flows of the hierarchy can be obtained as a particular case, namely $D_{m,0,0,0} = -\partial_m$ for $m > 0$.

**THEOREM 5.2.41.** *The additional flows are symmetries of the SKP$_2$ hierarchy, that is they commute with the MRSKP flows:*

$$[D_i, D_{m,k,l,p}] = 0 \ . \qquad (5.2.42)$$

**PROOF.** Using the expression of the flows in terms of the supersymmetric Radul map and Theorem 5.2.12 we have that

$$[\partial_i, D_{m,k,l,p}] = D_{W'(-[\partial^i, \Gamma^k D^l Q^p \partial^m]_\phi)} \ , \qquad (5.2.43)$$

with

$$[\partial^i, \Gamma^k D^l Q^p \partial^m]_\phi = D_{W'(\partial^i)} \Gamma^k D^l Q^p \partial^m - D_{W'(\Gamma^k D^l Q^p \partial^m)} \partial^i + [\partial^i, \Gamma^k D^l Q^p \partial^m]$$
$$= -[\partial_i - \partial^i, \Gamma^k D^l Q^p \partial^m] \ , \qquad (5.2.44)$$

which using Lemma 5.2.39 gives us the announced result. $\qquad\square$

**COROLLARY 5.2.45.** *The algebra of additional symmetries of the SKP$_2$ hierarchy given by (5.2.40) is isomorphic to the Lie algebra of* SDOP.

Proof. Indeed, the isomorphism is given by

$$z \mapsto -\partial \; , \qquad \xi \mapsto \tfrac{1}{2}(D-Q)\partial^{-1}$$
$$\partial_z \mapsto \Gamma \; , \qquad \partial_\xi \mapsto \tfrac{1}{2}(D+Q) \; . \tag{5.2.46}$$

$\square$

the jskp hierarchy

Following the same path as for MRSKP it is easily seen that the JSKP flows can be written in terms of the $W'$ map, (5.2.22), by

$$D_{2n}\phi = -W'(\partial^n) = -D_{W'(\partial^n)}\phi$$
$$D_{2n+1}\phi = -W'(\partial^n\partial_\theta) = -D_{W'(\partial^n\partial_\theta)}\phi \; . \tag{5.2.47}$$

Proposition 5.2.48. *The JSKP flows satisfy a commutative Lie superalgebra.*

Proof. This is already clear since $[\partial^n, \partial^m] = [\partial^n, \partial^m\partial_\theta] = [\partial^n\partial_\theta, \partial^m\partial_\theta] = 0$. $\square$

In this case we will look for additional symmetries generated by operators $\Gamma$ satisfying

$$[D_{2i} - \partial^i, \Gamma] = 0$$
$$[D_{2i+1} - \partial^i\partial_\theta, \Gamma] = 0 \; . \tag{5.2.49}$$

The two obvious solutions are, as expected, the even and odd derivations on the ring $\mathcal{S}$, $\partial$ and $\partial_\theta$. This means in particular that, unlike MRSKP, all the JSKP flows are also symmetries of the hierarchy. One has nevertheless more.

Lemma 5.2.50. *Let $S[t_i]$ be the extension ring of $S$ by the time variables $\{t_i\}$ and let*

$$\Gamma_0 = x + \sum_{j\geq 1} jt_{2j}\partial^{j-1} + \sum_{j\geq 1} jt_{2j+1}\partial^{j-1}\partial_\theta \tag{5.2.51}$$

$$\Gamma_1 = \theta + \sum_{j\geq 1} t_{2j-1}\partial^{j-1} \tag{5.2.52}$$

$$\Gamma_2 = x\partial_\theta + \sum_{j\geq 1} jt_{2j}\partial^{j-1}\partial_\theta \; . \tag{5.2.53}$$

*be formal infinite order differential operators in $S[t_i][[\partial, \partial_\theta]]$ of $\mathbb{Z}_2$ -degrees $|\Gamma_0| = 0$ and $|\Gamma_1| = |\Gamma_2| = 1$. These operators have the following properties:*

a) *$[D_{2i} - \partial^i, \Gamma_k] = 0$ and $[D_{2i+1} - \partial^i\partial_\theta, \Gamma_k] = 0$ for all $k = 0, 1, 2$;*
b) *$[\partial, \Gamma_1] = 0$, $[\partial_\theta, \Gamma_1] = 1$, $[\Gamma_1, \Gamma_1] = 0$;*
c) *$[\partial, \Gamma_2] = \partial_\theta$, $[\partial_\theta, \Gamma_2] = 0$, $[\Gamma_2, \Gamma_2] = 0$;*
d) *$[\partial, \Gamma_0] = 1$, $[\partial_\theta, \Gamma_0] = 0$, $[\Gamma_1, \Gamma_2] = \Gamma_0$.*

PROOF. It follows after a routine calculation. □

We can now define a three-parameter family of flows

$$D_{2m,k,l}\phi = W'(\Gamma_0^k \Gamma_1^l \partial^m)$$
$$D_{2m+1,k,l}\phi = W'(\Gamma_0^k \Gamma_1^l \partial^m \partial_\theta) \ , \qquad (5.2.54)$$

where $k \geq 0$, $l = 0, 1$, and $m \in \mathbb{Z}$. Here the original JSKP flows are a special case, $D_{m,0,0} = -D_m$ for $m \geq 0$, whereas the other ones represent the additional symmetries of the JSKP hierarchy.

THEOREM 5.2.55. *The additional flows are symmetries of the JSKP hierarchy, in other words they commute with the flows on the Volterra group.*

PROOF. We only have to use Theorem 5.2.12 and we obtain, for instance, for the even flows

$$[D_{2i}, D_{2m,k,l}] = -D_{W'([\partial^i, \Gamma_0^k \Gamma_1^l \partial^m]_\phi)} \ , \qquad (5.2.56)$$

where

$$[\partial^i, \Gamma_0^k \Gamma_1^l \partial^m]_\phi = D_{W'(\partial^i)} \Gamma_0^k \Gamma_1^l \partial^m + [\partial^i, \Gamma_0^k \Gamma_1^l \partial^m]$$
$$= -[D_{2i} - \partial^i, \Gamma_0^k \Gamma_1^l \partial^m]$$
$$= 0 \ . \qquad (5.2.57)$$

Analogous computations give us that

$$[D_{2i}, D_{2m+1,k,l}] = -D_{W'([\partial^i, \Gamma_0^k \Gamma_1^l \partial^m \partial_\theta]_\phi)} = 0 \ , \qquad (5.2.58)$$
$$[D_{2i+1}, D_{2m,k,l}] = -D_{W'([\partial^i \partial_\theta, \Gamma_0^k \Gamma_1^l \partial^m]_\phi)} = 0 \ , \qquad (5.2.59)$$
$$[D_{2i+1}, D_{2m+1,k,l}] = -D_{W'([\partial^i \partial_\theta, \Gamma_0^k \Gamma_1^l \partial^m \partial_\theta]_\phi)} = 0 \ , \qquad (5.2.60)$$

which finally proves the above statement. □

COROLLARY 5.2.61. *The Lie superalgebra of symmetries of the Jacobian SKP hierarchy is isomorphic to* SDOP *which is isomorphic (as a Lie algebra) to* SW$_{1+\infty}$.

PROOF. Let $\mathcal{A}$ be the Lie superalgebra of symmetries given by (5.2.54). It is generated via the Radul map by $\Gamma_0^k \Gamma_1^l \partial^m$ and $\Gamma_0^k \Gamma_1^l \partial^m \partial_\theta$ for $k \geq 0$, $l = 0, 1$ and $m \in \mathbb{Z}$. The isomorphism SDOP $\to \mathcal{A}$ is given explicitly by

$$z \mapsto -\partial \ , \qquad \xi \mapsto \partial_\theta$$
$$\partial_z \mapsto \Gamma_0 \ , \qquad \partial_\xi \mapsto \Gamma_1 \ . \qquad (5.2.62)$$

□

Of the flows $D_{m,k,l}$ defined by (5.2.54), all but the $D_{m,0,0}$ with $m \geq 0$ are additional symmetries. These additional symmetries are isomorphic to the direct sum of $\mathsf{SW}_\infty$ with the abelian algebra generated by the flows $D_{m,0,0}$ with $m < 0$. These flows are present only because the JSKP hierarchy is defined on the superVolterra group. If, as in the KP hierarchy, JSKP were defined on the space of Lax operators, these extra flows would not be present; for they act trivially on $L = \phi \partial \phi^{-1}$.

The isomorphism between the additional symmetries of all three SKP hierarchies deserves a final comment. The picture that begins to emerge is that the additional symmetries, although realized dynamically with explicit dependence on the times, are actually a kinematical property of the dynamical systems; that is, symmetries of the phase space in which the systems are defined.

# Chapter Six

# REDUCTIONS OF SKP HIERARCHIES

We have seen in Chapter Four how one can build up the Lax formalism for the supersymmetric integrable hierarchies by analogy with the nonsupersymmetric case. We concluded then that there is no unique supersymmetric KP hierarchy which enjoys all the properties we would expect it to do (for example to yield by reduction all the generalized sKdV hierarchies). Rather we were forced to define several SKP hierarchies, all of which contain KP as a particular reduction. A closer analysis of MRSKP and $SKP_2$ revealed moreover that important questions concerning these hierarchies still remain to be answered. We do not understand yet the hamiltonian structure of the MRSKP hierarchy, although it has been shown that the space of supersymmetric Lax operators admits a Poisson structure analogous to the second Gel'fand-Dickey bracket of the $n$-KdV hierarchies. On the other hand the $SKP_2$ hierarchy does not possess odd flows and its bihamiltonian structure does not seem to display a superconformal structure.

There exists nevertheless a particular reduction of $SKP_2$, the sKdV hierarchy, which gives us a glimpse of hope since its natural Poisson structure—the superVirasoro algebra—appeared via hamiltonian reduction from the Poisson structure of the $SKP_2$ hierarchy. Also a first attempt of unraveling the odd part of $SKP_2$ has been made in [**64**] where nonlocal conservation laws for the sKdV equation have been constructed. This particular example begs the question whether there exist more general reductions of the SKP hierarchies which possess the properties we would like them to have, such as locality, hamiltonian structure, explicit superconformal structure, odd flows...

In this chapter we will attempt to give answers to some of these questions. In particular we show that for the case of dressable Lax operators we can endow the $SKP_2$ hierarchy with odd flows. Following [**51**] we study the symmetric reduction of odd order sKdV hierarchies whose Poisson brackets define classical W-superalgebras. Then we consider in detail the $BSKP_2$ hierarchy which turns out to be local, hamiltonian and whose Poisson brackets display a manifest superconformal structure. The results in this chapter follow my paper [**65**] with E. Ramos.

## 6.1.  DRESSABILITY AND ODD FLOWS FOR SKP$_2$

Our main goal in this section is to define consistent odd flows for the SKP$_2$ hierarchy. In order to do so we will consider particular SKP$_2$ Lax operators that admit a square root, namely $\mathcal{L} = \Lambda^2$, where $\Lambda$ is the Lax operator of the MRSKP hierarchy. We will be able to induce, via MRSKP, nonlocal odd flows for dressable SKP$_2$ Lax operators.

An SKP$_2$ Lax operator $\mathcal{L}$ is called dressable, if there exists an even SΨDO

$$\phi = 1 + \sum_{\geq 1} A_i D^{-i} \ , \tag{6.1.1}$$

such that

$$\mathcal{L} = \phi \partial \phi^{-1} \ . \tag{6.1.2}$$

A simple computation reveals the following:

LEMMA 6.1.3.  *A necessary and sufficient condition for dressability is that $U_1 = U_2 = 0$.*                                                                          □

In the introduction to Chapter Four, we motivated the SKP$_2$ hierarchy by the fact that $\mathcal{L}$ does not admit in general a (unique) square root. Notice nevertheless that if we restrict ourselves to dressable $\mathcal{L}$'s then there exists

$$\mathcal{L}^{1/2} = \phi D \phi^{-1} \ . \tag{6.1.4}$$

In other words, the Lax operator of SKP$_2$ admits a square root which is nothing but the Lax operator of the MRSKP hierarchy. Clearly $\mathcal{L}^{1/2}$ will satisfy in this case the dressability condition for MRSKP. As proven in [**29**] the square root, if it exists, is not necessarily unique. Uniqueness can be achieved in our case if we further impose manifest supersymmetry as well as homogeneity with respect the natural grading. Moreover, if one works out explicitly the condition for the square root of $\mathcal{L}$ to exist, one obtains once more that $U_1$ and $U_2$ must necessarily vanish, and this is, as we have just shown, precisely the condition that $\mathcal{L}$ be dressable. One more remark is in order. One can easily convince oneself, by working out explicitly the square root and dressability conditions for SKP$_2$, that both the coefficients of $\mathcal{L}^{1/2}$ and $\phi$ are nonlocal in the $U_j$'s.

We are now in a position to define odd flows for SKP$_2$.

PROPOSITION 6.1.5.  *Provided we restrict ourselves to dressable operators, there is a one-to-one correspondence between flows in MRSKP and SKP$_2$.*

PROOF. Consider a generic flow of MRSKP:

$$D_t \Lambda = [P, \Lambda] \,, \tag{6.1.6}$$

with $P$ a certain $\mathsf{S\Psi DO}$. The flow on $\Lambda^2$ is then

$$\begin{aligned}
D_t \Lambda^2 &= (D_t \Lambda) \Lambda + (-)^{|P|} \Lambda (D_t \Lambda) \\
&= [P, \Lambda^2] \,; \tag{6.1.7}
\end{aligned}$$

in other words,

$$D_t \mathcal{L} = [P, \mathcal{L}] \,. \tag{6.1.8}$$

Conversely, consider a generic $\text{SKP}_2$ flow, of the form (6.1.8) and take into account the fact that $\mathcal{L} = \Lambda^2$. Then we get

$$(D_t \Lambda - [P, \Lambda]) \Lambda + (-)^{|P|} \Lambda (D_t \Lambda - [P, \Lambda]) = 0 \,. \tag{6.1.9}$$

Assume for a contradiction that the expression in the parenthesis does not vanish but rather that $D_t \Lambda - [P, \Lambda] = B_N D^N + B_{N-1} D^{N-1} + \dots$, for some $N \in \mathbb{Z}$. The leading coefficients must satisfy the following conditions:

$$B_N = 0 \quad \text{for } N \text{ odd},$$

and

$$\left. \begin{array}{c}
B'_{N-1} = 0 \\
B'_N + (-)^{|P|} 2 B_{N-1} = 0
\end{array} \right\} \quad \text{for } N \text{ even.} \tag{6.1.10}$$

Then provided we drop the constants of integration, the leading coefficient $B_N$ must vanish; whence

$$D_t \Lambda = [P, \Lambda] \,. \tag{6.1.11}$$

This proves the proposition. □

It is convenient to change now our notation of the $\text{SKP}_2$ flows, namely we will denote by $D_p$ the $p$-th flow of the hierarchy in such a way that $D_{2p} = \partial_p$.

COROLLARY 6.1.12. *The following are odd flows for the $\text{SKP}_2$ hierarchy:*

$$D_{2i-1} \mathcal{L} = -[\mathcal{L}_-^{i-\frac{1}{2}}, \mathcal{L}] \,. \qquad \qquad □$$

It is important to remark that although the odd flows are explicitly local in terms of the MRSKP variables, they are not local when written in terms of the $\text{SKP}_2$ variables. This was already pointed out by Dargis and Mathieu in [**64**] for the sKdV case. One could therefore ask whether there is no other way of providing the $\text{SKP}_2$ hierarchy with *local* odd flows. An explicit computation for the first few odd flows suggests that there is none.

## 6.2. THE SYMMETRIC REDUCTION

In this section we investigate the reduction of the supersymmetric Gel'fand–Dickey bracket induced by demanding that the Lax operator have a definite adjointness property. We basically follow [**51**], the difference being that we consider general pseudodifferential Lax operators. To motivate the definition of the adjoint, let us think of differential operators as acting on superfields with inner product

$$\langle U, V \rangle = \int_B UV \ . \tag{6.2.1}$$

If $L \in \mathcal{S}_+$ is a homogeneous differential operator, we define its adjoint $L^*$ by $\langle LU, V \rangle = (-)^{|L||U|} \langle U, L^* V \rangle$, for any homogeneous superfields $U, V$. The proof of the following proposition is routine.

PROPOSITION 6.2.2. $*$ *extends to an involution in the space* $\mathcal{S}$ *of* S$\Psi$DO*'s which obeys the following properties:*
(1) *For all* $P \in \mathcal{S}$, $(P^*)^* = P$
(2) *For all homogeneous* $P, Q \in \mathcal{S}$, $(PQ)^* = (-)^{|P||Q|} Q^* P^*$
(3) *If* $P \in \mathcal{S}$ *is homogeneous and invertible,* $(P^{-1})^* = (-)^{|P|}(P^*)^{-1}$.
(4) *For all* $p \in \mathbb{Z}$, $(D^p)^* = (-)^{\frac{p(p+1)}{2}} D^p$.
(5) *For all* $P \in \mathcal{S}$, $(P_\pm)^* = (P^*)_\pm$.
(6) *For all* $P \in \mathcal{S}$, $\operatorname{sres} P^* = \operatorname{sres} P$ *(in particular,* $\operatorname{Str} P^* = \operatorname{Str} P$*).*          $\square$

If a Lax operator $L = D^n + \cdots$ has a definite adjointness property, it is dictated by the first term. We shall say that $L$ is **symmetric** if $L^* = (-)^{n(n+1)/2} L$. We will show that the supersymmetric Gel'fand–Dickey bracket in the space $\mathfrak{M}_{2k+1}$ of Lax operators of a given odd order induces a Poisson bracket in the submanifold $\widetilde{\mathfrak{M}}_{2k+1}$ of symmetric Lax operators and that the induced fundamental Poisson brackets define a W-superalgebra extending the $N{=}1$ superVirasoro algebra.

In order to understand the constraints that the symmetry condition imposes on the coefficients $U_i$ of $L$ it is convenient to write $L$ in a manifestly symmetric way. In general, a symmetric Lax operator has the form

$$L = D^n + \tfrac{1}{2} \sum_{j \in I_n} \left\{ V_j \, , \, D^{n-j} \right\} \ , \tag{6.2.3}$$

where $\left\{ V_j \, , \, D^{n-j} \right\} = V_j D^{n-j} + (-)^{j(n-j)} D^{n-j} V_j$ is the graded anticommutator and the sum runs over the index set

$$I_n = \left\{ j = 1, 2, \dots \left| (-)^{(n-j)(n-j+1)/2} = (-)^{n(n+1)/2} \right. \right\} \ . \tag{6.2.4}$$

Equation (6.2.3) manifestly exhibits which of the fields $V_j$ are independent;

namely, if $n$ is even, we pick those $V_j$ with $j \equiv 0, 1 \mod 4$; and if $n$ is odd, then we take those $V_j$ with $j \equiv 0, 3 \mod 4$.

Some general facts readily emerge. If $n$ is odd, there is always an independent field of weight $\frac{3}{2}$ and, moreover, this is the field of smallest weight. One can show that its Poisson bracket is that of the classical $N{=}1$ superVirasoro algebra. For even $n$ the situation is radically different: there is never a field of weight $\frac{3}{2}$ but there is always a field of weight $\frac{1}{2}$.

In order to describe the induced Poisson bracket we first need to identify the vector fields and the one-forms on $\widetilde{\mathfrak{M}}_{2k+1}$ as subobjects of the corresponding objects in $\mathfrak{M}_{2k+1}$. The vector fields of $\widetilde{\mathfrak{M}}_{2k+1}$ will be parametrized by the deformations of $L$ that remain in $\widetilde{\mathfrak{M}}_{2k+1}$; that is, deformations of a symmetric Lax operator $L$ which keep it symmetric. These are clearly the differential operators of order at most $2k$ obeying the same symmetry property as $L$. As explained in Section 2.3, one-forms on $\widetilde{\mathfrak{M}}_{2k+1}$ must be chosen to be those one-forms on $\mathfrak{M}_{2k+1}$ which are mapped (via the Adler map) to vector fields tangent to $\widetilde{\mathfrak{M}}_{2k+1}$. In other words, one-forms on $\widetilde{\mathfrak{M}}_{2k+1}$ are S$\Psi$DO's $X = \sum_l D^{-l-1} X_l$ satisfying $J(X)^* = -(-)^k J(X)$. Computing this we find

$$
\begin{aligned}
J(X)^* &= [(LX)_+ L - L(XL)_+]^* \\
&= (-)^{|X|+1} \left[ L^*(LX)_+^* - (XL)_+^* L^* \right] \\
&= -\left[ L^*(X^*L^*)_+ - (L^*X^*)_+ L^* \right] \\
&= (LX^*)_+ L - L(X^*L)_+ \\
&= J(X^*) \ ;
\end{aligned}
\tag{6.2.5}
$$

whence $X$ must have the same symmetry properties of $L$, namely $X^* = -(-)^k X$, for it to be a one-form on $\widetilde{\mathfrak{M}}_{2k+1}$. It is easy to verify that these one-forms are non-degenerately paired with the vector fields tangent to $\widetilde{\mathfrak{M}}_{2k+1}$. In fact, since $\mathrm{Str}\, AX = \mathrm{Str}\, A^* X^*$ we see that the supertrace pairs up one-forms and vector fields of the same symmetry properties. Therefore the Poisson bracket of two functions $F = \int_B f$ and $G = \int_B g$ on $\widetilde{\mathfrak{M}}_{2k+1}$ is obtained from (4.1.28) (with $J$ given by (4.1.29)) by simply requiring that $dF$ and $dG$ have the correct symmetry properties: $(dF)^* = -(-)^k dF$ and the same for $dG$.

One can now explicitly compute the induced fundamental Poisson brackets on $\widetilde{\mathfrak{M}}_{2k+1}$. We have seen that the field of smallest weight is $V_3$, which has weight $\frac{3}{2}$. One can actually show that the induced fundamental Poisson bracket $\{V_3(X)\,,\,V_3(Y)\}$ defines a $N{=}1$ superVirasoro algebra. Indeed if we define the differential operators $\Omega_{ij}$ by

$$
\{V_i(X)\,,\,V_j(Y)\} = \Omega_{ij}\delta(X-Y) \ ,
\tag{6.2.6}
$$

where the $\Omega_{ij}$ are taken at the point $X$, we find

$$\Omega_{33} = \frac{k(k+1)}{4}D^5 + \frac{3}{2}V_3 D^2 + \frac{1}{2}V_3' D + V_3'' \ , \qquad (6.2.7)$$

whence, if we let $\mathbb{G} = V_3$ this will give rise to a classical version of the $N=1$ superVirasoro algebra

$$\{\mathbb{G}(X)\,,\,\mathbb{G}(Y)\} = \left[\frac{k(k+1)}{4}D^5 + \frac{3}{2}\mathbb{G}(X)D^2 + \frac{1}{2}\mathbb{G}'(X)D + \mathbb{G}''(X)\right]\delta(X-Y) \ . \qquad (6.2.8)$$

## 6.3.  THE BSKP$_2$ HIERARCHY

One of the remarkable features of the bihamiltonian structure for the KdV-type hierarchies is the fact that the second Gel'fand–Dickey bracket exhibits (after a standard reduction) a conformal structure. In the supersymmetric case however, and in spite of the fact that the SKP$_2$ hierarchy is bihamiltonian, its second Gel'fand–Dickey bracket does not explicitly exhibit a superconformal structure. On the other hand, its natural reduction to dressable Lax operators gives rise to a nonlocal induced hamiltonian structure [**66**], whose first reduced Poisson bracket is indeed the superVirasoro algebra. This encourages us to search for a local reduction of the SKP$_2$ hierarchy whose hamiltonian structure yields an nonlinear extension of the superVirasoro algebra.

In the introduction to this chapter we remarked the existence of a particular reduction of SKP$_2$ which displays these properties: the sKdV hierarchy is the unique reduction (of a fourth order Lax operator) of the SKP$_2$ hierarchy satisfying the constraint

$$L^* = DLD^{-1} \ , \qquad (6.3.1)$$

where $*$ is the involution in Proposition 6.2.2. On the other hand, this condition is nothing but the supersymmetric analogue of the constraint used in [**25**] to define the BKP hierarchy. This prompts us to consider the general reduction of SKP$_2$ defined by

$$\mathcal{L}^* = -D\mathcal{L}D^{-1} \ . \qquad (6.3.2)$$

We call it the BSKP$_2$ hierarchy and we denote by $\widetilde{\mathfrak{M}}_2$ the subspace of SKP$_2$ Lax operators satisfying this symmetry condition. This hierarchy is to be compared with the orthosymplectic SKP hierarchy studied by Ueno, Yamada, and Ikeda in [**67**]. The reduction (6.3.2) was shown in [**55**] to be consistent, as long as only half of the flows are considered. Let us for the sake of completeness recall this result.

PROPOSITION 6.3.3. *The condition* (6.3.2) *is equivalent to the conditions*

$$\operatorname{sres}(\mathcal{L}^{2n-1}D^{-1}) = 0 \ ,$$
$$2\operatorname{sres}\mathcal{L}^{2n} - \operatorname{sres}(\mathcal{L}^{2n}D^{-1})' = 0 \ . \qquad \Box$$

Let us now see what is the concrete form that the Lax operator must have for it to satisfy the constraint condition; in other words, which are the fields that survive the reduction. Since we have an infinite number of fields it is not practicable to try to explicitly work out the constraint at the level of the $U_j$'s. Instead, it is more convenient to write the Lax operator $\mathcal{L}$ in a more symmetrical form, by using a different basis.

In order to do this, as well as for later purposes, we define now a map $\varphi$ going from the space $\mathfrak{M}_1$ of MRSKP Lax operators to the space $\mathfrak{M}_2$ of SKP₂ Lax operators, given by

$$\mathcal{L} \equiv \varphi(\Lambda) = \Lambda D \ . \tag{6.3.4}$$

It is not difficult to see that $\varphi$ maps symmetric operators $\Lambda$ into BSKP₂ Lax operators $\mathcal{L}$ obeying (6.3.2). From Section 2, we know that the symmetric operator $\Lambda$ can be written as

$$\Lambda = D + \frac{1}{2} \sum_{\substack{j \equiv 0,3 \bmod 4 \\ j > 0}} \left\{ V_j \, , \, D^{1-j} \right\} \ . \tag{6.3.5}$$

It follows then that the BSKP₂ Lax operator can be written without loss of generality as

$$\mathcal{L} = D^2 + \frac{1}{2} \sum_{\substack{j \equiv 0,3 \bmod 4 \\ j > 0}} \left\{ V_j \, , \, D^{1-j} \right\} D \ . \tag{6.3.6}$$

It is sufficient to write down the first few terms of $\mathcal{L}$ to realize that this reduction sets the first two fields of $\mathcal{L}$ equal to zero. Hence $\mathcal{L}$ is dressable and we can apply the machinery developed in the previous section in order to define the odd flows associated to the half-integer powers of the Lax operator.

Let us now investigate which of the SKP₂ flows survive the reduction. From Section 1 we know that SKP₂ has both odd and even flows.

PROPOSITION 6.3.7. *Only those flows* $D_p$ *with* $p \equiv 2, 3 \bmod 4$ *are consistent with the reduction* (6.3.2).

PROOF. Let us first consider the even flows $D_{2p}$. Take the adjoint of the even SKP₂ flow given by (4.3.13) and use the condition (6.3.2). If follows already from the leading term that $n$ has to be an odd integer. In addition one gets that the following condition has to be satisfied:

$$(D\mathcal{L}^p D^{-1})_- = D\mathcal{L}^p_- D^{-1} \ . \tag{6.3.8}$$

It is a simple computational matter to check that for $A$ an arbitrary $\mathsf{S\Psi DO}$, $(DAD^{-1})_- = DA_- D^{-1}$ if and only if sres $AD^{-1} = 0$, which is precisely the case for $p$ odd by Proposition 6.3.3. In order to see which of the odd flows will survive one has to first analyze the MRSKP hierarchy under the reduction $\Lambda^* = -D\Lambda D^{-1}$. In this case one sees that condition (6.3.2) implies sres $\Lambda^{4k-1} D^{-1} = 0$, which is equivalent to

$$(D\Lambda^{4k-1} D^{-1})_- = D\Lambda^{4k-1}_- D^{-1} \ . \tag{6.3.9}$$

From this it follows that only the MRSKP flows $D_{4k-1}$ preserve the constraint. Therefore only the SKP$_2$ flows $D_{4k-1}$ preserve the constraint. In summary, we find that precisely the SKP$_2$

$$\begin{aligned} D_{4k-1}\mathcal{L} &= -[\mathcal{L}^{2k-\frac{1}{2}}_-, \mathcal{L}] \text{ and} \\ D_{4k-2}\mathcal{L} &= -[\mathcal{L}^{2k-1}_-, \mathcal{L}] \end{aligned} \tag{6.3.10}$$

will survive.                                                                □

One can trivially write down the first flow $D_2$ on any of our fields, $V_j$, $j \equiv 0, 3 \bmod 4$, $j > 0$, since $D_2\mathcal{L} = [\mathcal{L}_+, \mathcal{L}] = (\partial\mathcal{L})$.

Finally, let us remark that these flows span the following subalgebra of the algebra of flows:

$$[D_{4i-2}, D_{4j-2}] = [D_{4i-2}, D_{4j-1}] = 0 \quad \text{and} \quad [D_{4i-1}, D_{4j-1}] = 2D_{4i+4j-2} \ . \tag{6.3.11}$$

### HAMILTONIAN STRUCTURE

One would naively expect that the hamiltonian structure for BSKP$_2$ can be simply obtained by a suitable reduction of the hamiltonian structure of the SKP$_2$ hierarchy. Unfortunately this is not the case, since the constraints seem to be formally first-class. Although in the finite-dimensional case there exists a well-developed machinery to treat such constraints, here their infinite number forbids a similar analysis. Nevertheless, the solution to the problem is suggested by the connection (6.3.4) between the MRSKP and SKP$_2$ Lax operators. In a nutshell, we will pull back the Adler map from MRSKP to SKP$_2$ and we will check that the even flows are hamiltonian relative to the induced hamiltonian structure.

Let us consider again the map $\varphi$ from the space $\mathfrak{M}_1$ of MRSKP operators to the space $\mathfrak{M}_2$ of SKP$_2$ operators given by equation (6.3.4). The image under $\varphi$ of the subspace $\widetilde{\mathfrak{M}}_1$ defined by the condition $\Lambda^* = -\Lambda$ is precisely the space $\widetilde{\mathfrak{M}}_2$ of BSKP$_2$ Lax operators.

From the geometric formalism developed in Chapters Three and Four, we know that the Adler map can be understood as tensorial map from one-forms to vector fields. Therefore, we can use the map $\varphi$ to pull back the Adler map on $\widetilde{\mathfrak{M}}_1$ to $\widetilde{\mathfrak{M}}_2$. This is done by completing the following commutative square:

$$
\begin{array}{ccc}
T^*_\Lambda \widetilde{\mathfrak{M}}_1 & \xrightarrow{\ J\ } & T_\Lambda \widetilde{\mathfrak{M}}_1 \\[2pt]
\Big\uparrow {\scriptstyle \varphi^*} & & \Big\downarrow {\scriptstyle \varphi_*} \\[2pt]
T^*_\mathcal{L} \widetilde{\mathfrak{M}}_2 & \xdashrightarrow{\ J_B\ } & T_\mathcal{L} \widetilde{\mathfrak{M}}_2
\end{array}
$$

In other words, $J_B = \varphi_* \circ J \circ \varphi^*$. Let us compute this. If $X$ is any 1-form in $\widetilde{\mathfrak{M}}_2$, it obeys

$$X^* = (-)^{|X|} D X D^{-1} \ . \tag{6.3.12}$$

Its pull-back via $\varphi$ to a one-form on $\widetilde{\mathfrak{M}}_1$ is given by

$$Y = \varphi^*(X) = DX \ . \tag{6.3.13}$$

Notice that $Y^* = -Y$ and therefore it is indeed a one-form on $\widetilde{\mathfrak{M}}_1$. Similarly, if $A$ is a vector field on $\widetilde{\mathfrak{M}}_1$, its pushforward via $\varphi$ to a vector field on $\widetilde{\mathfrak{M}}_2$ is given by $\varphi_*(A) = AD$. Therefore, the induced hamiltonian map in $\widetilde{\mathfrak{M}}_2$ is given by

$$J_B(X) = (\mathcal{L}X)_+ \mathcal{L} - \mathcal{L} D^{-1} (DX\mathcal{L}D^{-1})_+ D \ . \tag{6.3.14}$$

In order to show that $J_B$ provides a hamiltonian structure for the BSKP$_2$ flows, let us first consider the following set of hamiltonian functions

$$H_{4k-2} = \frac{1}{2k-1} \operatorname{Str} \mathcal{L}^{2k-1} \ . \tag{6.3.15}$$

Their gradients are given by

$$dH_{4k-2} = \mathcal{L}^{2k-2} \ , \tag{6.3.16}$$

and they obey $(dH_{4k-2})^* = D dH_{4k-2} D^{-1}$, so they define one-forms on $\widetilde{\mathfrak{M}}_2$. From (6.3.14) and Proposition 6.3.3 it follows that

$$D_{4k-2} \mathcal{L} = J_B(dH_{4k-2}) \tag{6.3.17}$$

as required. To prove that the odd flows are hamiltonian with respect to $J_B$ is a much more delicate matter. Notice that the natural odd hamiltonians are

provided by

$$H_{4k-1} = \frac{1}{2k - \frac{1}{2}} \, \mathrm{Str} \, \mathcal{L}^{2k-\frac{1}{2}} \, , \qquad (6.3.18)$$

and their integrands are nonlocal in the $U_j$. Therefore the formalism developed for differential polynomials of the $U_j$ should be extended to integrodifferential polynomials. Nevertheless, if one proceeds formally, one obtains

$$dH_{4k-1} = \mathcal{L}^{2k-3/2} \, , \qquad (6.3.19)$$

which defines an odd one-form in $\widetilde{\mathfrak{M}}_2$; whence

$$D_{4k-1}\mathcal{L} = J_B(dH_{4k-1}) \, . \qquad (6.3.20)$$

that is, the odd flows are also hamiltonian.

In order to compute the Poisson bracket between two functions $F = \int_B f$ and $G = \int_B g$ on $\widetilde{\mathfrak{M}}_2$, we simply require that their differential have the correct symmetry properties; that is, $(dF)^* = (-)^{|dF|} DdFD^{-1}$ and the same for $dG$, and use (4.1.28) with $J_B$. Notice that despite its appearance the fundamental Poisson brackets induced by (6.3.14) in $\widetilde{\mathfrak{M}}_2$ are local. This can be seen most easily by realizing that at the component level the map $\varphi$ is simply an identity, therefore the locality of the fundamental Poisson bracket on $\widetilde{\mathfrak{M}}_1$ ensures the locality of the fundamental Poisson bracket among the $U_j$. Moreover this induced Poisson bracket defines a nonlinear extension of the $N = 1$ superVirasoro algebra ($\mathsf{W}_{BSKP}$) by fields of spin $k > 0$, where $2k \equiv 0, 3 \bmod 4$.

# Chapter Seven

# SUPERSYMMETRIC HIERARCHIES IN STRING THEORY

One of the most pleasant surprises that noncritical string theory has had in store for us is its relation with classical integrable hierarchies of the KP-type. The KdV hierarchy appeared unsuspectedly in the double scaling limit of the one-matrix model—a fact which recurs in the multimatrix models for the generalized KdV hierarchies, and which allows one to exactly compute correlation functions on arbitrary topology. Indeed, the partition function of the $(p-1)$-matrix model, which in its $q$th critical point describes $(p, q)$ conformal matter coupled with two-dimensional gravity, coincides with the $\tau$-function of the $p$-KdV hierarchy. This $\tau$ function has to be special in the sense that it has to fulfill an infinite set of W-constraints which excludes, for example, the polynomial soliton solutions. This success notwithstanding, the generalization of these techniques to the supersymmetric case is still an open problem and the precise relation, if any, with supersymmetric integrable hierarchies remains elusive.

In this chapter we discuss new supersymmetrizations of the generalized KdV hierarchies suggested by a new supersymmetric extension of the KdV hierarchy that has appeared in a matrix-model inspired approach to two-dimensional quantum supergravity. The resulting supersymmetric hierarchies are generically nonlocal, with the exception of the KdV and Boussinesq which turn out to be integrable and bihamiltonian. The first section is expository in nature, describing briefly the appearance of the KdV hierarchy in the hermitean one-matrix model. We follow the treatment in [**68**]. The next two sections describe the contents of my papers [**69**] and [**70**] with J.M. Figueroa-O'Farrill.

## 7.1. THE KDV HIERARCHY AND THE ONE-MATRIX MODEL

Let us start by defining the Hermitean one-matrix model. Consider therefore the zero-dimensional quantum field theory having as generator of Feynman diagrams the following partition function

$$Z(N, t) = \int dM \exp -N(\tfrac{1}{2} \operatorname{Tr} M^2 + t \operatorname{Tr} M^4) , \qquad (7.1.1)$$

where $M$ is a Hermitean $N \times N$ matrix and we have the standard measure $dM = \prod_i dM_i^i \prod_{i<j} d(Re M_i^j) d(Im M_i^j)$. Starting from this field-theoretical model one can immediately deduce the Feynman rules and and draw the corresponding Feynman diagrams. These will turn out to have a double-line structure due to the matrix-valued fields (see Fig. 1.3). By joining together the edges of the double lines in such a way that they form closed curves and by filling these circles with oriented discs, we can associate to any of these diagrams a Riemann surface $\Sigma$ together with a simplicial decomposition.

We are clearly interested in the continuum limit of this model. Therefore, in order to obtain the large $N$ behavior of the model, we need the power counting of $N$ for a given connected diagram. (Notice that $W(N,t) = \log Z(N,t)$ will generate all the connected diagrams.) Each vertex contributes a factor of $N$, each edge (propagator) a factor of $N^{-1}$ since the propagator is the inverse of the quadratic term, and each face a factor of $N$ due to the trace. Thus each connected diagram has an overall factor $N^{V-E+F} = N^{2-2h} = N^{\chi}$ (where $\chi$ is the Euler characteristic of the Riemann surface) and as a result $W(N,t)$ may be expanded as

$$W(N,t) = \sum_{h=0}^{\infty} N^{2-2h} W_h(t) \ , \tag{7.1.2}$$

with

$$W_h(t) = \sum_{n=0}^{\infty} t^n S(h,n) \ , \tag{7.1.3}$$

where $S(h,n)$ is the number of all possible quadrangulations of a surface of genus $h$ with $n$ squares.

Suppose we consider now, independently, a $D = 0$-dimensional string theory, that means a pure theory of surfaces with no coupling to additional "matter" degrees of freedom on the string worldsheet. The remarkable result is that if one computes the corresponding regularized partition function obtained by discretizing the Riemann surfaces it turns out that this can be identified with the free energy of the matrix model. Moreover one can recover the string theory in a suitably defined continuum limit, known as the **double scaling limit**.

In the continuum limit we are clearly interested in the case of an infinite number of squares, that is the limit in which the infinite tessellations dominate the sum (7.1.3). This occurs when $t \to t_c$ for which (7.1.3) ceases to converge. Given the large $n$ behaviour of $S(h,n) \sim e^{cn} n^{(\gamma-2)\chi/2-1} b_h$, where $c$, $\gamma$, $b_h$ are constants, the genus $h$ contribution starts to diverge when $t \to t_c = e^{-c}$. The double scaling limit is defined then by $N \to \infty$, $t \to t_c$, keeping fixed the "renormalized" string coupling constant $\lambda_s$, where

$$\lambda^{-1} = N \left( \frac{t-t_c}{t_c} \right)^{\frac{2-\gamma}{2}} \ . \tag{7.1.4}$$

Then the continuum limit of $W(N, t)$ will be given by

$$W_{\text{cont}} = \sum_{h=0}^{\infty} \lambda_s^{2-2h} b_h \Gamma\left(\frac{(\gamma - 2)\chi}{2}\right) . \tag{7.1.5}$$

ORTHOGONAL POLYNOMIALS AND THE KDV HIERARCHY

Our aim in this section is to see how in the process of computing the partition function of the one-matrix model the KdV hierarchy and the string equation appear. For this it will be useful to consider a slight generalization of (7.1.1), namely

$$Z(N, t) = \int dM \ e^{-N \operatorname{Tr} V(M, t)} , \tag{7.1.6}$$

where the general potential $V(M, t)$ is given by

$$V(M, t) = \frac{1}{2} \operatorname{Tr} M^2 + \sum_{k=0}^{\infty} t_k \operatorname{Tr} M^{2k} . \tag{7.1.7}$$

In other words we consider tessellations of genus $h$ Riemann surfaces by $2k$-gons of arbitrary $k$. This model is invariant under the unitary group $U(N)$ which acts like $M \mapsto U M U^\dagger$, leaving $dM$ and $V(M, t)$ unchanged. This allows one to bring any matrix $M$ to a diagonal form $\Lambda = \operatorname{diag}(\lambda_1, \lambda_2, \ldots, \lambda_N)$, such that the partition function becomes

$$Z(N, t) = \frac{\Omega_N}{N!} \int_{\mathbb{R}^N} \prod_{i=1}^{N} d\mu(\lambda_i) \prod_{i<j} (\lambda_i - \lambda_j)^2 , \tag{7.1.8}$$

where the new measure is given by $d\mu(\lambda_i) = d\lambda_i e^{-NV(\lambda_i, t)}$, with $V(\lambda_i, t) = \frac{1}{2}\lambda_i^2 + \frac{1}{2} \sum_{k=0}^{\infty} t_k \lambda_i^{2k}$ and $\Omega_N = \operatorname{Vol} U(N)$. This integral can be nicely factorized by introducing the **orthogonal polynomials** $\psi_n(\lambda)$, $n = 0, 1, 2, \ldots$ corresponding to this measure $\psi_n(\lambda) = \lambda + \ldots$ such that

$$\int d\mu \ \psi_n(\lambda)\psi_m(\lambda) = h_n \delta_{nm} , \tag{7.1.9}$$

obtaining in this fashion

$$Z(N, t) = \Omega_N \prod_{n=0}^{N-1} h_i(t) . \tag{7.1.10}$$

The way to compute the $h_i$'s is by introducing the infinite-dimensional ma-trices $P$ and $Q$ defined by

$$Q_{nm}h_m = \int d\mu \ \psi_n(\lambda)\lambda\psi_m(\lambda)$$

$$P_{nm}h_m = \int d\mu \ \psi_n(\lambda)\frac{d}{d\lambda}\psi_m(\lambda) \ , \qquad (7.1.11)$$

that obviously obey $[P, Q] = 1$, which is the discrete analogue of the string equation.

The remarkable fact is that in the double scaling limit the matrix $Q$ becomes a second-order differential operator of the form

$$Q = \partial^2 + u(x, t) \ , \qquad (7.1.12)$$

where the field $u(x,t) = 2\partial_x W(t)$. This differential operator not only looks very much like the Lax operator of the KdV hierarchy, but indeed satisfies the KdV flows

$$\frac{\partial Q}{\partial t_k} = [Q_+^{k/2}, Q] \ , \qquad (7.1.13)$$

whereas the string equation reads

$$1 = \sum_{k=1}^{\infty} kt_k[Q, Q_+^{k/2-1}] \ . \qquad (7.1.14)$$

What about the continuum limit of the matrix $P$? After much toil it follows that $P$ turns into the differential operator

$$P = -\tfrac{1}{2}(MQ^{-\frac{1}{2}})_+ \ , \qquad (7.1.15)$$

where $M$ is the dressed version $M = \phi\Gamma\phi^{-1}$ of the operator in (5.1.16).

$$\Gamma = \sum_{j=1}^{\infty} jt_j\partial^{j-1} \ . \qquad (7.1.16)$$

This is equivalent to saying that the partition function of the one-matrix model $Z(x,t) = \exp(W(x,t))$ is a $\tau$-function for the KdV hierarchy obeying—as a consequence of the string equation—an infinite set of constraints

$$L_m Z(x,t) = 0 \ , \qquad (7.1.17)$$

for $m \geq -1$, where the operators $L_m$ are differential polynomials in the $t_k$'s satisfying the Virasoro algebra

$$[L_n, L_m] = (n - m)L_{n+m} \; , \tag{7.1.18}$$

for any $n, m \geq -1$ or, more precisely, forming a maximal anomaly-free subalgebra of the Virasoro algebra.

## 7.2. INTEGRABLE HIERARCHIES IN SUPERMATRIX MODELS

### SUPERSYMMETRIC 'MATRIX' MODELS

Given the success of the matrix model approach to noncritical string theory and the most pleasant surprise of its relation with classical integrable hierarchies of the KdV type, it seems natural to try to construct a similar approach to noncritical superstrings. This nevertheless turns out to be a fairly difficult task. The generalization of these techniques to the supersymmetric case is still an open problem and the precise relation, if any, with supersymmetric integrable hierarchies remains elusive.

In order to circumvent the problems encountered in an earlier unsuccessful attempt ([**71**]) to define a theory of noncritical superstrings using supermatrices, a model was proposed in [**72**] in which one does away with the matrices all together, and takes as a starting point the integral over the would-be eigenvalues, which is the supersymmetric analogue of (7.1.8)

$$Z_s(N, t, \tau) \propto \int \prod_{i=1}^{N} d\mu(\lambda_i, \theta_i) \prod_{i<j} (\lambda_i - \lambda_j - \theta_i \theta_j)^2 \; , \tag{7.2.1}$$

where $\theta_i$ are odd variables and the measure is given by $d\mu(\lambda, \theta) = d\lambda \, d\theta \, e^{-V(\lambda, \theta)}$ with the 'potential' $V(\lambda, \theta) = \sum_{k \geq 0} (t_k + \tau_k \theta) \lambda^k$.

By imposing superVirasoro constraints—in analogy with the Virasoro constraints in the one-matrix model—correlation functions and critical exponents were calculated to first order in the topological expansion. Remarkably, they were found to coincide with those of certain superconformal matter coupled to $2-d$ supergravity. Recently, in [**73**], the model was solved for arbitrary genus. In the double scaling limit the analogue of the field $u = 2\partial^2 \log Z$ is now a pair $(u, \xi)$ with $u$ the 'body' of the two-point function of the puncture operator and $\xi$ the first fermionic scaling variable, satisfying

$$\partial^2 \log Z_s = u + \xi \partial^2 \xi \; . \tag{7.2.2}$$

Moreover the fields $u$ and $\xi$ were found to satisfy a whole hierarchy of flows which looked very much like a supersymmetric extension of the KdV hierarchy. Indeed the odd flows on $u$ are trivial

$$\frac{\partial u}{\partial \tau_k} = 0 \quad \forall k \ , \tag{7.2.3}$$

whereas the even flows are those of the KdV hierarchy:

$$\frac{\partial u}{\partial t_k} = R'_{k+1} \tag{7.2.4}$$

$$= \left[ \kappa^2 \partial^3 + 2u\partial + 2\partial u \right] \cdot R_k \ , \tag{7.2.5}$$

where the Gel'fand–Dickey polynomials $R_k = R_k(u)$ are the gradients of the conserved charges of the KdV hierarchy and $\kappa$ is the renormalized string coupling constant. The equality of (7.2.4) and (7.2.5) imply the celebrated Lenard relations between the $R_k$, which can be translated into a recursion relation for the flows:

$$\frac{\partial u}{\partial t_{n+1}} = \left[ \kappa^2 \partial^2 + 2u + 2\partial u \partial^{-1} \right] \cdot \frac{\partial u}{\partial t_n} \ . \tag{7.2.6}$$

Normalizing $R_0 = \frac{1}{2}$, we can compute all the other $R_k$ recursively: $R_1 = u$, $R_2 = \kappa^2 u'' + 3u^2$, and so on. In terms of the $R_k$, the commutativity of the KdV flows translates into

$$\frac{\partial R'_k}{\partial t_n} = \frac{\partial R'_{n+1}}{\partial t_{k-1}} \ , \tag{7.2.7}$$

an identity that, as we will see shortly, implies the invariance of the even flows under supersymmetry. From the analysis in [**73**], $\xi$ is given by

$$\xi = -\sum_{k \geq 0} \tau_k R_k \ , \tag{7.2.8}$$

wherefrom we can read how it evolves along the flows

$$\frac{\partial \xi}{\partial \tau_k} = -R_k \quad \text{and} \quad \frac{\partial \xi}{\partial t_n} = -\sum_{k \geq 0} \tau_k \frac{\partial R_k}{\partial t_n} \ . \tag{7.2.9}$$

The first nontrivial even flows were found in [**73**] to be

$$\frac{\partial u}{\partial t_1} = \kappa^2 u''' + 6uu' \quad \text{and} \quad \frac{\partial \xi}{\partial t_1} = \kappa^2 \xi''' + 6u\xi' \ , \tag{7.2.10}$$

whereas the odd flows were found to be

$$\frac{\partial u}{\partial \tau_1} = 0 \quad \text{and} \quad \frac{\partial \xi}{\partial \tau_1} = -u \ . \tag{7.2.11}$$

Notice that the the first equation in (7.2.10) is nothing but the KdV equation for $u$.

THE SKDV-B HIERARCHY

The purpose of this section is to identify, along the lines in [**69**], the hierarchy found in [**73**] and mention some of its immediate properties: conserved charges, bihamiltonian structure, and integrability. We will conclude that this hierarchy is simply a supersymmetric covariantization of the KdV hierarchy, and as such not very different from it.

It was observed already in [**73**] that the first even flow on $u$ and $\xi$ is invariant under the (global) supersymmetric transformations

$$\delta u = \xi' \quad \text{and} \quad \delta \xi = u \ . \tag{7.2.12}$$

In fact, as we will show in a moment, this continues to be the case for all the even flows. On the other hand, the odd flows are not supersymmetric, for whereas $\xi$ evolves, its supersymmetric partner $u$ does not. Nevertheless, one can modify the odd flows to make them supersymmetric. We will comment on this further on.

PROPOSITION 7.2.13. *The even flows are invariant under (7.2.12), while the odd flows satisfy*

$$\left[ \delta \, , \frac{\partial}{\partial \tau_n} \right] = -\frac{\partial}{\partial t_{n-1}} \ . \tag{7.2.14}$$

PROOF. We first consider the even flows:

$$\left[ \delta \, , \frac{\partial}{\partial t_n} \right] u = \left( \frac{\delta R_{n+1}}{\delta u} \cdot \xi' \right)' - \frac{\partial \xi'}{\partial t_n}$$

$$= -\left( \sum_{k \geq 0} \tau_k \frac{\partial R'_{n+1}}{\partial t_{k-1}} \right)' + \left( \sum_{k \geq 0} \tau_k \frac{\partial R'_k}{\partial t_n} \right)' = 0 \ .$$

Notice that we can rewrite the flows on $\xi$ in a simpler way:

$$\frac{\partial \xi}{\partial t_n} = \delta R_{n+1} \ . \tag{7.2.15}$$

From this, the analog result for $\xi$ follows trivially, because

$$\delta \frac{\partial \xi}{\partial t_n} = R'_{n+1} = \frac{\partial u}{\partial t_n} = \frac{\partial}{\partial t_n} \delta \xi \ . \tag{7.2.16}$$

On the other hand, for the odd flows we obtain for $u$

$$\left[ \delta \, , \frac{\partial}{\partial \tau_n} \right] u = \frac{\partial \xi'}{\partial \tau_n} = -\frac{\partial u}{\partial t_{n-1}} \ , \tag{7.2.17}$$

whereas for $\xi$ one has

$$\left[\delta, \frac{\partial}{\partial \tau_n}\right]\xi = -\delta R_n = -\frac{\partial \xi}{\partial t_{n-1}} \ , \tag{7.2.18}$$

where we have once again used (7.2.15).    □

Since the sKdV-B hierarchy is supersymmetric, one can express its flows in a way that makes this manifest, whereto we introduce the superfield $T = \xi + \theta u$, a function in a $(1|1)$ superspace. In superspace, the supersymmetry algebra is realized as supertranslations, which on superfields look like $\delta T = QT$, where $Q = \frac{\partial}{\partial \theta} - \theta \partial$. We will denote by $D$ the supercovariant derivative $D = \frac{\partial}{\partial \theta} + \theta \partial$ , which anticommutes with $Q$. One can recover the fields $u$ and $\xi$ by taking the appropriate projections: $u = DT|_{\theta=0}$ , $\xi = T|_{\theta=0}$.

Rewriting both equations in (7.2.10) as a single equation on the superfield $T$, we find

$$\frac{\partial T}{\partial t_1} = \kappa^2 T^{[6]} + 6T'T'' \ , \tag{7.2.19}$$

where $^{[\ ]}$ denotes differentiation with respect to $D$. Note that we are using the convention that on a superfield $'$ denotes derivative with respect to $D$, whereas on components it denotes derivative with respect to $\partial$. This should cause no confusion. Now notice that if we differentiate both sides of the equation once more with respect to $D$, we get

$$\frac{\partial T'}{\partial t_1} = \kappa^2 T^{[7]} + 6T'T''' \ , \tag{7.2.20}$$

which is nothing but the KdV equation (*cf.* the first equation in (7.2.10)) for the superfield $T' = u + \theta\xi'$. In fact, as we now show, this continues to be the case for all the other equations of the hierarchy; whence we will be able to conclude that the sKdV-B hierarchy is essentially equivalent to the KdV hierarchy.

This may require some explanation. The abstract KdV hierarchy is defined as the hierarchy of isospectral deformations of the Lax operator $L = \kappa^2 \partial^2 + u$, where $u$ is simply a commuting variable generating a differential ring. Particular representations of this abstract KdV hierarchy are obtained by letting $u$ be, for instance, a smooth function on the circle or a rapidly decaying smooth function on the real line. A more exotic representation can be defined by taking $u$ to be an even superfield, say, $T'$. We claim that the hierarchy so obtained is precisely sKdV-B. For notational convenience we will denote by KdV($T'$) the KdV hierarchy with $T'$ as the basic variable, and reserve KdV for when the basic variable is $u$.

Consecutive flows in both the KdV($T'$) and sKdV-B hierarchies are related by a recursion relation. This means that knowing the first flow one can obtain all the others by repeated application of a recursion operator. We have seen that the first flows of both hierarchies agree, thus all we need to show in order to prove the equivalence is that the recursion operators are the same.

The recursion relation for the flows of the KdV($T'$) hierarchy can be read off from (7.2.6) and is given by

$$\frac{\partial T'}{\partial t_{n+1}} = \left[ \kappa^2 \partial^2 + 2\partial T' \partial^{-1} + 2T' \right] \cdot \frac{\partial T'}{\partial t_n} . \qquad (7.2.21)$$

Stripping off a $D$ from both sides, we can rewrite this as

$$\frac{\partial T}{\partial t_{n+1}} = \left[ \kappa^2 \partial^2 + 2DT'D^{-1} + 2D^{-1}T'D \right] \cdot \frac{\partial T}{\partial t_n} , \qquad (7.2.22)$$

which in components reads

$$\begin{pmatrix} \frac{\partial \xi}{\partial t_{n+1}} \\ \frac{\partial u}{\partial t_{n+1}} \end{pmatrix} = \begin{pmatrix} \kappa^2 \partial^2 + 2u + 2\partial^{-1}u\partial & 2\partial\xi\partial^{-1} - 2\partial^{-1}\xi\partial \\ 0 & \kappa^2 \partial^2 + 2u + 2\partial u\partial^{-1} \end{pmatrix} \cdot \begin{pmatrix} \frac{\partial \xi}{\partial t_n} \\ \frac{\partial u}{\partial t_n} \end{pmatrix} , \quad (7.2.23)$$

and this, in turn, agrees with the recursion relation (40) in **[73]**. Thus, we conclude that the flows of the two hierarchies agree.

BIHAMILTONIAN STRUCTURE AND INTEGRABILITY

It was shown in **[32]** that sKdV-type reductions of the SKP$_2$ hierarchy are bihamiltonian: the two structures being given by the supersymmetric analogs of the Gel'fand–Dickey brackets constructed in **[18]**. In particular, the hierarchy associated to the operator $D^4 + U_1 D^3 + U_2 D^2 + U_3 D + U_4$ is bihamiltonian, and so is its reduction $U_1 = U_2 = U_4 = 0$ to sKdV. It would thus seem reasonable to expect that the sKdV-B hierarchy, which is obtained as the reduction $U_1 = U_2 = U_3 = 0$ and $U_4 = T'$, would inherit a bihamiltonian structure in this fashion. However, this turns out not to be the case: it is easy to show that setting $U_1 = U_2 = U_3 = 0$ collapses the rest of the phase space.

We can nevertheless exhibit a bihamiltonian structure for sKdV-B exploiting its equivalence with KdV($T'$). We first rewrite the analogs of (7.2.4) and (7.2.5) for KdV($T'$):

$$\frac{\partial T'}{\partial t_k} = \partial \cdot \frac{\delta H_{k+1}^{\mathrm{KdV}}}{\delta u} \bigg|_{u=T'} \qquad (7.2.24)$$

$$= \left[ \kappa^2 \partial^3 + 2T'\partial + 2\partial T' \right] \cdot \frac{\delta H_k^{\mathrm{KdV}}}{\delta u} \bigg|_{u=T'} . \qquad (7.2.25)$$

For $H_k^{\text{KdV}} = \int h_k(u)$, we have that

$$
\begin{aligned}
\left.\frac{\delta H_k^{\text{KdV}}}{\delta u}\right|_{u=T'} &= \sum_{i \geq 0} (\partial^i)^* \cdot \left.\frac{\partial h_k}{\partial u^{(i)}}\right|_{u=T'} \\
&= \sum_{i \geq 0} (D^{2i})^* \cdot \frac{\partial h_k}{\partial T^{[2i+1]}} \\
&= -D^{-1} \sum_{i \geq 0} (D^{2i+1})^* \cdot \frac{\partial h_k}{\partial T^{[2i+1]}}
\end{aligned}
$$

Since $h_k(T')$ only depends on the odd $D$-derivatives of $T$ we may add for free the contribution of the even derivatives, and we obtain

$$
\left.\frac{\delta H_k^{\text{KdV}}}{\delta u}\right|_{u=T'} = -D^{-1} \cdot \sum_{i \geq 0} (D^i)^* \frac{\partial h_k}{\partial T^{[i]}} = D^{-1} \cdot \frac{\delta H_k^{\text{sKdV-B}}}{\delta T} \tag{7.2.26}
$$

for $H_k^{\text{sKdV-B}} = \int_B h_k(T')$. We can thus rewrite (7.2.24) and (7.2.25) as follows

$$
\begin{aligned}
\frac{\partial T}{\partial t_k} &= \frac{\delta H_{k+1}^{\text{sKdV-B}}}{\delta T} \\
&= \left[ \kappa^2 \partial^2 + 2D^{-1}T'D + 2DT'D^{-1} \right] \cdot \frac{\delta H_k^{\text{sKdV-B}}}{\delta T} \ .
\end{aligned} \tag{7.2.27}
$$

These equations look already to be in hamiltonian form, with Poisson structures $J_1 = 1$ and $J_2 = \kappa^2 \partial^2 + 2D^{-1}T'D + 2DT'D^{-1}$. Notice that $J_1$ satisfies the Jacobi identities trivially, since it is constant. It may seem at first odd that it is not antisymmetric—but this is nothing new in supersymmetric hierarchies, which can have both even and odd Poisson structures. The second structure $J_2$ may not seem obviously Poisson, but it is not hard to show that the Jacobi identities are satisfied. Notice that $J_2$ also defines odd Poisson brackets which are moreover nonlocal. This is again nothing new in supersymmetric hierarchies: the first Poisson structure of sKdV is also nonlocal; although the flows, just like the ones here, are local. Notice, parenthetically, that as expected $J_2 J_1^{-1}$ coincides with the recursion operator (7.2.22) for sKdV-B.

Finally, notice that $J_1$ can be obtained from $J_2$ by shifting $T' \mapsto T' + \lambda$. Since $J_2$ is Poisson for any $T$, it follows that $J_1$ and $J_2$ are coordinated. Usual arguments now imply that the conserved charges are in involution relative to both Poisson structures. In summary, sKdV-B is an integrable bihamiltonian supersymmetric hierarchy.

### SKDV-B AS A REDUCTION OF SKP-TYPE HIERARCHIES

Since the sKdV-B flows are given by the isospectral deformations of the Lax operator $L = \kappa^2 \partial^2 + T'$, it is easy to see that sKdV-B is but a particular reduction of the $SKP_2$ hierarchy. First of all it is clear that $L$ has a unique square root of the form

$$L^{1/2} = \kappa\partial + \sum_{i \geq 1} A_i(T')\partial^{1-i} \ , \tag{7.2.28}$$

where the $A_i(T')$ are $\partial$-differential polynomials in $T'$. In terms of $L^{1/2}$, the flows defining sKdV-B are given by

$$\frac{\partial L^{1/2}}{\partial t_n} \propto \left[ L_+^{n-1/2}, L^{1/2} \right] \ . \tag{7.2.29}$$

Notice that $L^{1/2}$ is a special case of the Lax operator $\mathcal{L} = \kappa\partial + \sum_{k \geq 1} B_k(T)D^{2-k}$ for the $SKP_2$ hierarchy, which was treated in Section 4.2 ($\kappa$ aside). Here, the $B_k(T)$ are $D$-differential polynomials in $T$. Moreover the $SKP_2$ flows are given by (4.3.13) which agree (after relabeling and rescaling the times) with (7.2.29). In other words, the submanifold of $SKP_2$ operators of the form (7.2.28) is preserved by the $SKP_2$ flows and, moreover, these flows agree with the ones defining sKdV-B.

Furthermore, since the Lax operator $L = \kappa^2\partial^2 + T'$ can be 'undressed', one can map the sKdV-B hierarchy into the even part of the MRSKP hierarchy or, equivalently, the Jacobian SKP hierarchy. To this effect, let us define an element $\phi$ of the Volterra group by $L = \phi\kappa^2\partial^2\phi^{-1}$. In terms of $\phi$, the sKdV-B flows can be written as (up to $\kappa$ factors)

$$\frac{\partial \phi}{\partial t_n} \propto -(\phi\partial^{2n+1}\phi^{-1})_-\phi \ . \tag{7.2.30}$$

This equation is then the one defining the even flows of the SKP hierarchy, when we think of $\phi$ as an element of the larger superVolterra group.

### SOME REMARKS ON ODD FLOWS

Although as proven in Proposition 7.2.13 the odd flows are not supersymmetric, it is possible to modify them in such a way that they are. First of all notice that the explicit expression (7.2.8) of $\xi$ as a function of the odd times and the $R_k$ can only be reconciled with its transformation law (7.2.12) under supersymmetry, if $\tau_1$ transforms under supersymmetry. To see this, let us plug

(7.2.8) into the second equation of (7.2.12):

$$
\begin{aligned}
u &= -\sum_{k\geq 0}(\delta\tau_k)R_k + \sum_{k\geq 0}\tau_k\delta R_k \\
&= -\sum_{k\geq 0}(\delta\tau_k)R_k - \sum_{k\geq 0}\tau_k\frac{\partial}{\partial t_{n-1}}\sum_{\ell\geq 0}\tau_\ell R_\ell \quad \text{by (7.2.15) and (7.2.8)} \\
&= -\sum_{k\geq 0}(\delta\tau_k)R_k - \sum_{k,\ell\geq 0}\tau_k\tau_\ell\frac{\partial R_\ell}{\partial t_{n-1}} \\
&= -\sum_{k\geq 0}(\delta\tau_k)R_k \qquad\qquad\qquad\qquad\qquad \text{by (7.2.7)}
\end{aligned}
$$

which implies that

$$
\delta\tau_k = -\delta_{k,1} \; . \tag{7.2.31}
$$

Consider now the flows given by

$$
D_n \equiv \frac{\partial}{\partial\tau_n} - \tau_1\frac{\partial}{\partial t_{n-1}} \; . \tag{7.2.32}
$$

From (7.2.31) and (7.2.14) it follows that these flows are supersymmetric. It is moreover obvious that they commute with the even flows, and that all $D_{n\neq 1}$ (anti)commute among themselves. The remaining algebra of flows is

$$
D_1^2 = -\partial \quad\text{and}\quad [D_1\,,\,D_n] = -\frac{\partial}{\partial t_{n-1}} \quad \forall n > 1 \; , \tag{7.2.33}
$$

where we have used the fact that $\frac{\partial}{\partial t_0} = \partial$. This defines a supersymmetric extension of the sKdV-B hierarchy by odd flows.

It now remains to find a representation of the above algebra of flows in superspace. The main obstacle lies in that the $D_n$ explicitly depend on $\tau_1$ which, as (7.2.31) suggests, should be represented as $-\theta$. It is easy to check that the representation induced from $\delta \mapsto Q$ and $\tau_1 \mapsto -\theta$ is inconsistent, and we have thus far been unable to find a consistent superspace representation for the odd flows.

Alternatively one could try to induce odd flows via the embedding of the sKdV-B hierarchy into the even part of (a reduction of) the Jacobian SKP hierarchy. Nevertheless, via (7.2.30), we can understand these flows as flows in the superVolterra group. It is easy to see that the flows of neither of the two hierarchies preserve the Volterra subgroup where $\phi$ lives.

## 7.3. NEW SUPERSYMMETRIC KDV HIERARCHIES

We have seen in the previous section how the new supersymmetric extension of the KdV hierarchy that has appeared in the context a matrix-model-inspired approach to $2d$ quantum supergravity is but the KdV hierarchy in disguise—the KdV variable being replaced by an even superfield.

This result raises the question whether this supersymmetrization works for all the generalized KdV hierarchies. This question is interesting in view of its applications to noncritical superstrings, as well as from the the general theory of supersymmetric integrable systems. As evinced in Chapter Three, one can actually prove [**70**] that the supersymmetrization in [**69**] works only in the case of the Boussinesq hierarchy, whereas a different and—in a sense—more natural supersymmetrization works for all cases. These more general supersymmetric hierarchies are in a sense not new, since one can prove that they are particular reductions of the known supersymmetric KP hierarchies. Nevertheless their bi-hamiltonian structures do not arise in this way, and the conserved charges are constructed in a novel fashion that has features which make it interesting in its own right.

To understand the idea behind these new supersymmetrizations, let us briefly recall the main features of the generalized KdV hierarchies. The $n$-KdV hierarchy is defined as the isospectral flows of the differential operator $L = \partial^n + \sum_{i \geq 2} u_i \partial^{n-i}$. The flows are given by equations of the form

$$\frac{\partial u_i}{\partial t_j} = P_{ij}(u) \tag{7.3.1}$$

where the $P_{ij}$ are differential polynomials in the $\{u_i\}$. These flows are then extended as evolutionary derivations—*i.e.*, derivations commuting with $\partial$—to the whole differential ring $R[u]$ generated by the $\{u_i\}$. Therefore, formally, the $n$-KdV hierarchy is defined on any differential ring which is freely generated by abstract variables $\{u_i\}$. One can go a long way along this formal path. First of all, one can prove that the flows commute. Furthermore, using the formal calculus of variations, one can then define hamiltonian structures, construct conserved charges in involution, and prove the formal integrability of the hierarchies. It is only when discussing solutions of the evolution equations that one is forced to choose a concrete realization for the differential ring $R[u]$ as a subring, say, of the (rapidly decaying, periodic,...) smooth functions on the real line.

As we have seen, the supersymmetric extension of the KdV hierarchy ($n = 2$) discovered in [**73**]—hereafter referred to as sKdV-B—is obtained by replacing the KdV variable $u$ by the even superfield $U' = u + \theta \xi'$. Since $U'$ freely generates a differential ring, we are well within the domain of the formal KdV hierarchy and, in particular, this means that all the above mentioned results carry over.

There are two caveats, however. First of all, we want to interpret these flows as those from a supersymmetric hierarchy. This means that we cannot simply take the conserved charges to be the ones that would follow from the KdV hierarchy with $U'$ replacing $u$, since these still have $\theta$-dependence. In fact, each KdV conserved charge furnish us now with two conserved charges, since both the $\theta$-dependent and the $\theta$-independent parts are separately conserved. Only one of them, however, is invariant under supersymmetry and is the one that we would understand as the supersymmetric conserved charge. The second caveat is that since it is $\xi'$ that enters in the superfield, the evolution equations will be equations for $\xi'$. We must then make sure that the resulting equations for $\xi$ are indeed local.

As we showed previously neither of these two problems prevent the supersymmetrization of the KdV hierarchy, and one can prove along similar lines that neither are the analogous problems present for the supersymmetrization of the Boussinesq ($n{=}3$) hierarchy. For $n > 3$, however, the resulting equations for the superpartners of the $u_i$ are not in general local and we are forced to conclude that the supersymmetrization does not work.

One may wonder why it is that we replace the $u_i$ by $U'_i$ and not simply by even superfields $V_i = u_i + \theta\sigma_i$. From the point of view of supersymmetric integrable systems, of course, there is no reason not to consider these hierarchies which, in fact, appear much more natural [**70**]. However, they do not seem to the ones that are interesting in view of their applications to superstring theory.